\PassOptionsToPackage{table}{xcolor}

\documentclass[acmtog, nonacm]{acmart}

\usepackage{booktabs}
\usepackage{rotating}
\usepackage{wrapfig,graphicx}
\usepackage{soul}
\usepackage{svg}
\usepackage{epigraph} 
\usepackage{enumitem}
\usepackage{fancyvrb}

\theoremstyle{definition}

\usepackage{multirow}
\usepackage{amsmath}
\usepackage{tabularx}
\usepackage{xcolor}
\usepackage{tcolorbox}
\usepackage{array}

\usepackage{algorithm}
\usepackage{algpseudocode}
\usepackage{subcaption}
\usepackage{microtype}
\usepackage{hyperref}

\DeclareMathOperator*{\argmin}{arg\,min}

\begin{document}

\title[FabHacks]{FabHacks: Transform Everyday Objects into Functional Fixtures}

\author{Yuxuan Mei}
\affiliation{
  \institution{University of Washington}
  \city{Seattle}
  \country{USA}}

\author{Benjamin Jones}
\affiliation{
  \institution{University of Washington}
  \city{Seattle}
  \country{USA}}
  
\author{Dan Cascaval}
\affiliation{
  \institution{University of Washington}
  \city{Seattle}
  \country{USA}}

  \author{Jennifer Mankoff}
\affiliation{
  \institution{University of Washington}
  \city{Seattle}
  \country{USA}}

\author{Etienne Vouga}
\affiliation{
  \institution{The University of Texas at Austin}
  \city{Austin}
  \country{USA}}
  
\author{Adriana Schulz}
\affiliation{
  \institution{University of Washington}
  \city{Seattle}
  \country{USA}}

\renewcommand{\shortauthors}{Mei et al.}

\begin{abstract}
Storage, organizing, and decorating are an important part of home design.  While one can buy commercial items for many of these tasks, this can be costly, and re-use is more sustainable. An alternative is a ``home hack'', a functional assembly that can be constructed from existing household items. However, coming up with such hacks requires combining objects to make a physically valid design, which might be difficult to test if they are large, require nailing or screwing something to the wall, or the designer has mobility limitations.

In this work, we present a design and visualization system for creating workable functional assemblies, \textit{FabHacks}, which is based on a solver-aided domain-specific language (S-DSL) \textit{FabHaL}. By analyzing existing home hacks shared online, we create a design abstraction for connecting household items using predefined types of connections. We provide a UI for FabHaL that can be used to design assemblies that fulfill a given specification. Our system leverages a physics-based solver that takes an assembly design and finds its expected physical configuration. Our validation includes a user study showing that users can create assemblies successfully using our UI and explore a range of designs.
\end{abstract}



\begin{teaserfigure}
  \centering
  \vspace{-10pt}
  \includegraphics[width=\textwidth]{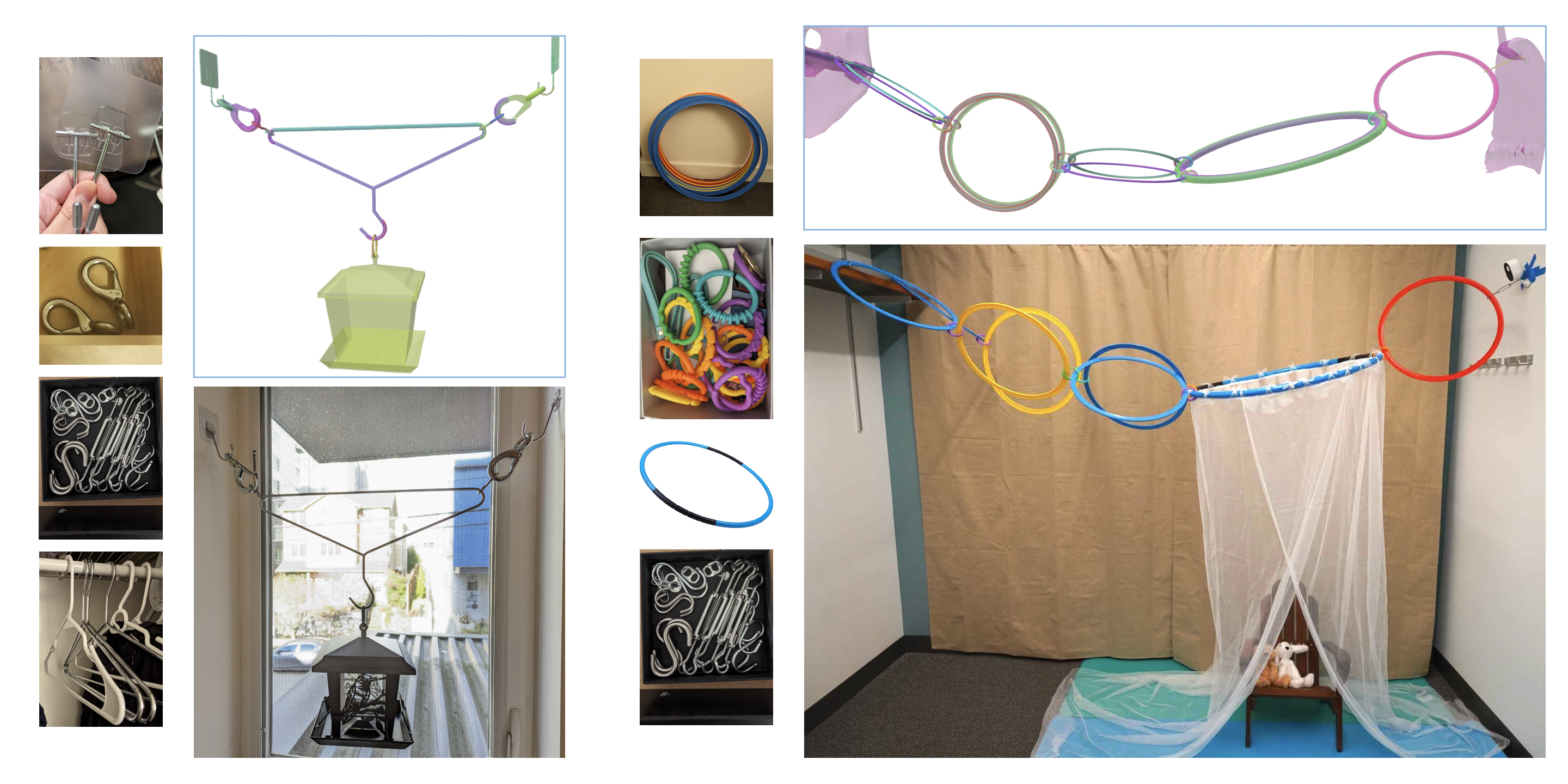}
  \vspace{-20pt}
  \caption{We created FabHacks, a design system for ``home hacks'' built from repurposed everyday objects. The system is built on FabHaL, our domain-specific language for representing rigid fixture hacks. This solver-aided DSL is equipped with verification and solving functionality to help the user finalize their designs. Here we show two hacks, each with the set of everyday items to build it, the solved configuration from our system, and the design fabricated in the real world. Left: the birdfeeder hanging hack made of S-hooks, eyehooks, sticky hooks and a hanger. Right: the reading nook hack made of obstacle rings, toy ringlinks, S-hooks, turnbuckles and a hula hoop; the environment for the reading nook hack was scanned and calibrated with the PolyCam mobile application.}
  \label{fig:teaser}
\end{teaserfigure}

\maketitle
\section{Introduction}\label{sec:intro}

\epigraph{In nature nothing is lost, nothing is created, everything is transformed.}{\textit{Antoine Laurent de Lavoisier}}

Everyday life presents all kinds of challenges that we are constantly trying to solve, from common wear and tear like stubborn stains on the stovetop, a loose outlet, to a cluttered space like a full countertop or a messy desk. As a result, we are bombarded by advertisements for the latest  cleaning or organizational tool or storage solution that promises to improve our lives. It is very tempting to just make one more purchase in a world of next-day-delivery, but this agglomeration of yet more products is wasteful, costly, unsustainable, and oftentimes unnecessary. 

Instead, a thriving subculture is growing on the Internet of sharing ``home hacks'' that repurpose common household items into cost-effective and environmentally-friendly solutions. For example, rubber bands can be used to bind tissues on tongs, which can then be used to clean blinds; or multiple hangers can be chained together with soda can tabs to make more effective use of the vertical space in a closet. We analyzed the space of home hacks (see the full analysis in Appendix~\ref{appendix:analysis}) and found that they can be divided into two categories based on their functionalities. One category, such as the blinds-cleaning tool, makes creative reuse of a single item to change the shape or feel of an existing object,  allowing for better grasping or easier interaction. The other, including the chained hangers, usually involves assembling multiple items into a structure that holds something at a specific location and orientation relative to the environment. We term the latter ``fixture hacks'' because their goal is to build an assembly that can fix some target object in an environment.
Our analysis found that rigid undeformed fixtures (i.e., made up of rigid parts that are combined together but not deformed or modified destructively) are typically used in fixture hacks, so in this work, we focus on this well-scoped subset.

Replicating fixture hacks at home might be straightforward with step-by-step instructions but inventing new hacks requires insight, creativity, experimentation, access to all of the parts, and access to the home environment. Furthermore, fixture hacks often involve multiple objects that interlock and mechanically interact, and gravity can affect a design's stability, making physical prototyping necessary to designing a hack. However, physical prototyping is not always possible, not only for people with limited mobility, but also in situations where not all parts are available, or prototyping would alter one's home permanently. For example, suppose a user wants to add baskets to their closet for storage, and\begin{wrapfigure}{l}{130pt}
    \vspace{-15pt}
    \centering
    \includegraphics[width=150pt]{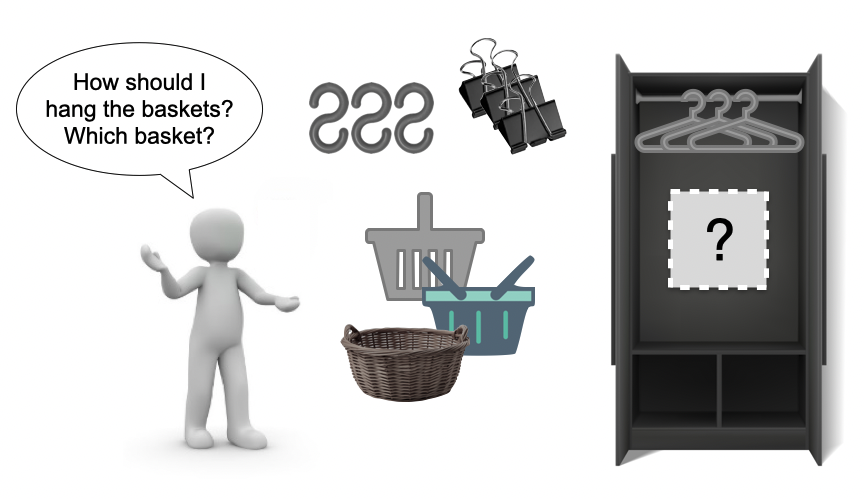}
    \vspace{-23pt}
\end{wrapfigure}wants to hang the baskets using a combination of surplus items such as hangers, hooks, or clips (see inset). Choosing the right basket size and shape to purchase is a chicken-and-egg problem: the user wants confidence that the assembly will fit and hold together, which is difficult to determine without physically prototyping the assembly to discover what works. Even with the right baskets on hand, if setting up the baskets may cause irreversible changes on the wall or closet surface, the user might prefer to have a design that they are satisfied with before drilling a hole. 

A key contribution of this work is a representation of fixture hacks that is well-suited to computational design. We use this representation to build a novel design system \emph{FabHacks} built on top of \emph{FabHaL} (short for FabHacks Language), a Solver-aided Domain-Specific Language (S-DSL) for designing home hacks that allow users to experiment with designs virtually and simulate their designs under gravity.

\begin{figure*}[hbt!]
\centering
\includegraphics[width=0.49\linewidth]{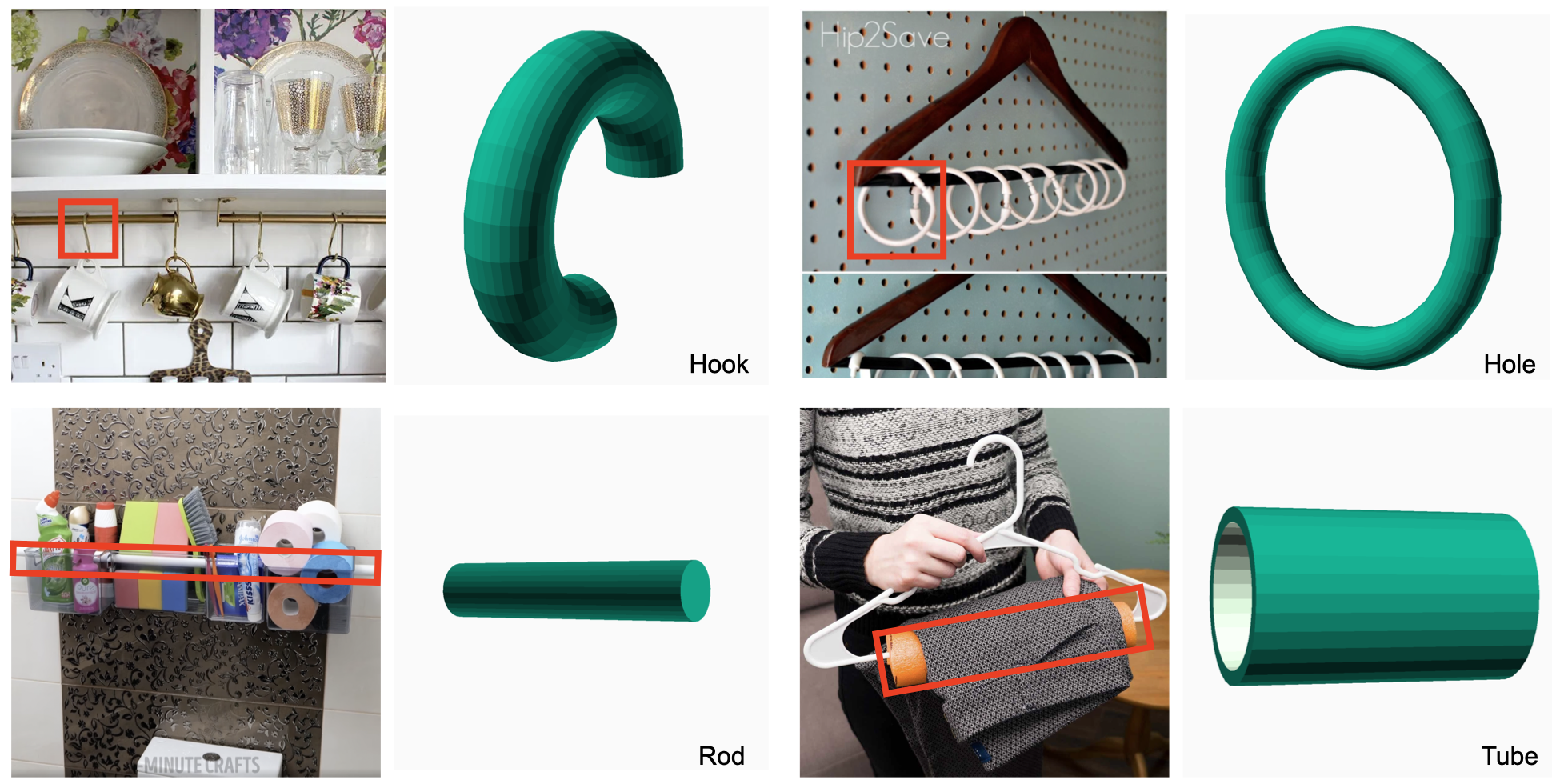}
\includegraphics[width=0.49\linewidth]{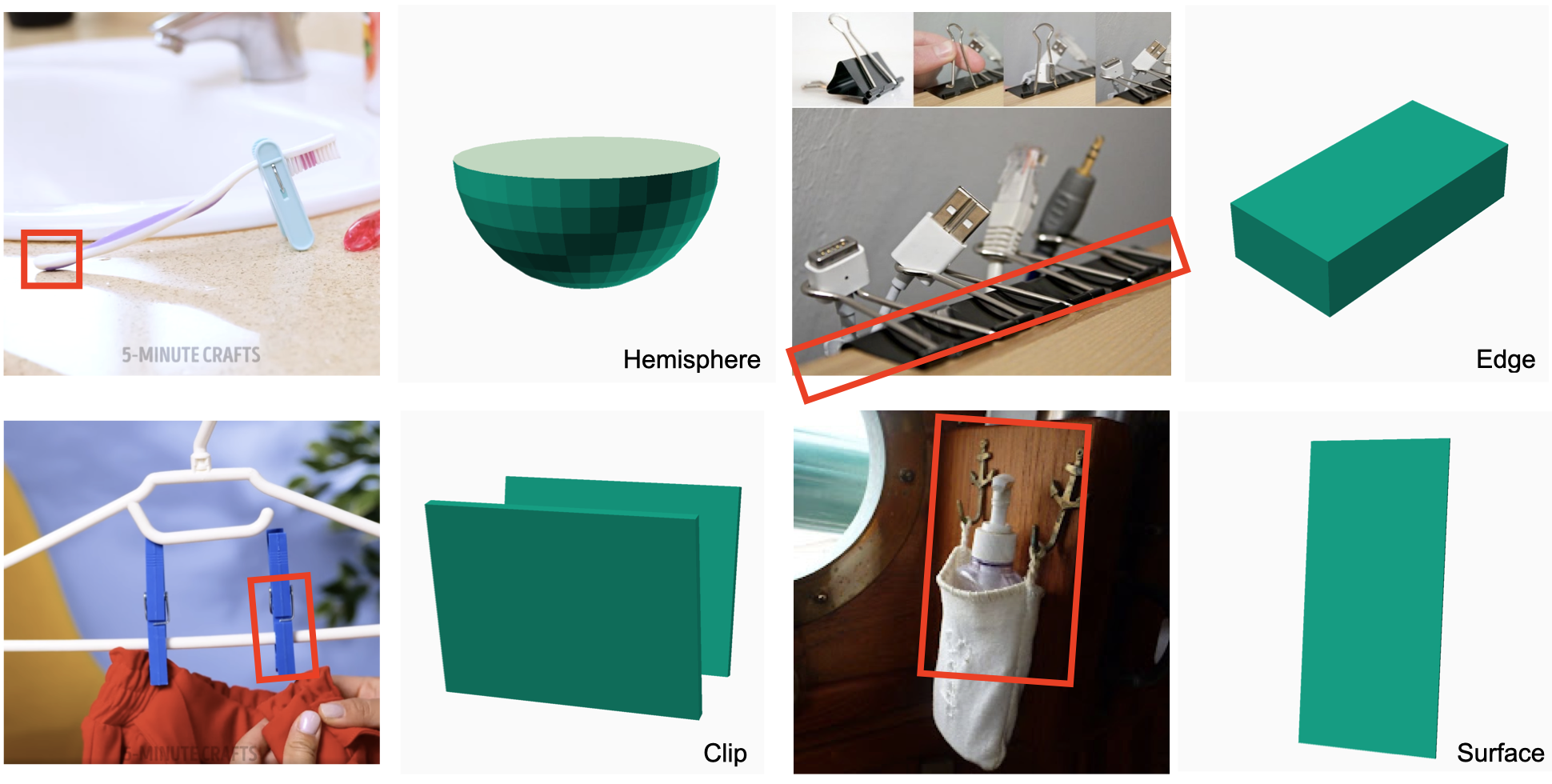}
\vspace{-10pt}
\caption{We analyzed 24 rigid undeformed fixture hacks and extracted eight connector primitive types found on objects in those hacks. Each connector is shown next to an example hack where it appears. The eight example hacks (left to right, top to bottom) are cup hanger (No.24), scarf organizer (No.22), toothbrush holder (No.7), charger holder (No.18), bathroom organizer (No.11), nonslip hanger (No.3), pants hanger (No.8), soap bottle bag (No.1) as numbered in Table~\ref{tab:24hacks}.}
\label{fig:hacks_primitives}
\end{figure*}

The main insight that informed our DSL design is that despite the variety of objects involved in rigid fixture hacks, these objects attach to each other via a small number of common types of \emph{connector primitive} (Figure~\ref{fig:hacks_primitives}) on each object. For example, the handle on a mug, the top hook on a hanger, and the handle on a basket can all be represented using a ``hook'' primitive as shown in Figure~\ref{fig:parametricPrimitives}. The connector primitive concept is thus an important component in FabHaL: these symbolized categories abstract away the complex low-level geometry of each object that is irrelevant to how objects can be combined or to the overall function of the combined assembly.

We also identified rules governing the interaction of pairs of connector primitives. Such behavior is local to the pair of primitives that forms the connection. For example, a hook can slide along a rod \begin{wrapfigure}{l}{120pt}
    \captionsetup{margin=0cm}
    \vspace{-12pt}
    \centering
    \includegraphics[width=140pt]{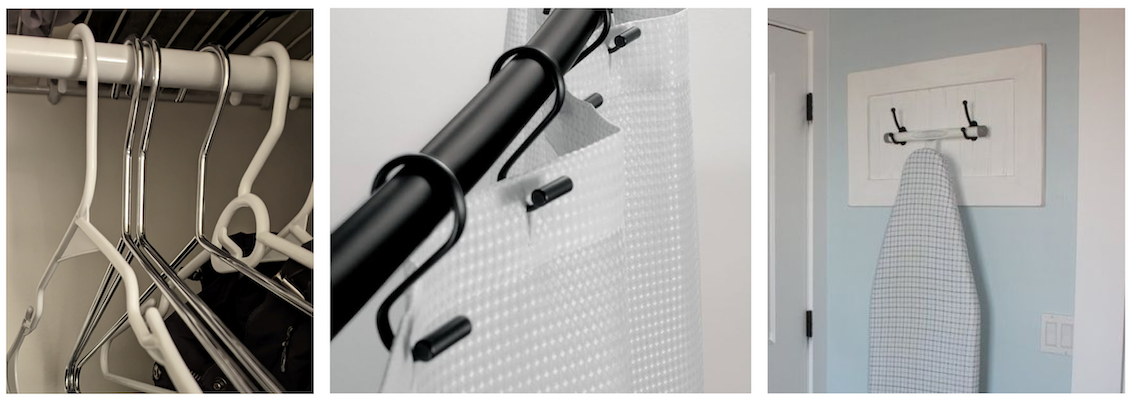}
    \vspace{-20pt}
    \caption{Three examples of rod-hook connections.}
    \label{fig:rodhook}
    \vspace{-15pt}
\end{wrapfigure}and flex about it, and this behavior is valid regardless of whether the rod is part of a closet, a shower, or an ironing board (see inset Figure~\ref{fig:rodhook}). We parameterize all possible ways two connector primitives can touch in our DSL. To further reduce the need for users to reason about low-level details, our DSL also enforces geometric compatibility constraints on two primitives being connected. For example, for a hook to connect to a rod, the hook's hoop radius must be larger than the rod's radius. We discuss in more detail the connection behavior and compatibility constraints for each pair of connector primitive types in Section~\ref{sec:DSL}. 

We assume that the goal of any fixture hack is to hold a \emph{target object} in place at a specific position and orientation in space. We also assume that every hack is attached to a fixed \emph{environment} (table, wall, ceiling hook, etc.). To define an assembly in FabHaL, users specify (1) the environment, target object, and target object's location and orientation; (2) the list of objects that comprise the hack; and (3) which pairs of connector primitives on each part (or target object or environment) should attach to each other. Note that the FabHaL program does not pin down each part's placement in space, and is therefore a \emph{partial specification}~\cite{rosette2013,mangpo2019swizzle} that is later completed by the solver, which solves for the final configurations of all the parts in the hack assembly under the effect of gravity, subject to all of the connection constraints. Our fixture hack representation allows users to experiment with hack designs at a high level, focusing on which connector primitives to connect; the low-level details of \emph{how} they connect are abstracted away by the DSL and handled by the underlying solver. FabHaL as a programming language also supports parametric program creation and (with the help of the solver) users can easily search over a family of designs.

\paragraph{Advantages over CAD} In principle, CAD software can be used to model and test new hack designs. However, modeling complex assemblies with today's computational tools requires a high degree of CAD expertise, and does not allow users to focus on high-level design details. Even after an assembly has been digitized as a CAD model, analyzing whether it holds together under gravity requires other tools. The closest CAD equivalent to our approach is a \textit{mate connector}: a local coordinate system on a CAD part or surface that defines how to orient a part relative to another part or surface. However, mate connectors are usually single-origin coordinate systems and have limited degrees of freedom. They are more suitable for representing a mechanical assembly where parts fit snugly over each other, leaving only a few degrees of freedom for the overall motion of the assembly. In contrast, the everyday hacks that inspired our system (see Figure~\ref{fig:24hacks}) consist of many loose connections, like a hook dangling over a rod, or a ring with a much greater radius than the hook it's attached to. Our connector primitives can more easily represent and model these loose connections, as we establish via a user study. 

Moreover, if the object geometry comes in other formats such as point clouds from scans, voxels, or inaccurate STLs, users need to create a B-rep model from these inputs before they can specify mate connectors in CAD tools. FabHacks can accept any format and simply requires it be tagged with paramaterized connector primitives. For example, in the reading nook hack in Figure~\ref{fig:teaser}, right, we scanned the room and used it as the geometry for the environment in this hack design. We created an OnShape plugin (Figure~\ref{fig:parametricPrimitives}, top) that can be used to tag the connector primitives on geometry that comes in various formats. 

In summary, our contributions include:
\begin{itemize}
\itemsep0em
    \item a representation for fixture hacks based on eight common types of connector primitives extracted from an analysis on the space of home hacks (Appendix~\ref{appendix:analysis});
    \item a solver-aided DSL (Section~\ref{sec:DSL}) that effectively captures the domain knowledge of modeling connections and simulation and thus allows users to focus on the high-level design;
    \item a design interface (Section~\ref{subsec:interactions}) that builds on top of the S-DSL and is validated with user study results (Section~\ref{subsec:userstudy}).
\end{itemize}

We validate our system design with a UI-based user study, where we compare our user interface to a traditional CAD tool for modeling home hacks and find it to be more efficient and intuitive to use, and that it can produce models which are more faithful to the real-life assemblies.
\section{Related Work}

Assembly design is important in manufacturing industries. Various tools have been developed for this task, including computer-aided design tools. Existing CAD tools~\cite{business_onshape_nodate, noauthor_3d_nodate} contain constructs for defining {\em assemblies} of parts using mate constraints. To use these, users must select both the relevant parts of two objects that should be mated and a type of mate; a solver will then yield a valid result under the constraints. This frees the user from having to manually create coordinate systems on part geometries and define constraints on them to position two parts.

However, mates are tricky to reason about: multiple different mate types between two parts could appear to encode the same kinematics, only to be shown different later in the design process when another part is added that further constrains the existing degrees of freedom. 
In addition, mates require the user to carefully specify the position and orientation of each part and tightly align them. Recent research~\cite{jones_automate:_2021} provides mating suggestions but some level of modeling expertise is still required. Moreover, as discussed in the \nameref{sec:intro}, the tightly-aligned constraints between parts common in CAD assemblies do not always adequately model the ``looser'' connection constraints in fixture hacks. 

Lastly, performance analysis is also important during assembly design. Existing CAD tools are primarily concerned with analyzing the kinematics of mechanical assemblies and evaluating whether they achieve the desired concerted motion. In contrast, evaluating the performance of rigid fixture hacks that we focus on means measuring their stability as a hanging assembly under gravity. This type of simulation-based analysis is either completely separate from current CAD design tools or exists with the CAD tool as part of a software suite that requires additional expertise to use.

In this work, we propose FabHacks based on the FabHaL DSL that addresses the specific challenges posed by the domain of fixture hack design. We survey recent research related to our approach.

\paragraph{Solver-aided DSLs}
DSLs have proven effective at abstracting away expert knowledge and allowing non-experts to create valid designs but they are, by definition, designed for a specific domain of applications. Several works~\cite{zhao_robogrammar:_2020, jones_shapeassembly:_2020} have used DSLs for geometric modeling in specific domains like simulated terrestrial robots and cuboid-based 3D shapes; they define a DSL and try to synthesize programs in the DSL given some specific objectives. DSLs can also be used for specifying designs and fabrication plans for carpentry~\cite{wu_carpentry_2019,zhao_co-optimization_2022} and program synthesis techniques can be introduced to help with the optimization.

In this work, we also propose a DSL (FabHaL) specifically for fixture hacks. FabHaL imitates the paradigm of {\em solver-aided languages}, where a user can partially specify a program (vastly reducing the search space) while leaving certain sections (such as expressions, or parameters) abstract~\cite{rosette2013}. An external solver is then invoked to concretize the partially specified program into a complete one, which can then be executed to verify the result. This paradigm has proven useful in a number of domains in the programming languages community such as program deobfuscation~\cite{jha2010oracle}, synthesizing GPU kernels~\cite{mangpo2019swizzle}, and validating and planning biology experiments~\cite{fisher2014biology}. The same technique has also been used in user interface designs~\cite{hottelier_programming_2014} for resolving potential conflicts introduced in the constraints of a layout design and in mathematical diagram designs~\cite{ye_penrose:_2020} for automatically placing visual elements given a user-defined specification. In our case, users can specify the skeleton of connections between primitives while leaving the precise placements of parts to be filled in by a solver and the solver could also provide feedback to users, such as informing the user of whether a connection is valid.

\paragraph{Generative Design of Connectors}
Existing works on modeling connections or creating connections involve generating new connection geometry. Koyama et al.~\shortcite{koyama_autoconnect:_2015} propose a tool for automatically generating structures that could be 3D printed given a user specification to hold or connect two objects. Hofmann et al. ~\shortcite{hofmann_greater_2018} also generate connections between objects and support the specification of assembly information and constraints affecting the assembly, but do not automate solving for those constraints. In addition, both works focus on manufacturing new parts, in contrast to our focus on exclusively reusing existing objects.

\paragraph{Sustainability in Design and Fabrication}
Sustainability considerations have become increasingly prevalent in our everyday lives and in fabrication research communities~\cite{yan_future_2023}. Our work explores the general question of how to fabricate more sustainably. In this space, prior work explored how fabrication can reduce waste through using 3D printing to fix broken objects~\cite{teibrich_patching_2015,lamb_automated_2019} and reusing materials, such as plastic bags~\cite{choi_therms-up!:_2021} and yarns~\cite{wu_unfabricate_2020}. Another line of work is to augment existing objects with fabrication to achieve repurposing~\cite{davidoff_mechanical_2011,ramakers_retrofab:_2016,guo_facade:_2017}, such as by generating structures for re-interfacing with robot arms, legacy physical interfaces, or appliances. Chen et al.~\shortcite{chen_encore:_2015,chen_reprise:_2016,chen_medley:_2018} use 3D printing to augment existing objects with additional functionality (some involving mechanisms), while Arabi et al.~\shortcite{arabi_augmenting_2022,arabi_mobiot:_2022} and Li et al.~\shortcite{li_robiot:_2019,li_romeo:_2020,li_roman:_2022} focus more on augmenting robots using everyday objects or mechanisms to help robots manipulate objects.

Our research looks at how to use rigid everyday objects of any shape without modifications to build a hanging fixture. Our work stands apart from the above literature landscape in that we consider how \emph{multiple} objects fit together into an assembly; the above work instead augment one specific object to allow robotic manipulation or to create a mechanism. (For example, none of the above work could be used to design the hanging bird feeder in Figure~\ref{fig:teaser} that makes use of several different parts.)

\begin{figure*}[ht]
\centering
\includegraphics[width=\linewidth]{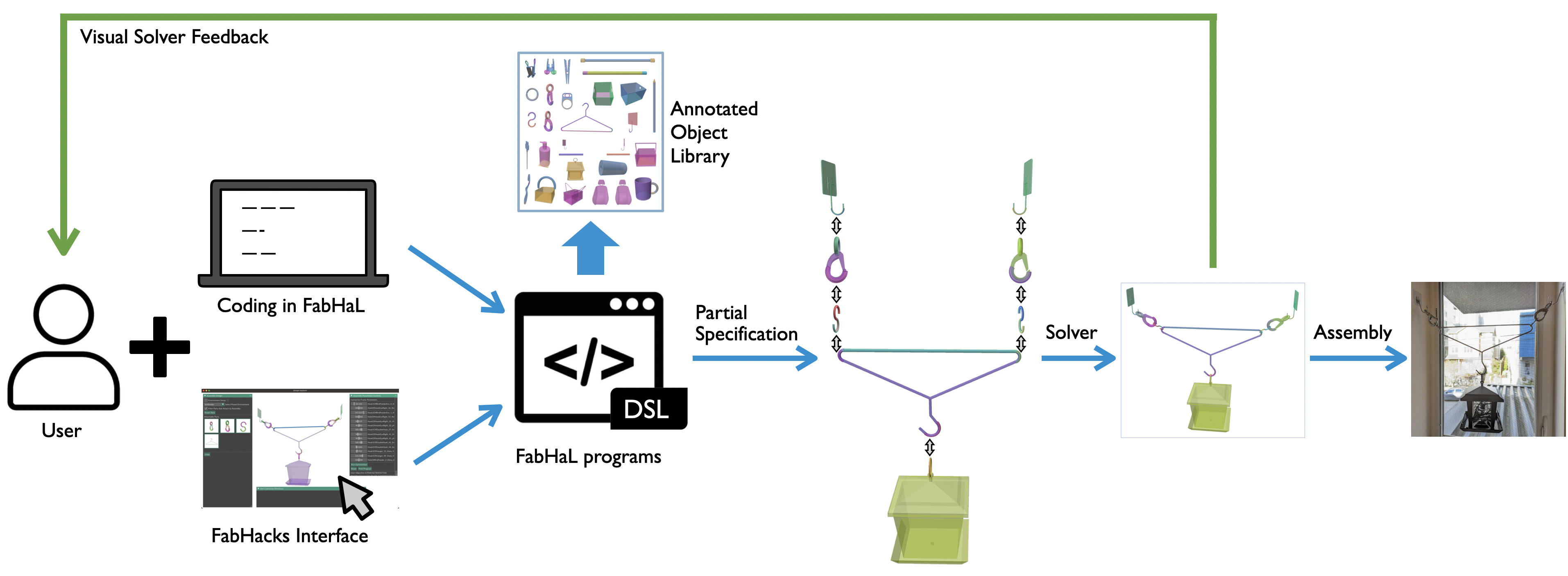}
\vspace{-20pt}
\caption{The overview of the FabHacks system. The user can either directly code in FabHaL or use the UI to create programs. FabHaL programs build on top of an annotated object library and the FabHaL DSL. They are partial specifications of parts and the connections between them, and the 3D configurations of the parts are then completed by the automatic solver. Users can get visual feedback from the program viewer after this solve and use the feedback to iterate on the design. Finally, when satisfied with the design the user fabricates the hack in the real world.}
\label{fig:overview}
\end{figure*}

\section{System Overview}\label{sec:overview}

Since our motivation for the FabHacks system is to allow users to create hack designs without requiring domain knowledge of geometric modeling or simulation, we embed the domain knowledge within the components of our system. Consider as an example a novice user designing a birdfeeder to hang between two hooks using FabHacks (see Figure~\ref{fig:overview}).

\begin{figure}[ht]
\centering
\includegraphics[width=\linewidth]{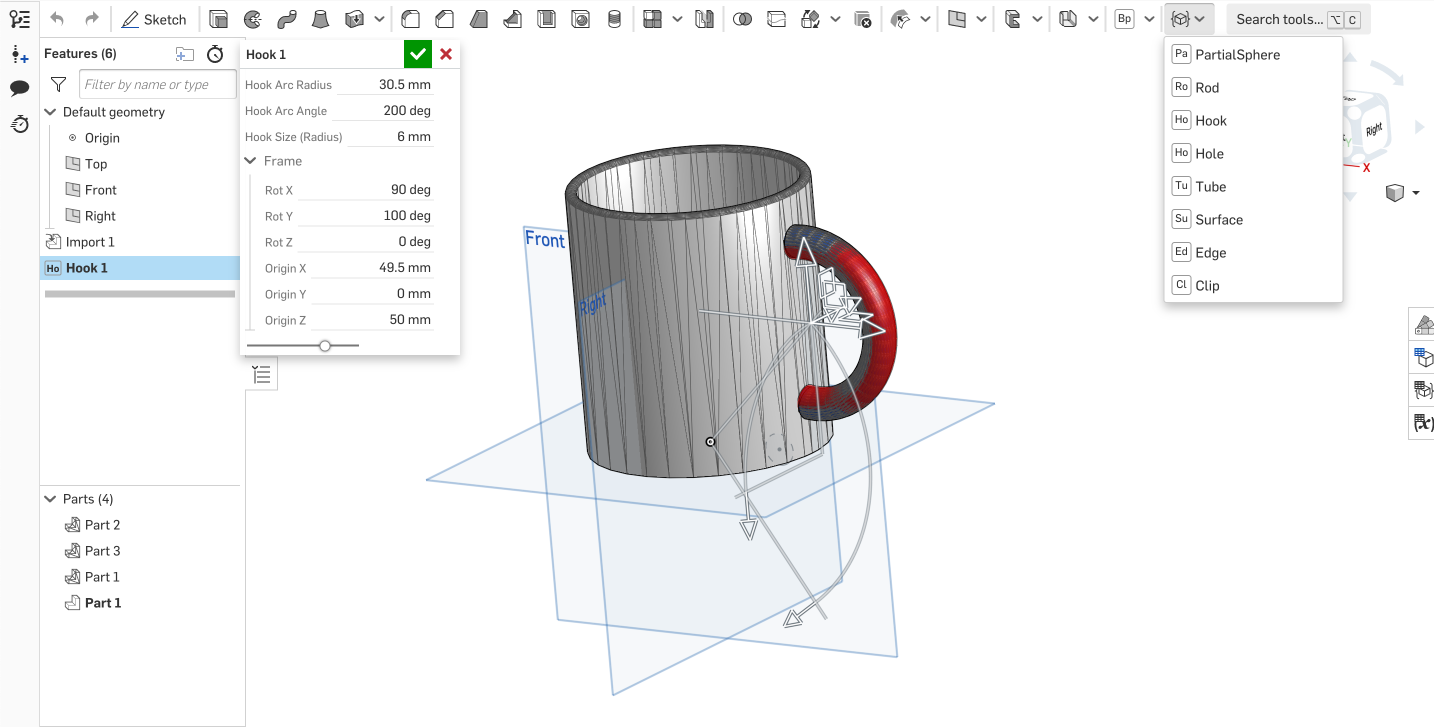}
\includegraphics[width=0.28\linewidth]{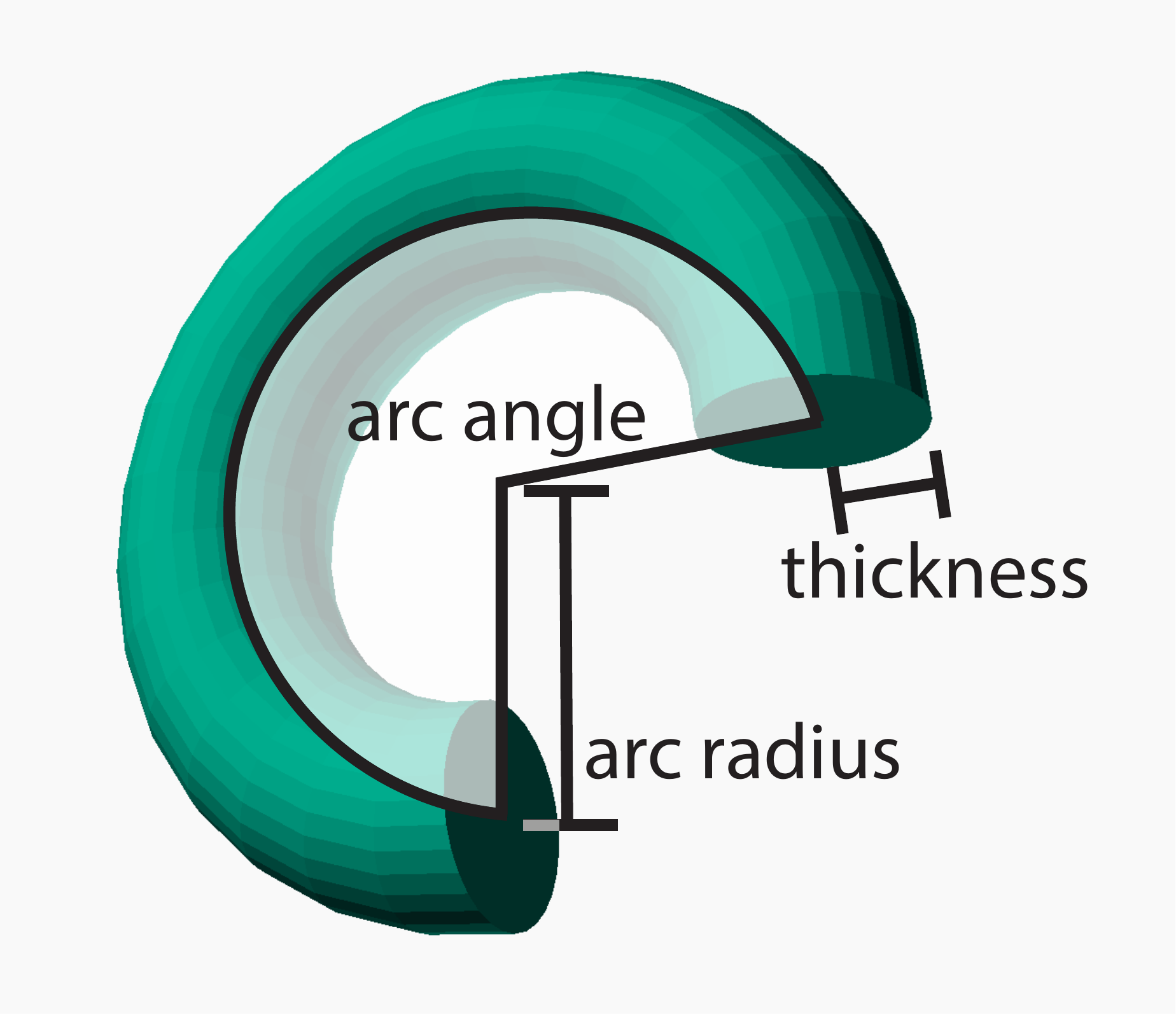}
\includegraphics[width=0.31\linewidth]{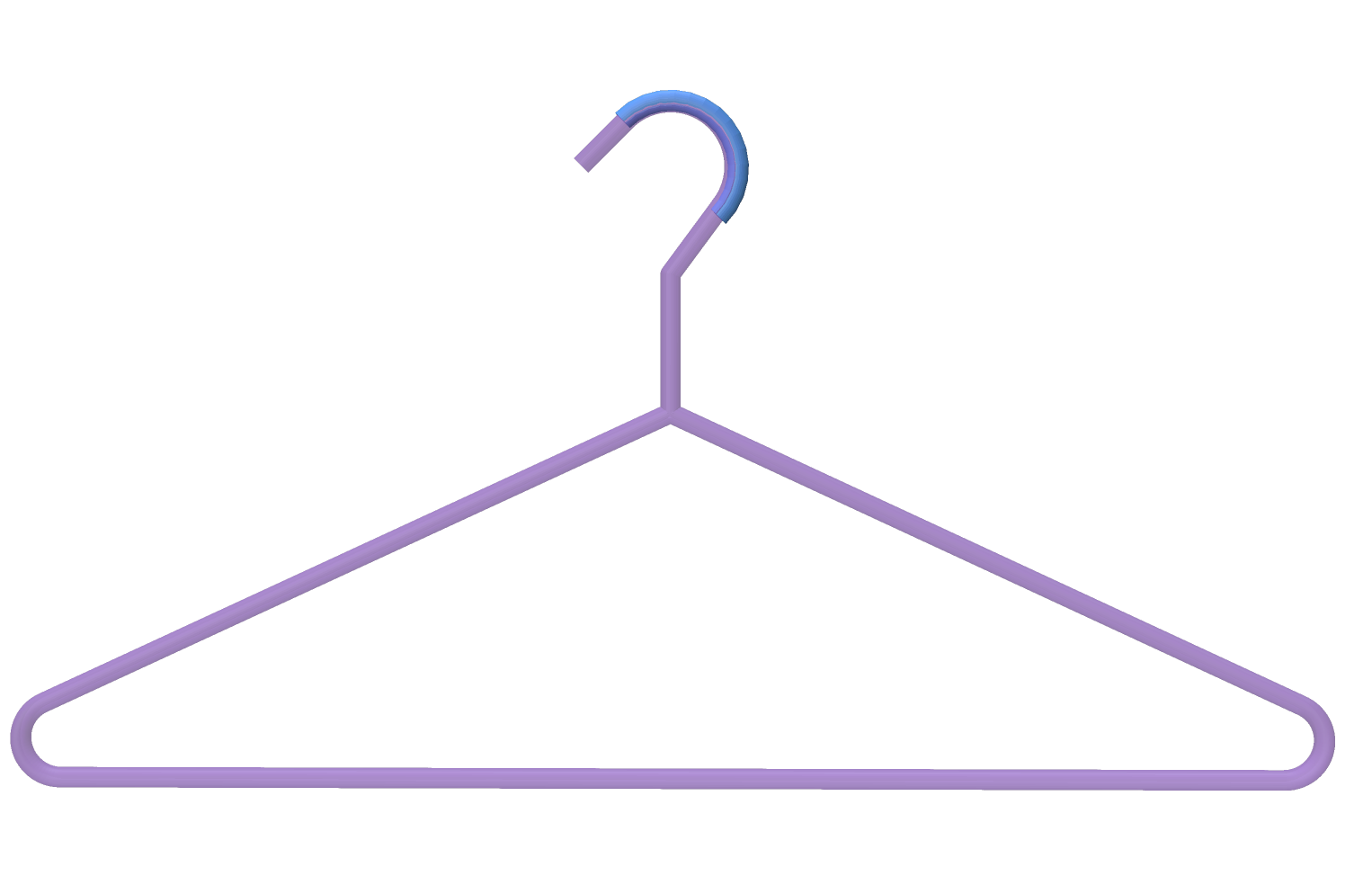}
\includegraphics[width=0.15\linewidth]{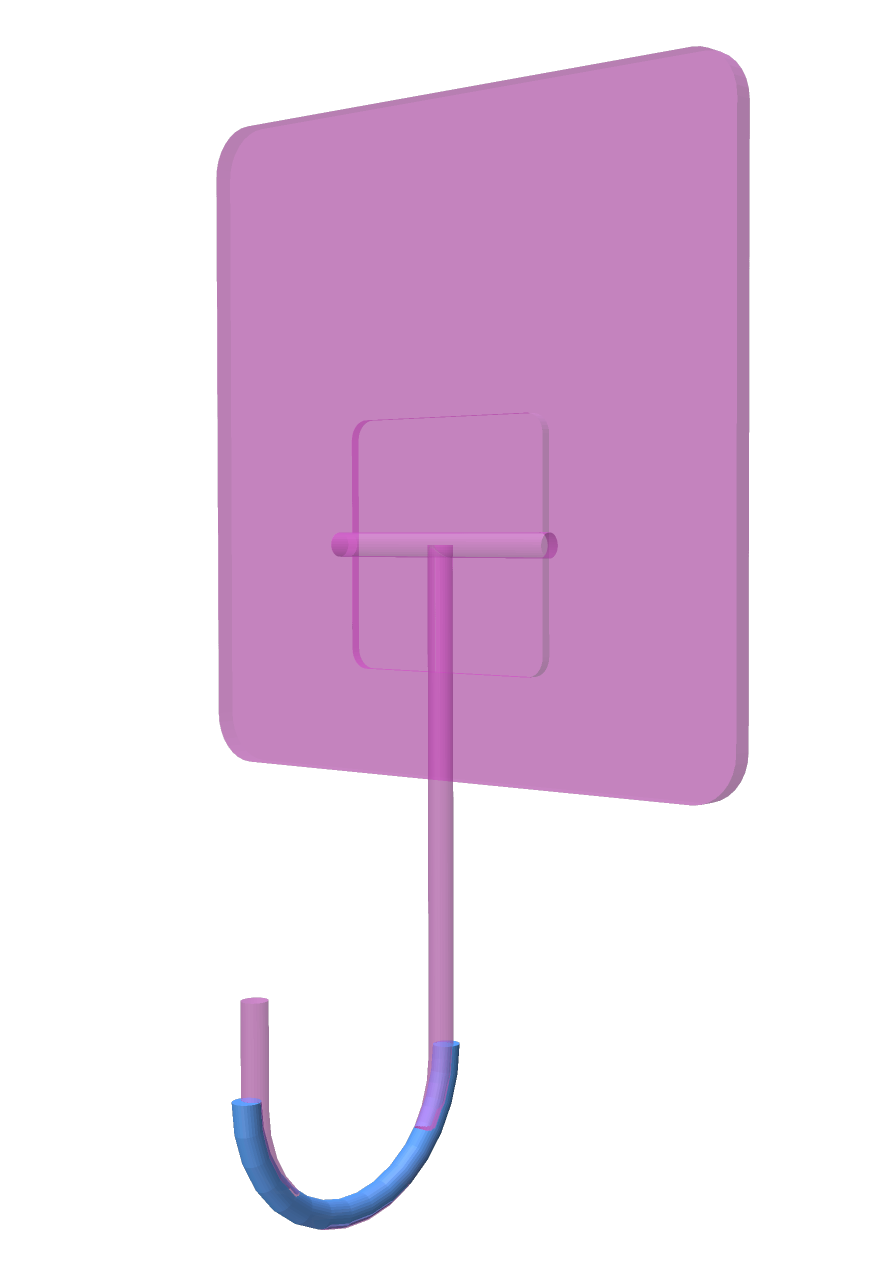}
\includegraphics[width=0.21\linewidth]{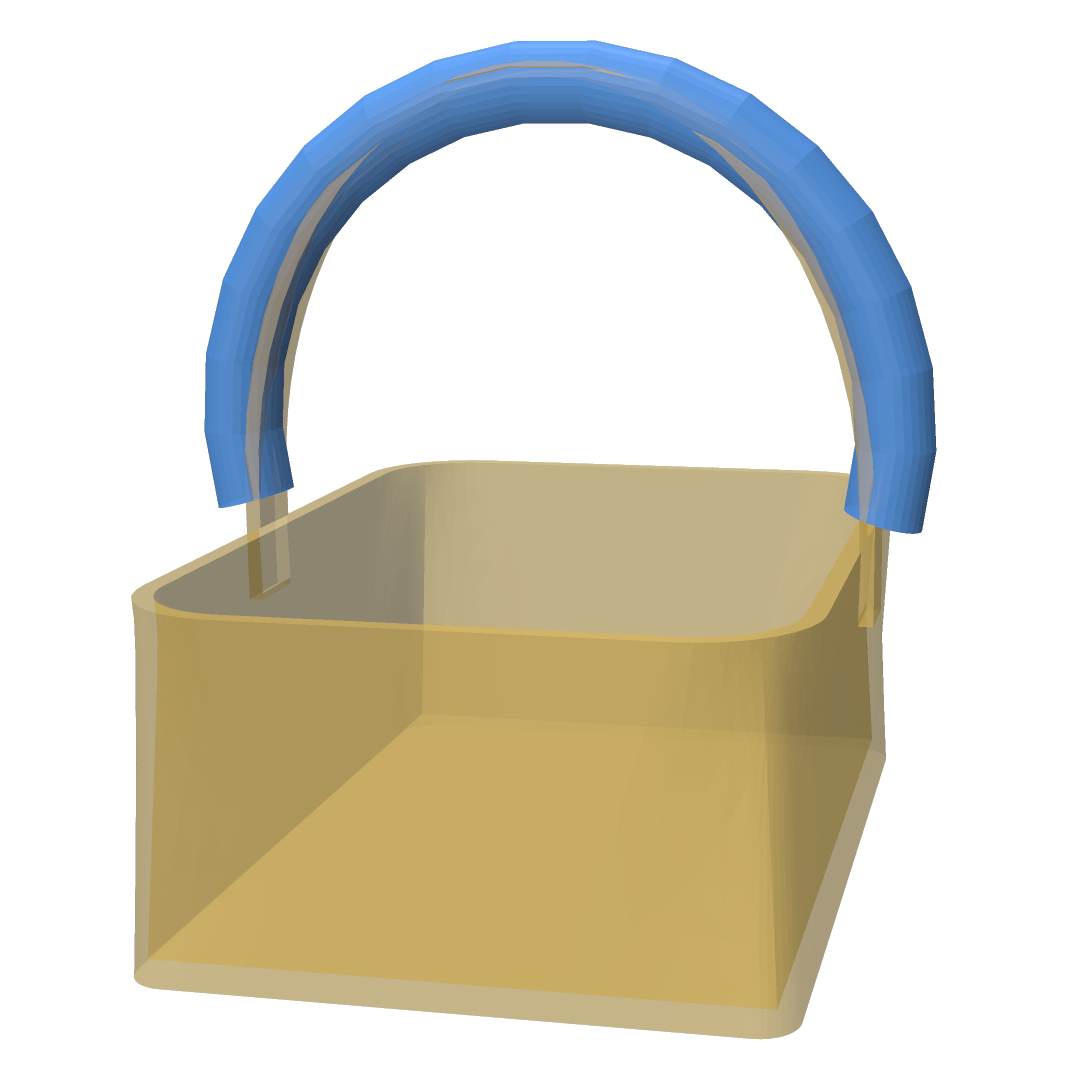}
\caption{Top: the OnShape plugin for tagging 3D models with the eight connector primitives. Bottom: an example showing how we defined the hook shape parametrically with its arc angle, arc radius, and thickness, and three parts tagged with a hook primitive each with different parameters.}
\label{fig:parametricPrimitives}
\end{figure}

\paragraph{Annotated Object Library} First, the user selects from the \emph{Annotated Object Library} the parts they would like to repurpose into their home hack. The Library contains 3D models of a variety of everyday objects, each annotated with the eight types of connector primitives. We call these annotated objects in the Library ``parts''. In addition to labeling regions of a part with a connector primitive type (such as ``hook''), the annotations specify the values of any connector-primitive-type-specific parameters needed to define the geometry of the primitive. For example, we show in Figure~\ref{fig:parametricPrimitives}, bottom, three example parts that have been annotated with a hook primitive, each parametrized to match the exact radius and thickness of the hook geometry in that part. 

We stress that the user does \emph{not} typically need to do any 3D modeling or annotation themselves, but rather can select parts from the predefined library. All examples in this work use a proof-of-concept library of 47 parts: 22 parts that can be used to model the fixed environment or the target object that is to be held fixed in place by the hack, and 25 everyday objects rich in connector primitives that are promising for use as components of a home hack. To build this database we extended the OnShape CAD modeling system's API to support part annotation. Our plugin (see Figure~\ref{fig:parametricPrimitives}, top) allows a user to import a 3D model of a part and add connector primitives. In this way, the Annotated Object Library can be extended (either because the user wishes to include a bespoke part in their hack that doesn't have a good match in the existing library, or as part of a community effort to expand the library). When a primitive is added to a part, parameters are set interactively to ensure the connector aligns with the part. 

\paragraph{FabHaL} Next, the user assembles parts into a hack design using FabHaL, our solver-aided domain-specific language. A FabHaL program is a partial specification of a home hack design: a sequence of instructions to attach a specific connector primitive on one part to a specific connector primitive on another. Users can also write \emph{parameterized} FabHaL programs; for example, they can specify that a hack should include a chain with an unknown number $N$ of links. The solver will search automatically over $N$ for valid hack designs (see Section~{\ref{subsec:programmatic}}). We describe the FabHaL language in more detail in Section~\ref{sec:DSL}.

Users have two ways of interacting with FabHaL to create hack designs: either directly writing programs in the FabHaL language, or by using the FabHacks graphical interface to click on two connectors of two parts to connect them. In either case, note that the user does not need to write down any kinematic constraints: these are inferred automatically by FabHacks from the part annotations.

\paragraph{Solver-aided Evaluation} Finally, the user asks FabHacks to realize the hack design in 3D space using a constrained optimization solver (see Section~\ref{subsec:solver}). Our solver checks whether the part connections are feasible and, if so, relaxes the 3D positions of the parts under gravity and presents the final, solved configuration visually to the user. Problems with the design are also reported to the user (such as infeasible connections or parts that would fall off the assembly if relaxed under gravity). In response to this feedback, the user can make iterative improvements to the FabHaL program and solve again.

\section{An S-DSL for FabHacks}\label{sec:DSL}

\begin{figure}
\centering
\begin{subfigure}{0.7\linewidth}
\centering
\includegraphics[width=\textwidth]{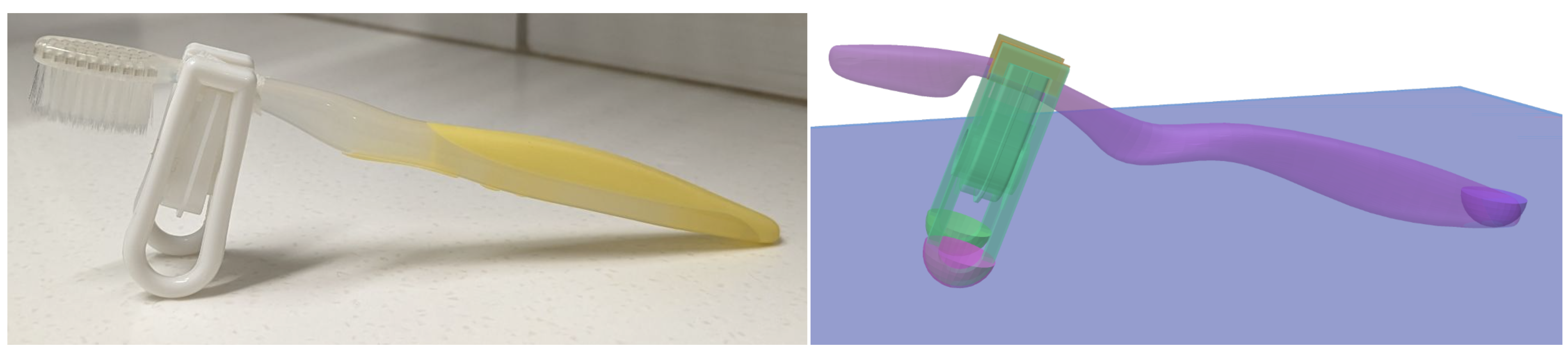}
\caption{Toothbrush Holder}
\label{fig:ex1}
\end{subfigure}
\hfill
\begin{subfigure}{0.25\linewidth}
\centering
\includegraphics[width=\textwidth]{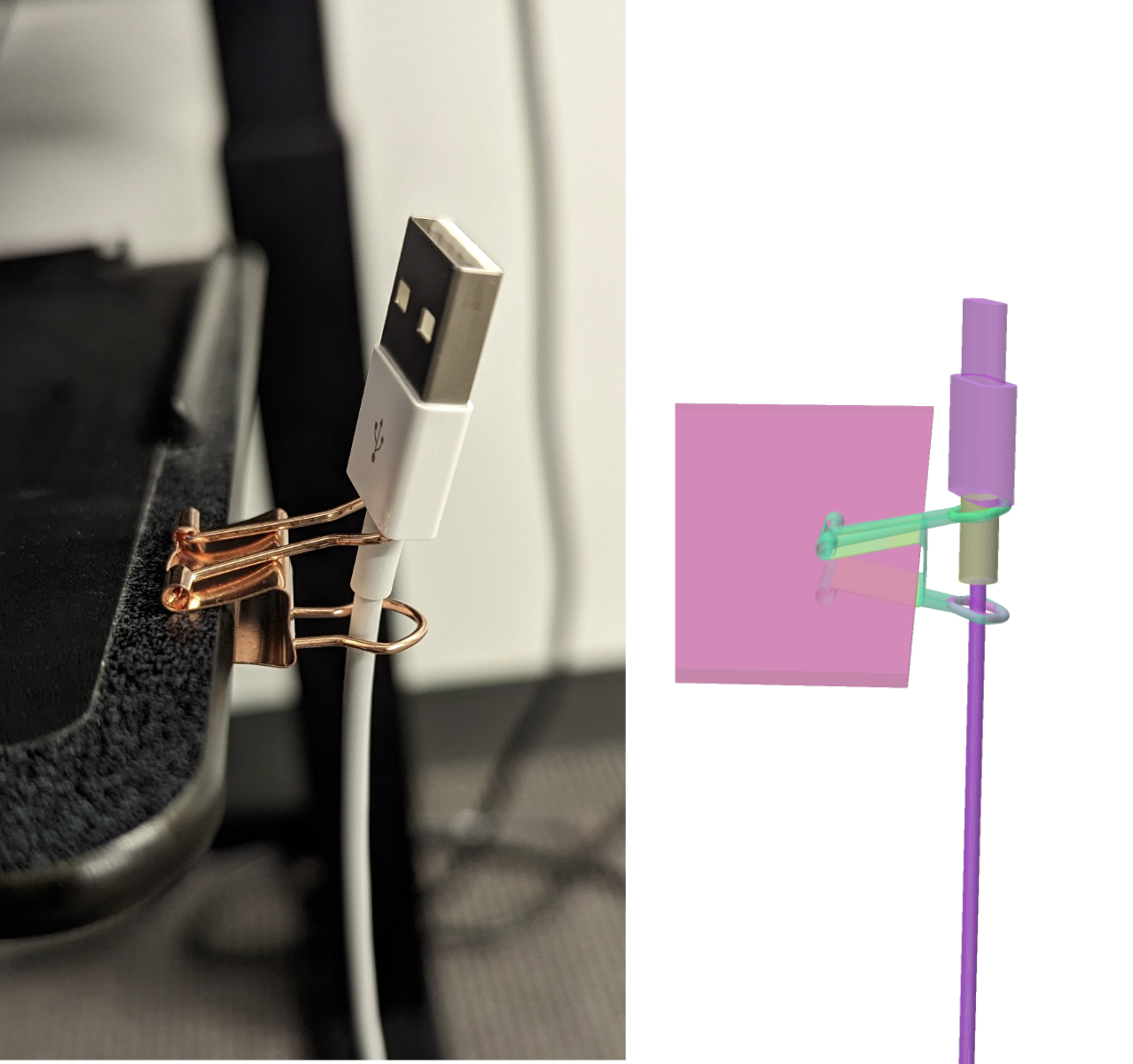}
\caption{Charger Holder}
\label{fig:ex2}
\end{subfigure}
\\
\begin{subfigure}{0.6\linewidth}
\centering
\includegraphics[width=\textwidth]{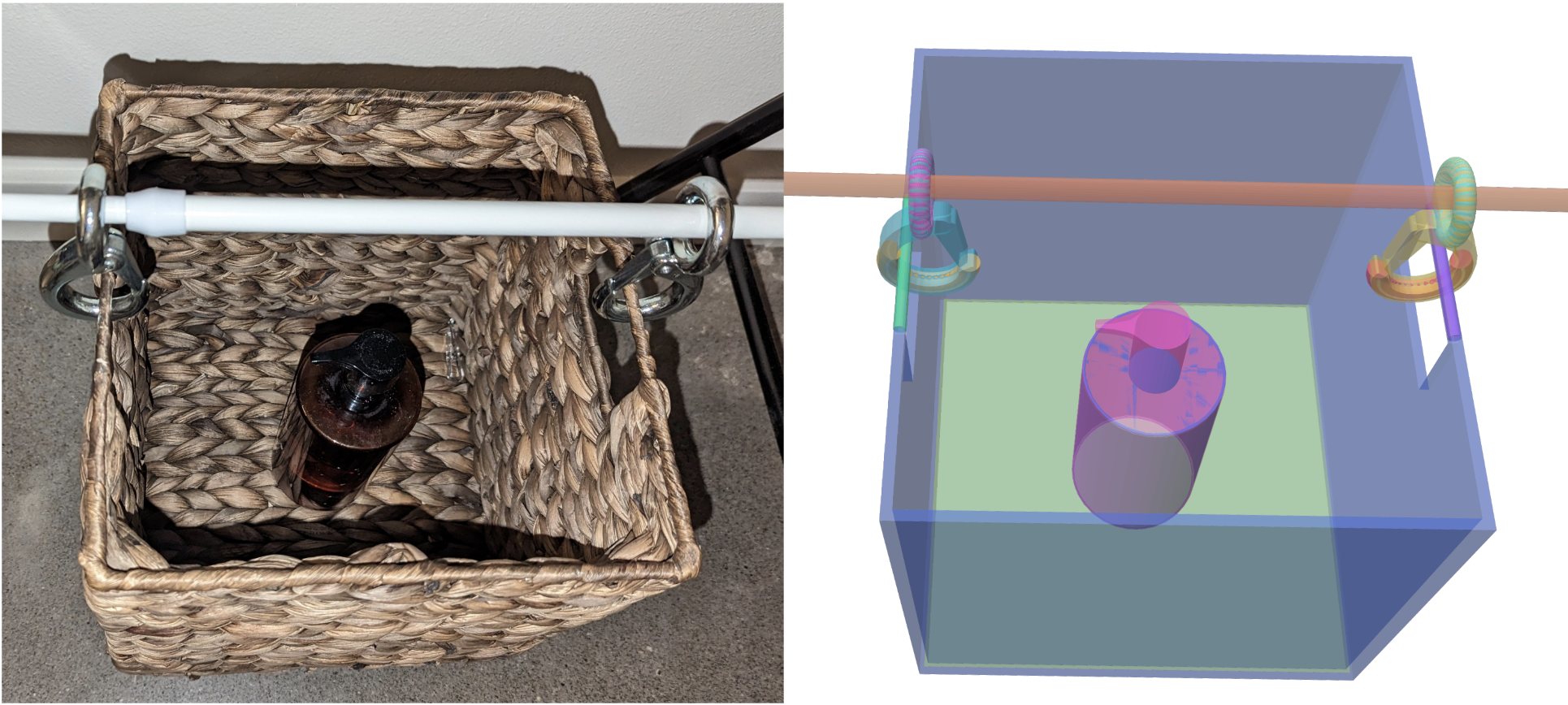}
\caption{Soap Bottle Holder}
\label{fig:ex3}
\end{subfigure}
\hfill
\begin{subfigure}{0.35\linewidth}
\centering
\includegraphics[width=\textwidth]{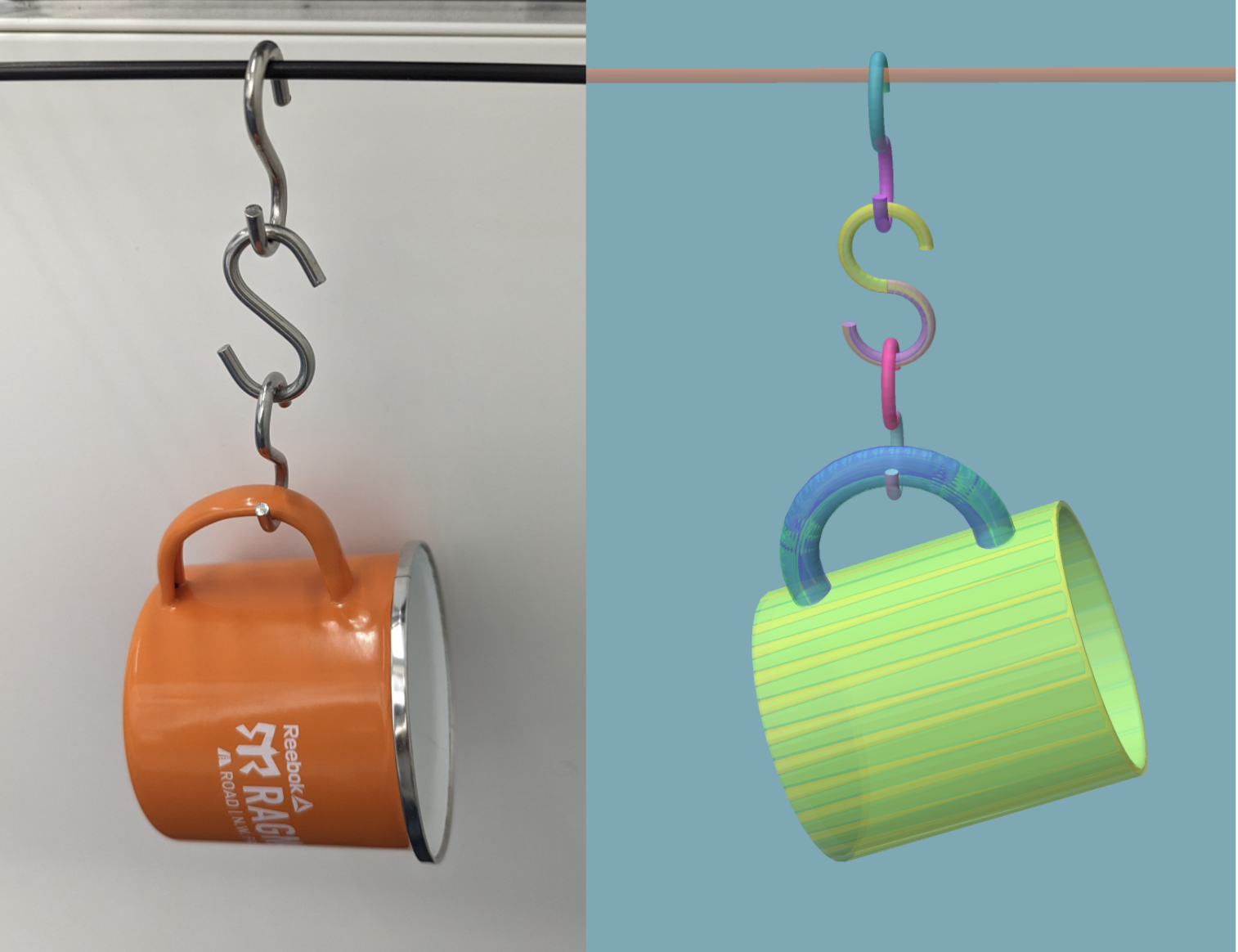}
\caption{Mug Hanger}
\label{fig:ex5}
\end{subfigure}
\\
\begin{subfigure}{0.525\linewidth}
\centering
\includegraphics[width=\textwidth]{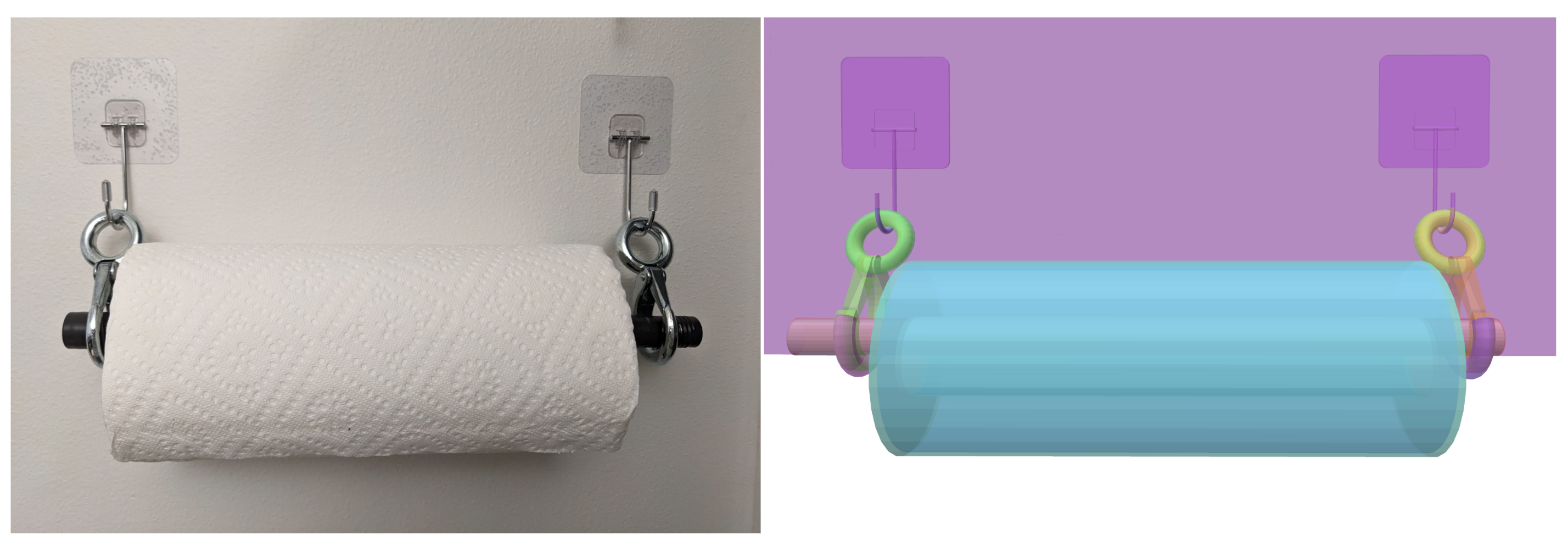}
\caption{Paper Towel Holder}
\label{fig:ex12}
\end{subfigure}
\hfill
\begin{subfigure}{0.425\linewidth}
\centering
\includegraphics[width=\textwidth]{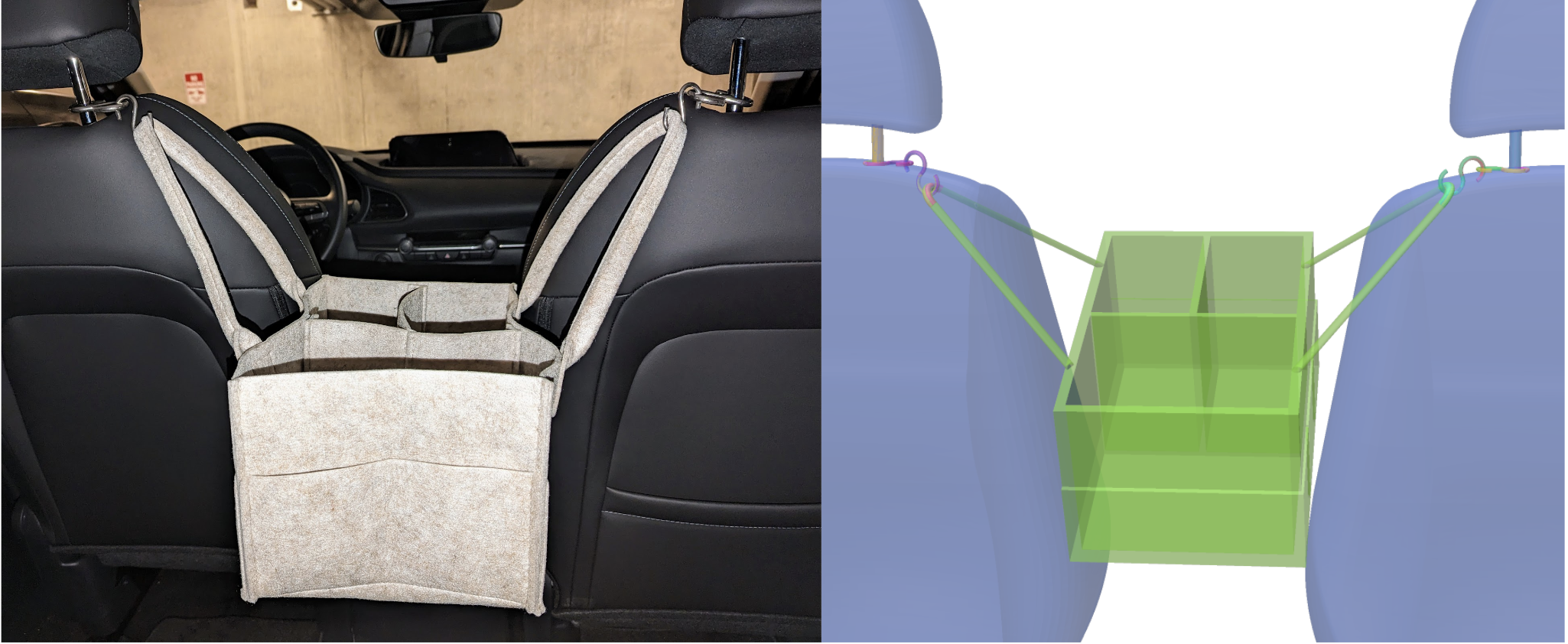}
\caption{Diaper Caddy}
\label{fig:ex13}
\end{subfigure}
\vspace{-5pt}
\caption{Six hacks created by directly programming in FabHaL, with photos and renderings of the corresponding programs (see Appendix~\ref{appendix:programs}) in our viewer.}
\label{fig:gallery}
\end{figure}

In this section, we introduce the S-DSL FabHaL for representing rigid fixture hacks. Figure~\ref{fig:gallery} shows some example designs that can be represented in this language.

The design decisions for FabHaL are two-fold. 

First, the language design of FabHaL is motivated by our analysis of home hacks (see Appendix ~\ref{appendix:analysis}), which found that objects in home ``fixture hacks'' are typically connected via a small number of common shapes, which we term \emph{connector primitives}. We will define connector primitives in detail in Section~\ref{subsec:connprimitives}.

The language design is also guided by our goal to use the DSL as a vessel for domain knowledge. Our hope is that this DSL allows users without any prior experience in modeling or simulation to design fixture hacks. Therefore, being straightforward and succinct is an important desideratum. To achieve this, we choose to introduce a solver to complete a partial specification of the hack design where the user only needs to specify the configuration for a target object and its environment, and which connector primitive connects with which. We will introduce the simple syntax and example usage in Section~\ref{subsec:langconstructs}, and the solver behind it in Section~\ref{subsec:solver}.

Lastly, we discuss the implementation choices and show an example application of a parametrized FabHaL program in Section~\ref{subsec:programmatic}.

\subsection{Connector Primitives}\label{subsec:connprimitives}

\begin{table}[ht]
\centering
\begin{tabular}{ c | c c c c c c c c }
 & hook & hole & hemi. & edge & rod & tube & clip & surf.  \\
 \hline
 hook & \checkmark & \checkmark & \cellcolor[HTML]{C0C0C0}&\cellcolor[HTML]{C0C0C0} & \checkmark & \checkmark & \cellcolor[HTML]{C0C0C0}& \cellcolor[HTML]{C0C0C0}\\
 hole & \cellcolor[HTML]{5A5A5A} & \cellcolor[HTML]{C0C0C0} & \cellcolor[HTML]{C0C0C0}&\cellcolor[HTML]{C0C0C0} & \checkmark & \checkmark & \cellcolor[HTML]{C0C0C0}&\cellcolor[HTML]{C0C0C0} \\
 hemi. & \cellcolor[HTML]{5A5A5A}& \cellcolor[HTML]{5A5A5A}& \cellcolor[HTML]{C0C0C0} & \cellcolor[HTML]{C0C0C0}& \cellcolor[HTML]{C0C0C0}&\cellcolor[HTML]{C0C0C0} &\cellcolor[HTML]{C0C0C0} & \checkmark\\
 edge & \cellcolor[HTML]{5A5A5A}& \cellcolor[HTML]{5A5A5A}& \cellcolor[HTML]{5A5A5A}& \cellcolor[HTML]{C0C0C0}& \cellcolor[HTML]{C0C0C0}&\cellcolor[HTML]{C0C0C0} & \checkmark & \cellcolor[HTML]{C0C0C0}\\
 rod & \cellcolor[HTML]{5A5A5A}& \cellcolor[HTML]{5A5A5A}& \cellcolor[HTML]{5A5A5A}& \cellcolor[HTML]{5A5A5A}& \cellcolor[HTML]{C0C0C0} & \checkmark & \checkmark & \cellcolor[HTML]{C0C0C0}\\
 tube & \cellcolor[HTML]{5A5A5A}& \cellcolor[HTML]{5A5A5A}& \cellcolor[HTML]{5A5A5A}& \cellcolor[HTML]{5A5A5A}& \cellcolor[HTML]{5A5A5A}& \checkmark & \checkmark & \cellcolor[HTML]{C0C0C0}\\
 clip & \cellcolor[HTML]{5A5A5A}& \cellcolor[HTML]{5A5A5A}& \cellcolor[HTML]{5A5A5A}& \cellcolor[HTML]{5A5A5A}& \cellcolor[HTML]{5A5A5A}& \cellcolor[HTML]{5A5A5A}& \cellcolor[HTML]{C0C0C0}& \cellcolor[HTML]{C0C0C0}\\
 surf. & \cellcolor[HTML]{5A5A5A}& \cellcolor[HTML]{5A5A5A}& \cellcolor[HTML]{5A5A5A}& \cellcolor[HTML]{5A5A5A}& \cellcolor[HTML]{5A5A5A}& \cellcolor[HTML]{5A5A5A}& \cellcolor[HTML]{5A5A5A}& \checkmark
\end{tabular}
\caption{A table showing which pairs of primitives can be connected. A checkmark means that connection is currently allowed by the DSL and a light grey cell means it is not. We ignore the lower-triangular region (dark grey) as it is redundant with the upper-triangular region.}
\vspace{-10pt}
\label{tab:connectivity}
\end{table}

\begin{table}[ht]
\centering
\begin{tabular}{c |p{1.5in} | >{\centering\arraybackslash}m{4em}}
    \textbf{Primitive} & \textbf{Shape Parameters} & \textbf{Example}\\
    \hline
    hook & arc angle, arc radius, thickness  & \includegraphics[height=3.5em]{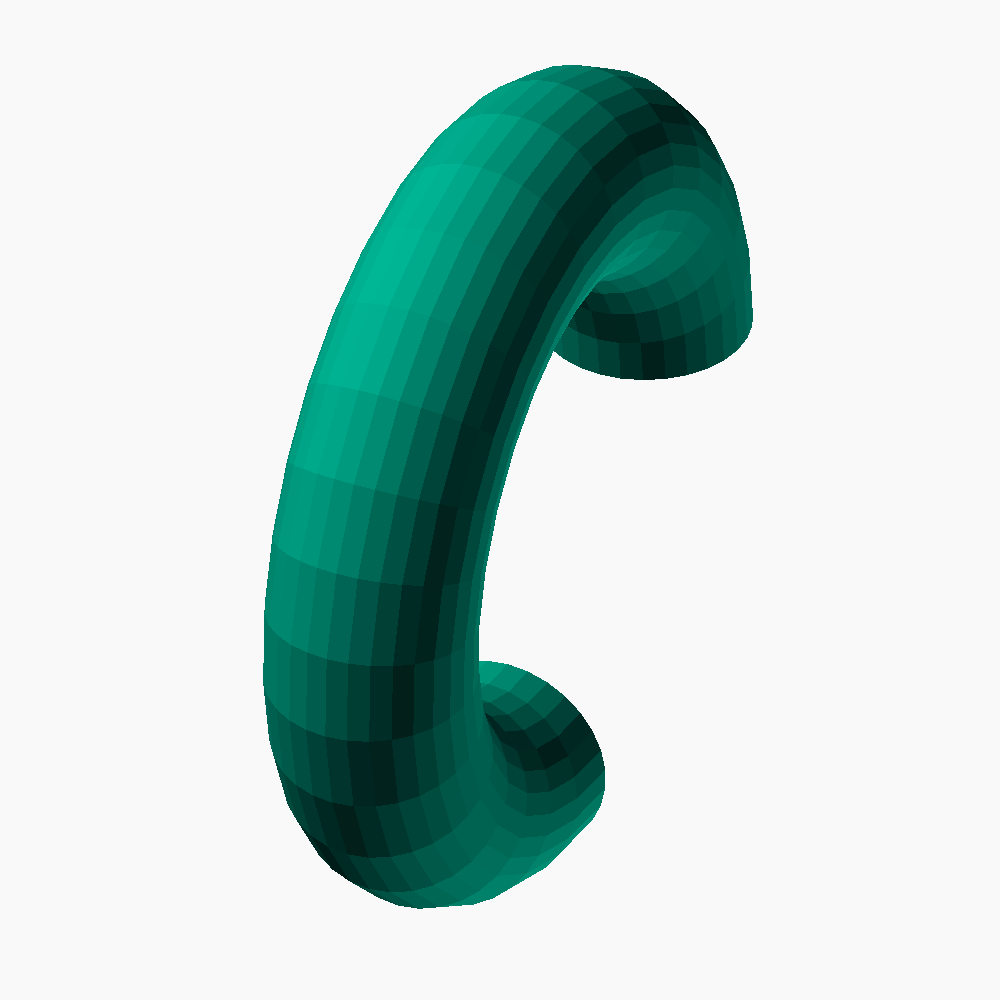}\\
    hole & arc radius, thickness  & \includegraphics[height=3.5em]{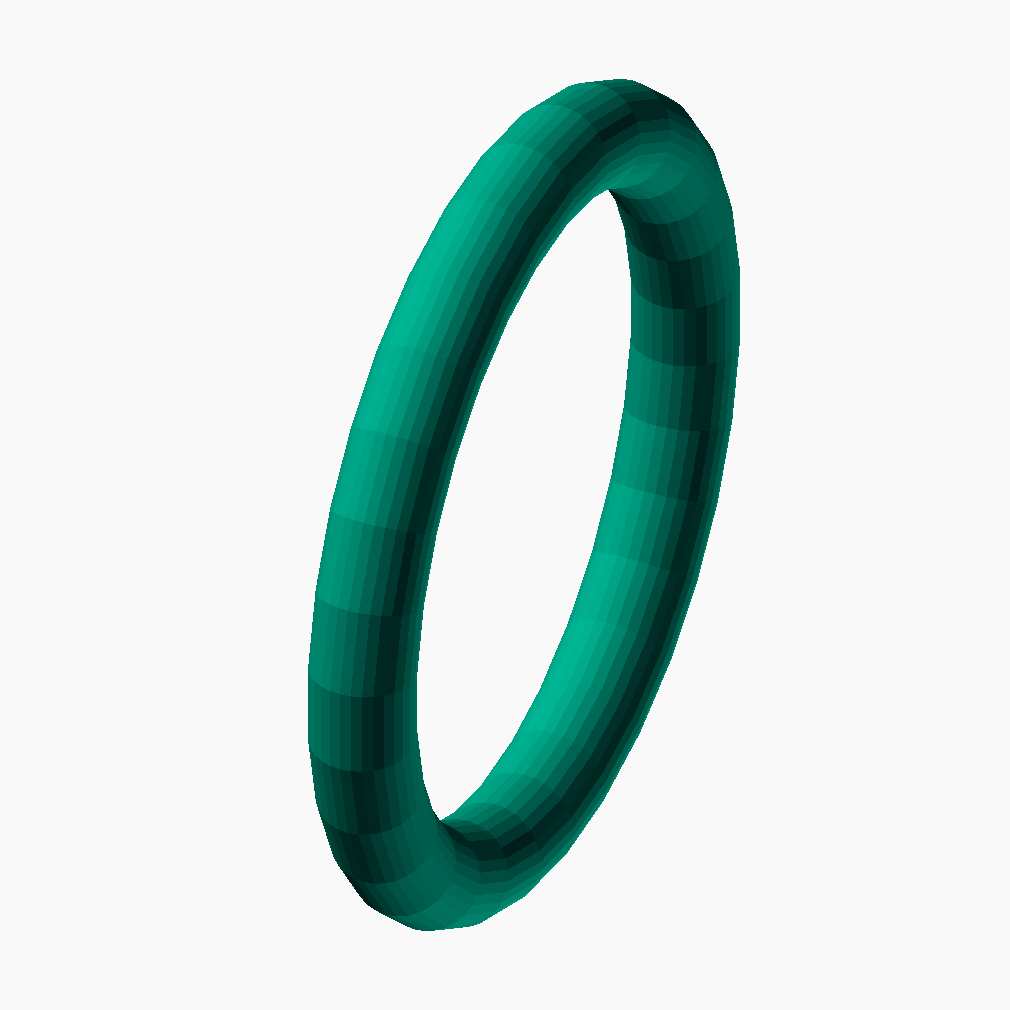}\\
    hemisphere & radius  & \includegraphics[height=3.5em]{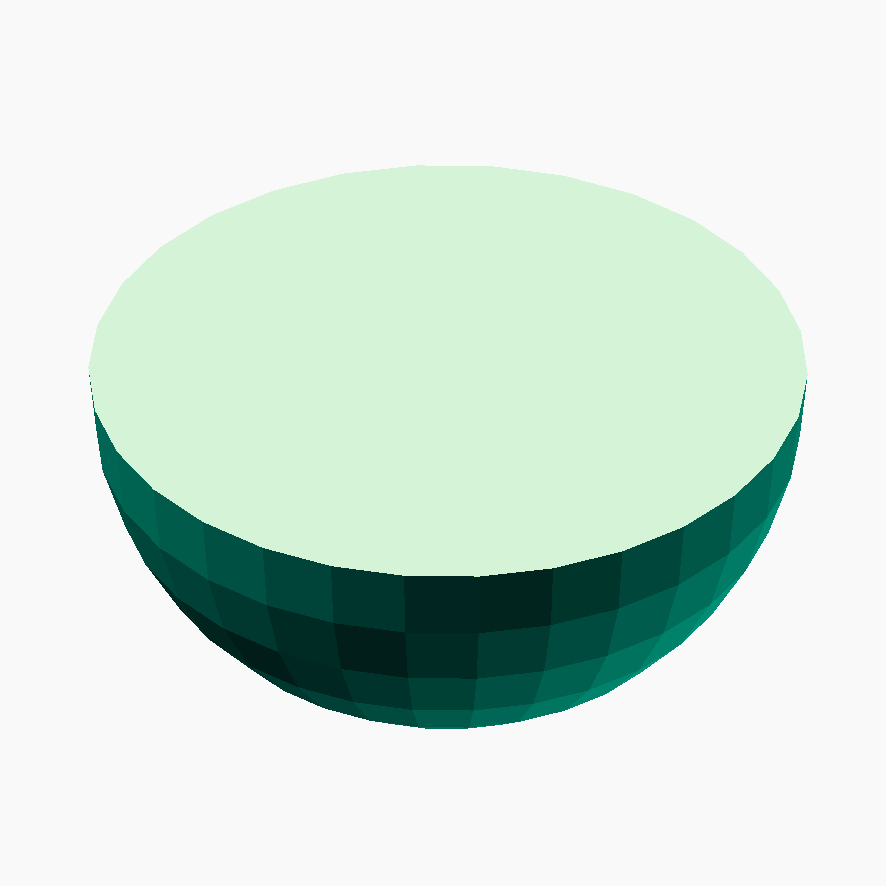}\\
    edge & width, length, height  & \includegraphics[height=3.5em]{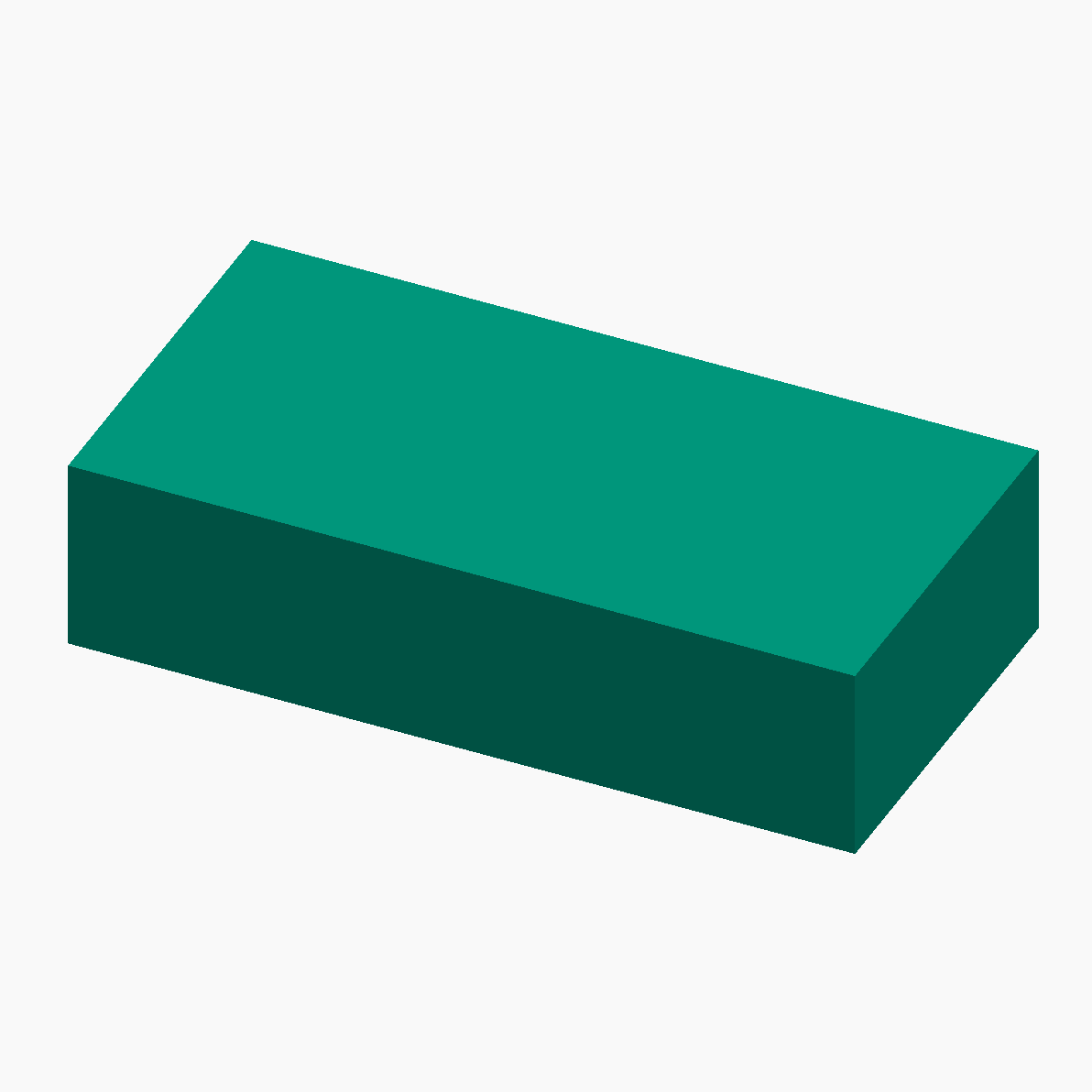}\\
    rod & radius, length  & \includegraphics[height=3.5em]{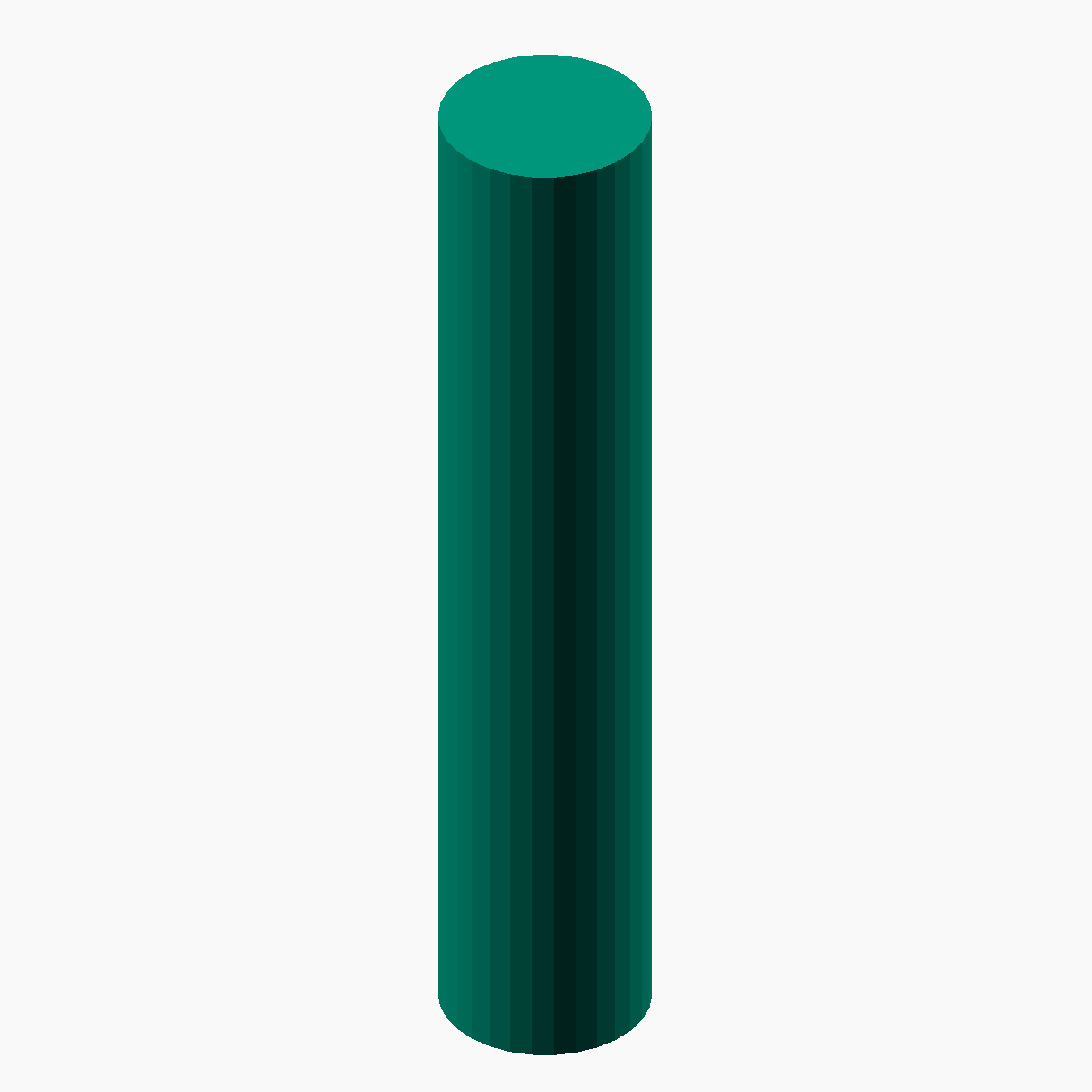} \\
    tube & inner radius, thickness, length  & \includegraphics[height=3.5em]{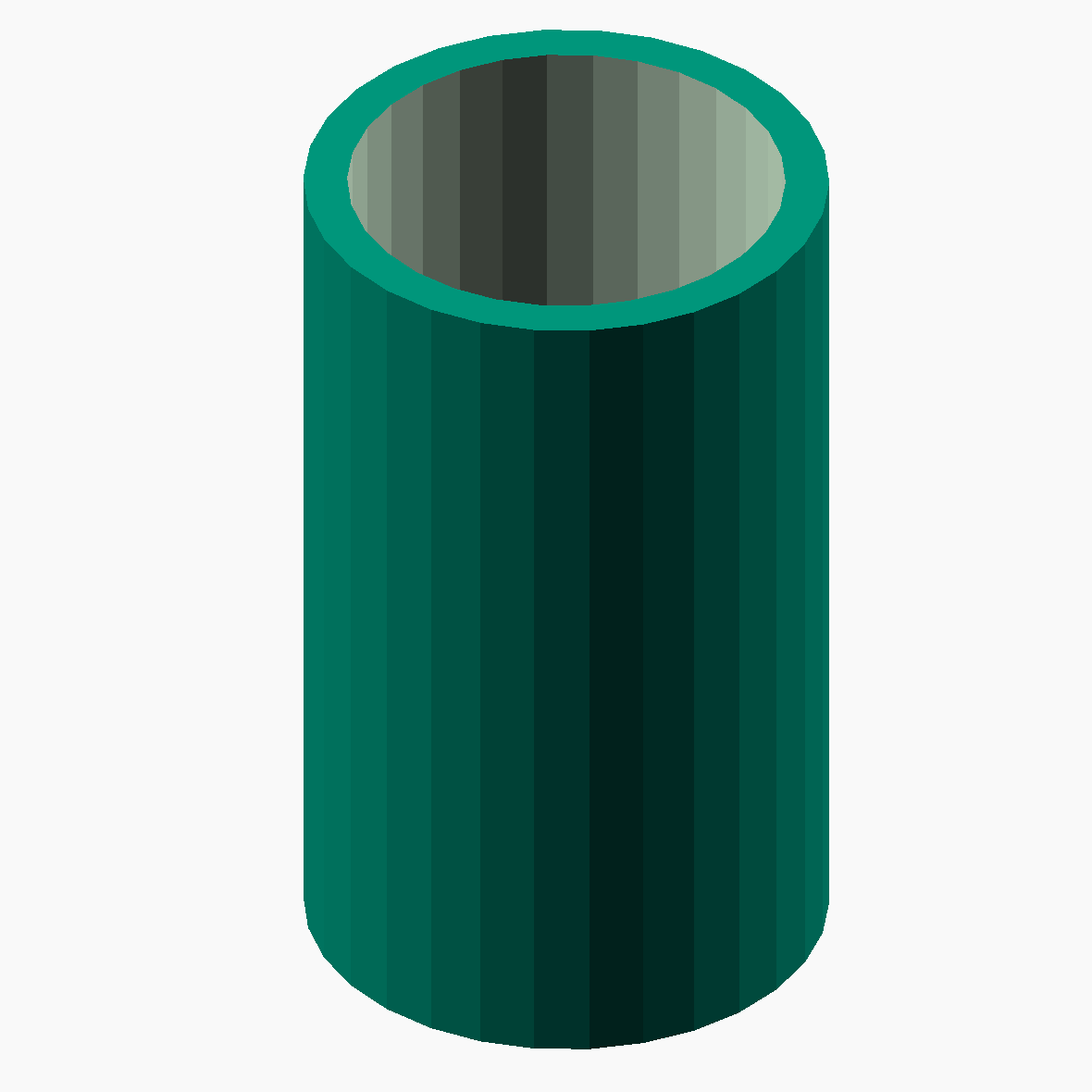} \\
    clip & width, height, base distance, open gap, thickness & \includegraphics[height=3.5em]{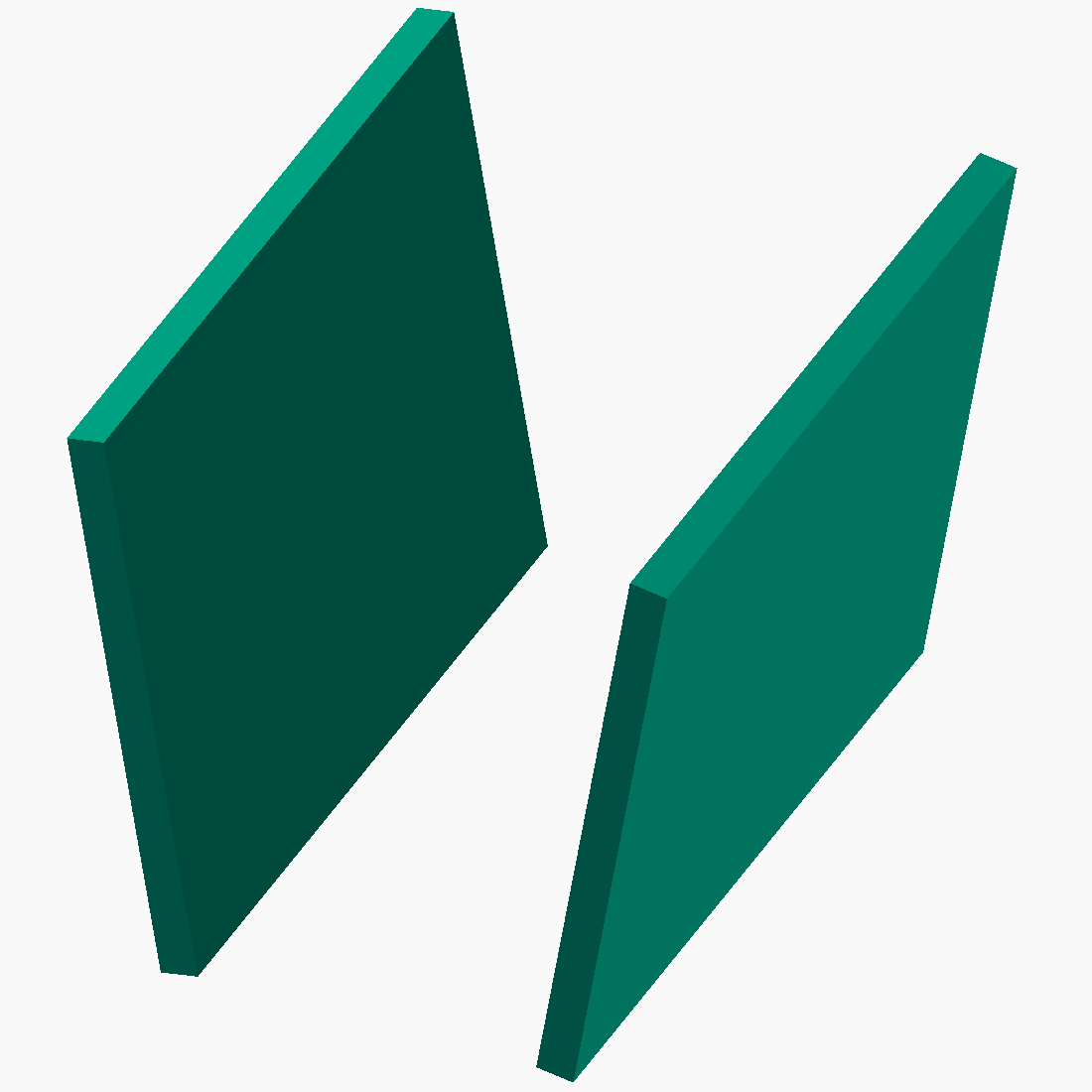}\\
    surface & width, length  & \includegraphics[height=3.5em]{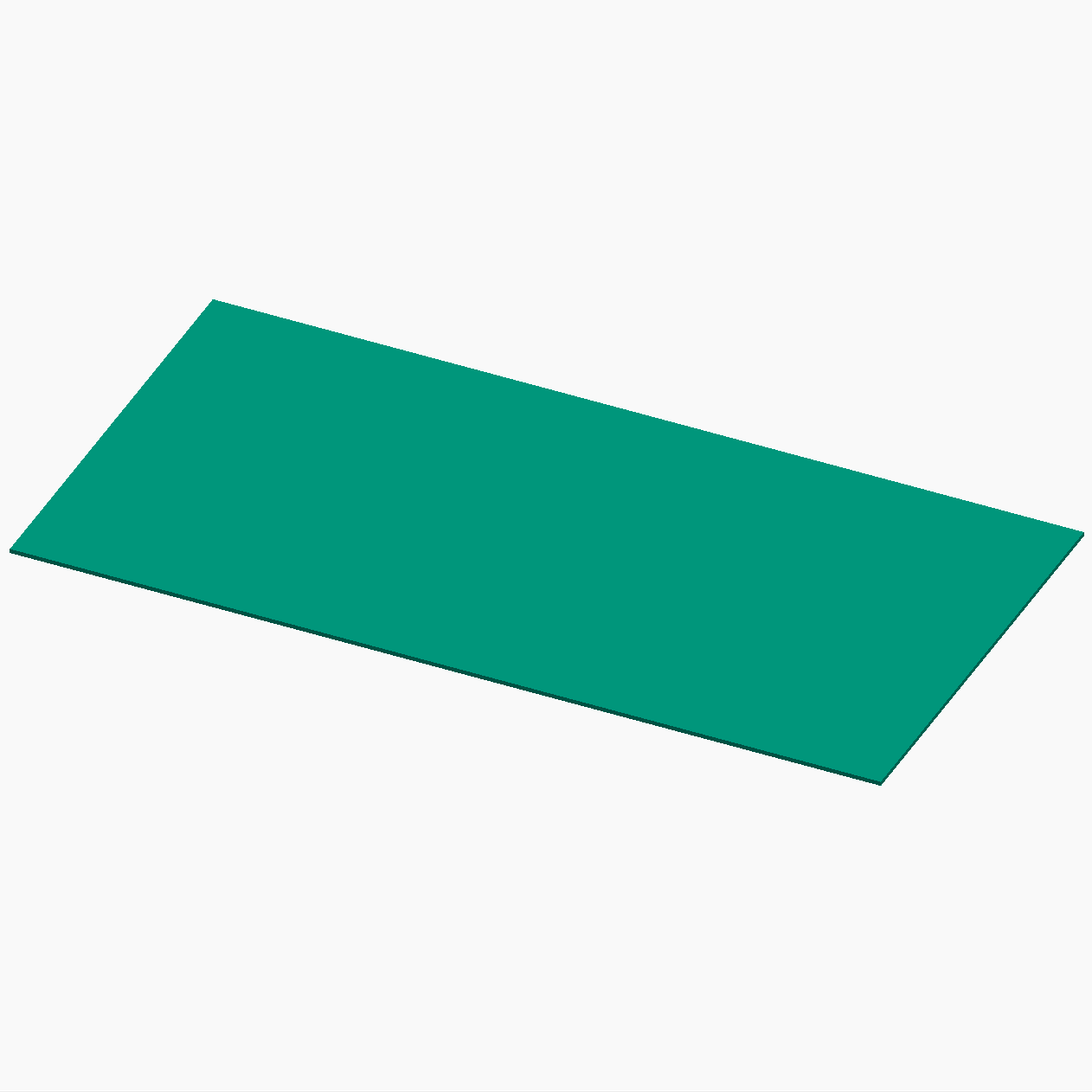}
\end{tabular}
\caption{We parametrize each primitive with its corresponding shape parameters, and show an example of the primitive.}
\vspace{-10pt}
\label{tab:shapeparams}
\end{table}

FabHaL includes eight types of primitives: a hook, a rod, a hole, a tube, a hemisphere, a clip, an edge, and a surface, which can be assigned to a wide variety of objects as described in \nameref{sec:overview} (Section~\ref{sec:overview}). We summarize the connectivity between these primitives in Table~\ref{tab:connectivity} and what shape parameters we use to parametrize a primitive's geometry in Table~\ref{tab:shapeparams}. Next, we explain in detail how we model the connection behavior between pairs of connector primitives, as well as the information associated with each primitive that will be used by the solver for verifying and finalizing the configuration of a hack design.

\paragraph{Connector frames} Our analysis of home hacks (Appendix~\ref{appendix:analysis}) found that the connection behavior between parts is local to the pair of primitives that forms the connection. Take as an example the rod-hook connection (inset Figure~\ref{fig:rodhook}): a hook can slide along a rod and flex about it, regardless of whether this rod is from a closet, a shower, or an ironing board. To represent such connection behavior mathematically so that we can formulate it as part of the constrained optimization in the solver, we need to establish the concept of a \texttt{Frame}.

In FabHaL, we define \texttt{Frame} to be a position vector $(x,y,z)$ and yaw-pitch-roll intrinsic Euler angles. Frames can be used to represent a single-origin coordinate system (similar to mate connectors in CAD), or the 3D configuration of a geometric entity (a primitive or a part). We use frames to represent the connection points on primitives and call them \emph{connector frames}. The connector frames of a primitive can be computed from its base frame and shape parameters (obtained from the part annotations), and some additional degrees of freedom \begin{wrapfigure}{l}{70pt}
    \vspace{-15pt}
    \centering
    \includegraphics[width=90pt]{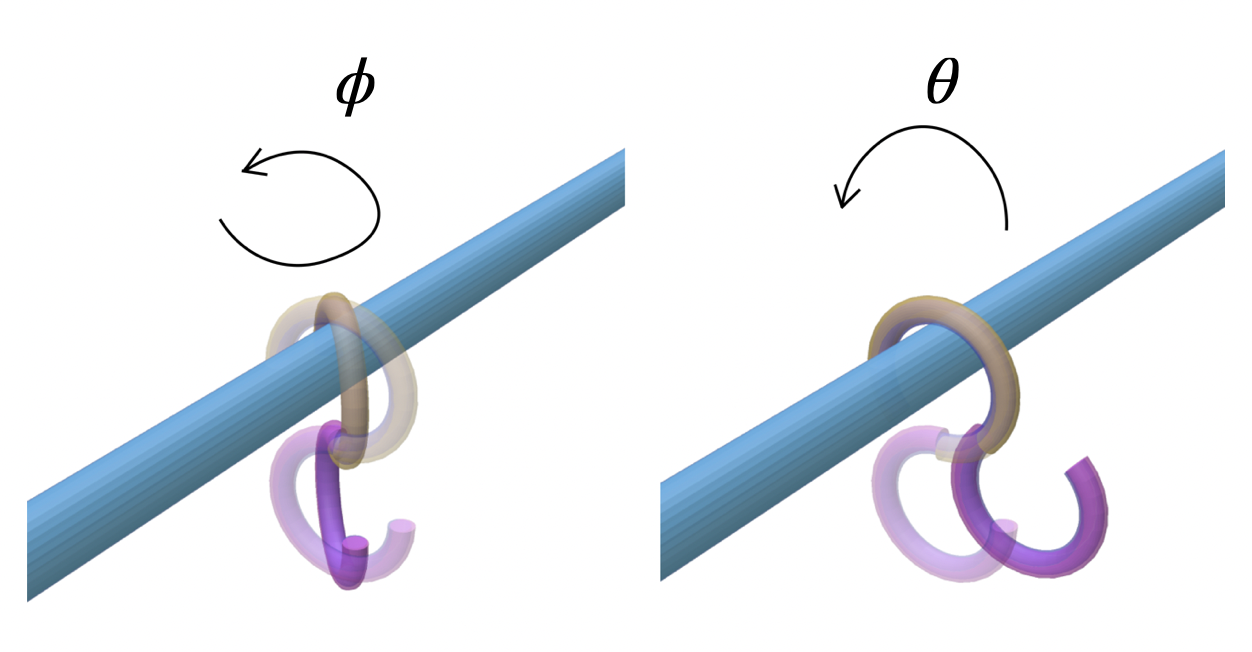}
    \vspace{-25pt}
\end{wrapfigure}specific to its type. For example, a hook primitive has two additional degrees of freedom, $\theta$ and $\phi$, parameterizing the location and orientation of the point of contact (see inset). In FabHaL, the additional degrees of freedom and the information on how to use these DoFs to compute the parametric connector frames are associated directly with each connector primitive.

\paragraph{Alignment offsets} When two primitives are connected, their connector frames need to be coincident in position, but the orientation may have some offset. Based on our analysis, this orientation offset is common to a pair of connectable primitives. For example, \begin{wrapfigure}{l}{15pt}
    \vspace{-17pt}
    \centering
    \includegraphics[width=40pt]{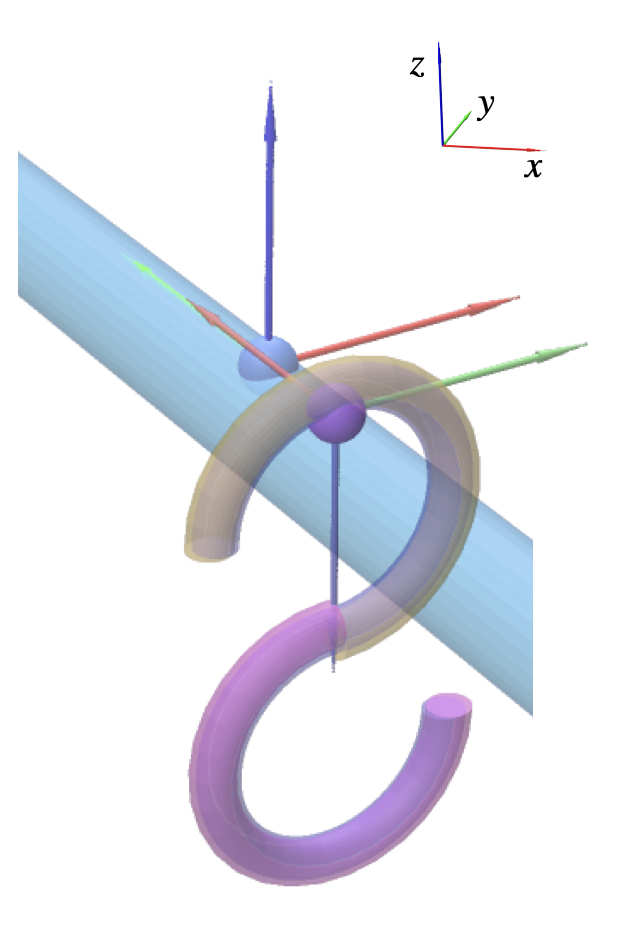}
    \vspace{-26pt}
\end{wrapfigure}as shown in the inset figure, when a rod and a hook connect, their connector frames are offset by a rotation of $[180^{\circ}, 0^{\circ}, 90^{\circ}]$ in yaw-pitch-roll intrinsic Euler angles. (Here the frames are intentionally placed to be not coincident at their origins to better display the orientation offset.) We call the offset rotation between two primitives' connector frames an \emph{alignment offset}.

With the connector frames and alignment offsets defined for each pair of connector primitives, we can represent the connection behavior precisely with respect to the degrees of freedom associated with each primitive. Even with a small number of categories of connector primitives, we can capture a wide range of possible connections that appear in hacks. This set of primitives and associated alignment offsets is also easily extensible.

In addition to the theoretically allowed connectivity between primitives (Table~\ref{tab:connectivity}), two primitives need to be physically compatible before they can be connected. We encode two pieces of additional information in connector primitives so that users do not need to reason about this lower-level detail.

\paragraph{Closed primitives} Two primitives with no openings cannot connect because there is no valid motion path to create the connection. Among the eight connector primitives, the hole primitive is always closed. In addition, primitives that aren't generically closed could be inaccessible in the context of the geometry of the part containing it. For example, the handle of the basket in the bottom-right of Figure~\ref{fig:parametricPrimitives} is tagged as a hook, which can connect to a hole according to Table~\ref{tab:connectivity}. But as an integral part of the basket, it is part of closed geometry; thus, a hole primitive without an opening cannot connect to this hook. We allow tagging of individual primitives like the basket handle as \emph{closed primitives} when annotating parts for the Annotated Object Library and our solver checks that designs do not attempt to connect two closed primitives to each other.

\paragraph{Critical dimensions} 
Primitives might not be able to supply enough physical space for a connection. For example, a one-to-one connection between a rod and a hook is only possible if the hook's hoop radius is greater than the rod's radius. For a multi-to-one connection between several hooks and tubes and a single rod, the hooks and tubes might fully occupy the length of the rod. Then no new connection can be made with the rod because there is no more available space on it.

To keep track of available physical space on primitives, we specify a \emph{critical dimension} for connector primitives that can have multiple connections (the eight primitives except hemisphere and clip). The \emph{available critical dimensions} refer to the dimension of a primitive that can be occupied when a new connection is made between itself and another primitive. For example, the critical dimension of a rod primitive is its length, and when a hook connects to this rod primitive, its available length is reduced by the width of the hook. The hook's critical dimension---the hoop radius---is also reduced by the rod's radius.

\subsection{Language Constructs and Hack Construction}\label{subsec:langconstructs}

\begin{figure}[ht]
\centering
\includegraphics[width=0.9\linewidth]{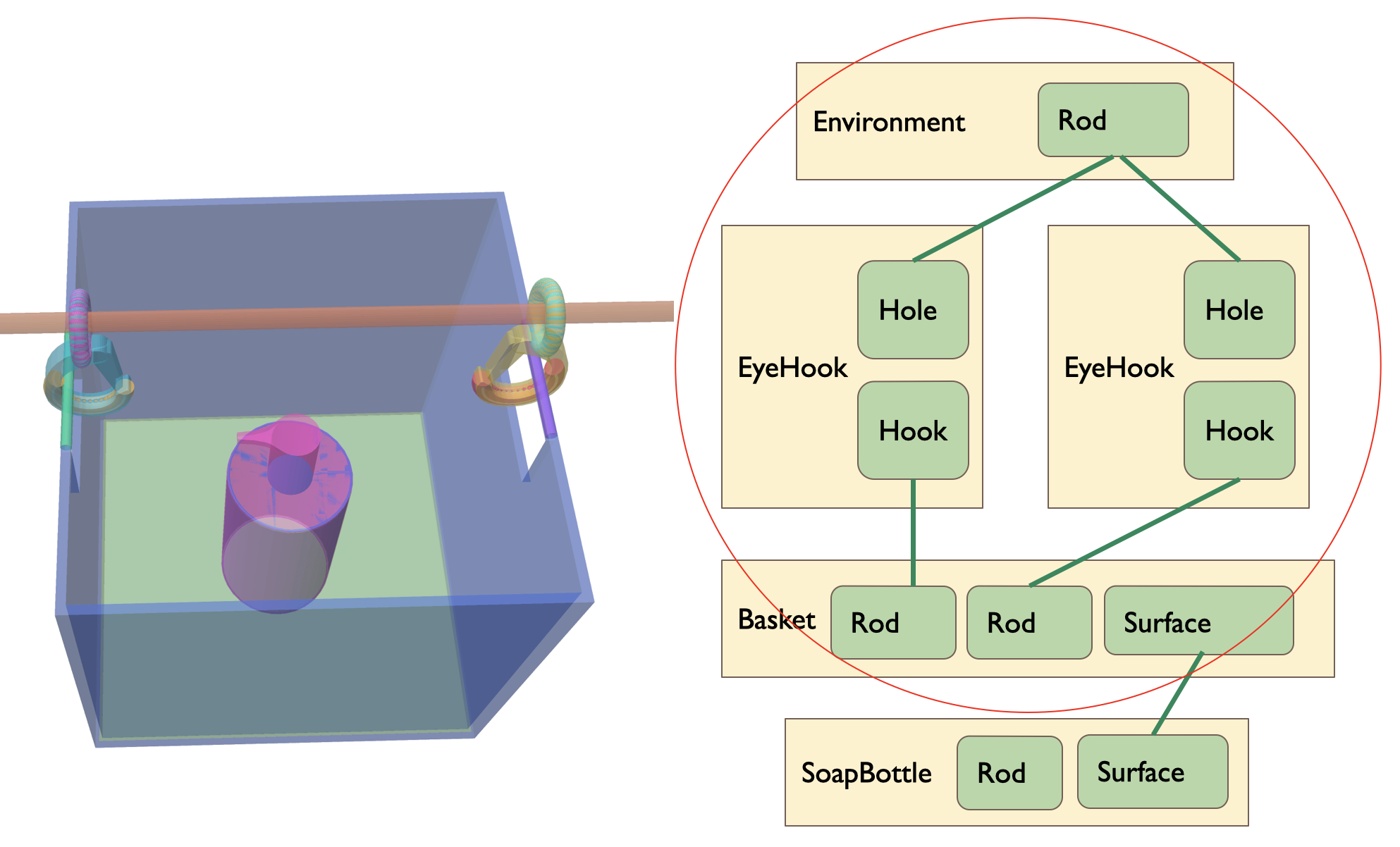}
\vspace{-10pt}
\caption{An assembly with a cycle: a basket is connected to a rod via two eyehooks, forming a cycle (in the red circle) between the basket and the environment. On the right, yellow rectangles represent parts and green rounded rectangles represent primitives.}
\label{fig:loop}
\end{figure}

To represent a hack, we must connect \textit{parts} (annotated objects from the Annotated Object Library) using their connector \textit{primitives}. These connected parts form a graph (see Figure~\ref{fig:loop}) that we call an \textit{assembly} (i.e., a hack).

Two special parts in an \texttt{Assembly} are assumed to be fixed in place: the part representing the \textit{environment} the assembly is attached to, and the \textit{target part}, a part that is meant to be fixed relative to the environment and whose configuration is used as a target for the solver. For example, the clip in Figure~\ref{fig:ex1} is resting on a table, supporting a toothbrush. The table is the environment and can be represented using a surface primitive with a fixed position and orientation. The toothbrush is the target part that we want to fix above the table.

Our DSL exposes three operations needed to create an \texttt{Assembly}:
\begin{itemize}
\itemsep0em
    \item \texttt{add(part, frame)}
    \item \texttt{end\_with(part, frame)}
    \item \texttt{connect(part1.primitive, part2.primitive)}
\end{itemize}

\texttt{add} is used for specifying the environment part with a fixed configuration (frame); and \texttt{end\_with} is for specifying the target part's configuration (frame). \texttt{connect} takes two primitives as arguments regardless of order and determines whether each \texttt{Part} is already part of the \texttt{Assembly} or is newly-introduced. It has two optional \begin{wrapfigure}{l}{30pt}
    \vspace{-13pt}
    \centering
    \includegraphics[width=50pt]{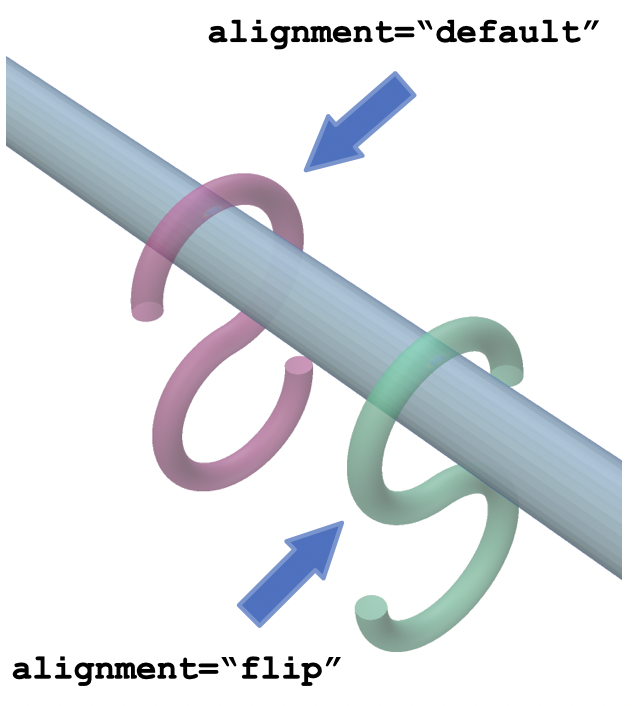}
    \vspace{-25pt}
\end{wrapfigure}parameters: (1) \texttt{alignment} (either ``flip'' or ``default'') to indicate an orientation flip, e.g., a hook can hang on a rod coming from both sides of the rod as shown in the inset; (2) \texttt{is\_fixed}, a boolean value that indicates that the degrees of freedom involved in the connection should be held constant during solver-aided evaluation (because the design involves, for example, taping connectors together).

If both connected parts are already part of the assembly, this connection will create a cycle in the graph representation of the connected parts (see Figure~\ref{fig:loop}). Not all \texttt{connect} operations will be physically realizable and we will discuss how the solver verifies whether a connection can be made in Section~\ref{subsec:solver}.

\begin{figure}[ht]
\centering
\includegraphics[width=0.9\linewidth]{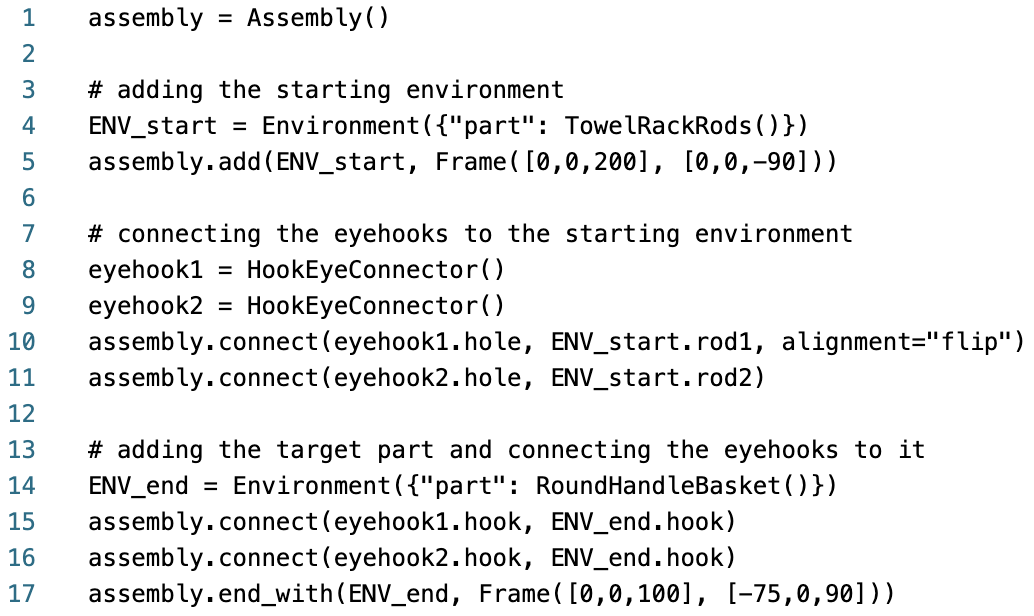}
\includegraphics[width=0.9\linewidth]{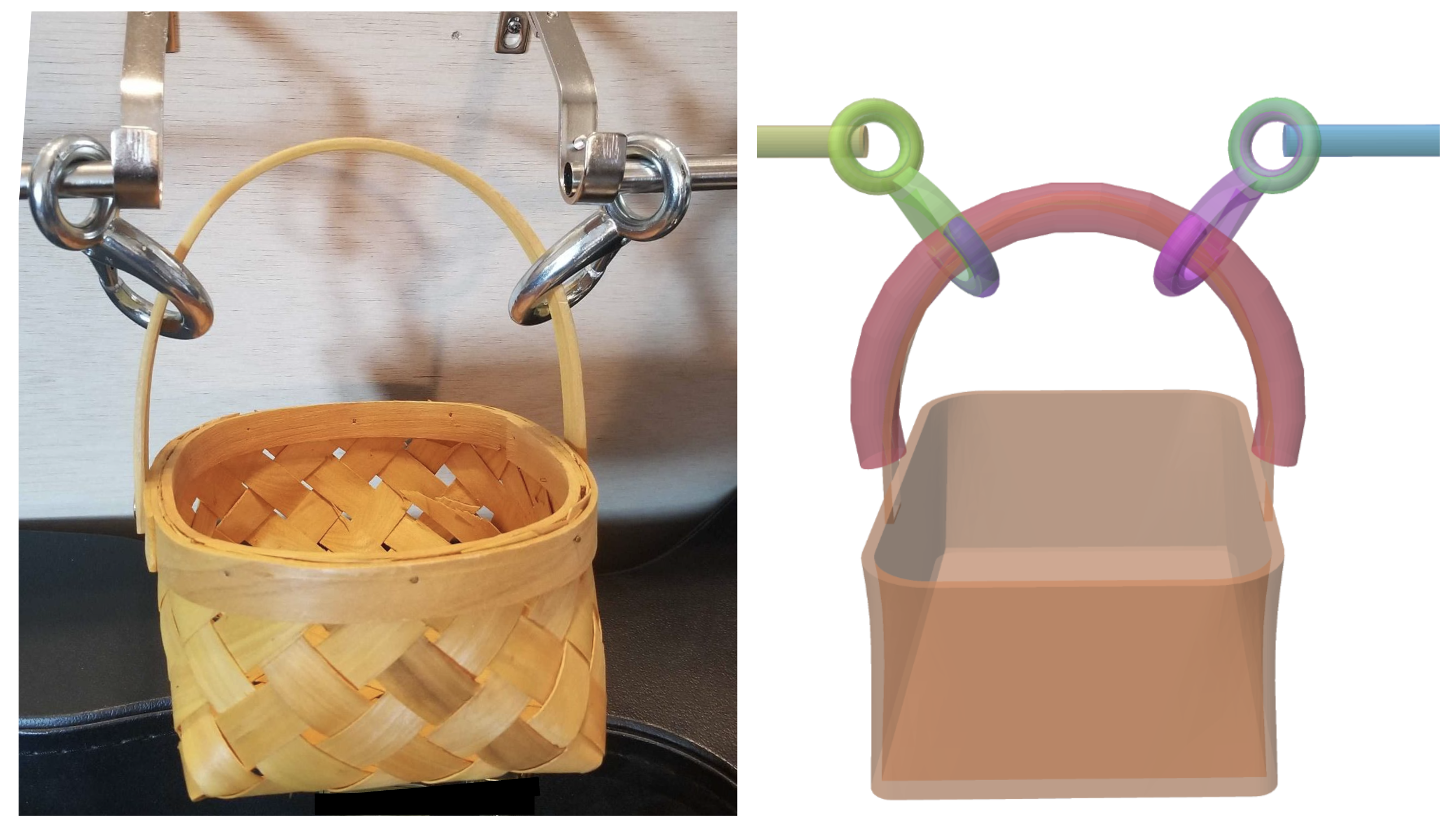}
\vspace{-10pt}
\caption{An example program in FabHaL (top), with the corresponding assembly solved for and rendered by our system (bottom right) and a physically reproduced design (bottom left).}
\label{fig:dslExample}
\end{figure}

Figure~\ref{fig:dslExample} shows an example program in our DSL. This fixture hack hangs a basket with a round handle in between two rods. In this program, we first initialize an \texttt{Assembly}, followed by the environment initialization. Then we use \texttt{connect} to add two eyehooks to the two rods by connecting the eyehook's eye to the rod. Finally, we initialize the target part and connect the hook part of the eyehooks to the handle of the basket. More example programs can be found in Appendix~\ref{appendix:programs}.

\subsection{Solver-aided Evaluation}\label{subsec:solver}

The core advantage of FabHaL is its ability to simplify the representation of an \texttt{Assembly} to a graph of connected \texttt{Parts}, leaving the job of calculating the placement of parts to the solver.

In our solver, we model an \texttt{Assembly} using a reduced representation of a kinematic rigid body chain, which is a common practice in fields like rigid body mechanics and robotics~\cite{featherstone_calculation_1983}.

The solver handles not only the simulation (\ref{subsubsec:twostepsolver}) but also the pre-checks (\ref{subsubsec:verifyattach}, \ref{subsubsec:cycle}) that check  whether the \texttt{connect()} operations can be physically realized. We will begin with the pre-checks before we discuss the simulation of the assembly under gravity.

\subsubsection{Verify connect()} \label{subsubsec:verifyattach}

There are two potential issues when a connection is being made between two parts.

First, a connection cannot be made between two primitives that cannot be joined together according to the primitive connectivity table (Table~\ref{tab:connectivity}), such as a rod to another rod, or when they are two \emph{closed primitives}.

Second, the solver needs to check whether the \emph{available critical dimension} of a primitive is enough for what is needed for a new connection. Based on the primitives' critical dimensions, we add constraints to the parameters of the connector primitive that has multiple connections. For example, when two hooks connect to the same rod, two sets of parameters that decide where along the rod the hooks connect to will be created. Suppose that the hooks each have widths $w_1, w_2$, the rod has length $l$, and the two connection parameters indicating the position of the hooks along the length of the rod are $t_1, t_2 \in [0,1]$. Then this ``multi-connection'' constraint $|t_1 - t_2| \cdot l \geq \frac{w_1+w_2}{2}$ will be created and included in the solving process. We represent this constraint as a soft penalty as follows:
\begin{equation*}
C_{m, f} = 0 \text{ if } f \geq 0, \,C_{m,f} = f^2 \text{ if } f < 0
\end{equation*}
where $f = |t_1 - t_2| \cdot l - \frac{w_1+w_2}{2}$. We will use the symbol $C_m$ to represent the sum of all multi-connection constraint penalties.

\subsubsection{Additionally verify connect() that creates cycles} \label{subsubsec:cycle}

A \texttt{connect} operation creates at least one cycle in the graph representation of the assembly if it is between two parts that are already part of the assembly (see Figure~\ref{fig:loop}). Such cycles require explicitly modeling constraints over the configurations of the parts that are being connected. Therefore, in addition to the pre-checks on connector types and available critical dimensions, we also need to check whether we can find a set of values for the degrees of freedom that satisfy these constraints.

For every cycle, we model six constraints measuring the failure of the connector frames on the two primitives being \texttt{connect}ed to match each other. An assembly with $n$ cycles is feasible if valid values of the connection parameters exist along the cycles that satisfy $6n$ equality constraints of the form $f_i(\mathbf{x}) = \vec{0}, i \in [1..n]$, where $\mathbf{x}$ is a vector of all the degrees of freedom (DoFs) in the assembly and $f_i\left(\mathbf{x}\right) \in \mathbb{R}^6$ measures the failure of the $i$th cycle to close up. We minimize the sum of constraints residuals $C(\mathbf{x}) = \sum_{i=1}^n \|f_i(\mathbf{x})\|^2$ subject to bound constraints on the DoFs, $\mathbf{x}_{\mathrm{min}} \leq \mathbf{x} \leq \mathbf{x}_{\mathrm{max}}$. In the example assemblies we created, most of the time $n=1$. We use the Powell method~\shortcite{2020SciPy-NMeth} to minimize $C$ and declare the assembly feasible (and thus the \texttt{connect} successful) if the solver succeeds in finding parameters with $C(\mathbf{x}) \leq 10^{-6}$. Since the success of the minimization depends on the initialization of the assembly parameters and can get stuck in local minima, we repeat the optimization $T$ times starting from different random initial guesses. We terminate early if a solution is found. We observed that $T=16$ works well in practice.

\paragraph{Geometric Quick Reject} Before we actually run a full optimization to find a system configuration that satisfies the connection constraints, we also utilize some precomputed information about the parts and primitives to perform a quick geometric check. 

Our geometric check is based on the triangle inequality: $k$ line segments of length $\ell_1 \geq \ell_2 \geq \cdots \geq \ell_k$ cannot be arranged into a closed loop in 3D unless $\ell_1 \leq \sum_{i=2}^k \ell_i$. To apply this principle to our problem, we note that since each part $i$ in a part cycle is rigid, we can bound the Euclidean distance $e_i \in [e_i^-, e_i^+]$ between the point \begin{wrapfigure}{l}{65pt}
    \vspace{-13pt}
    \centering
    \includegraphics[width=85pt]{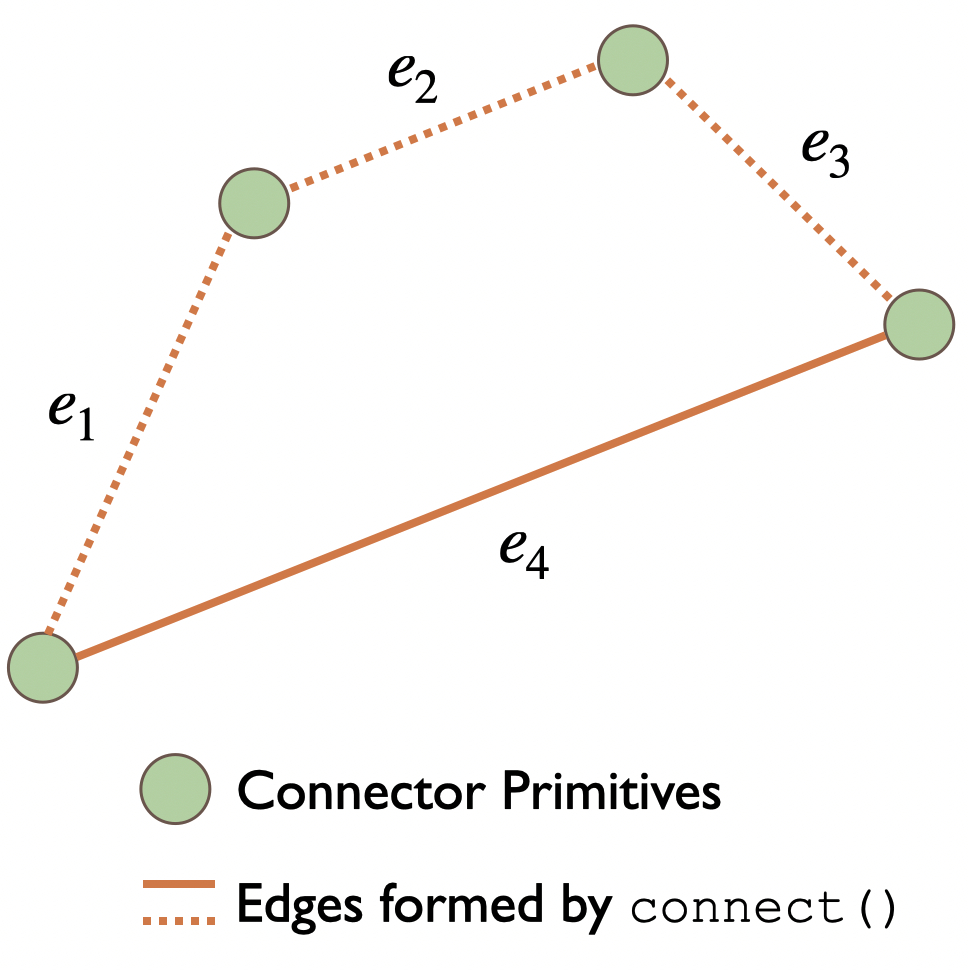}
    \vspace{-27pt}
\end{wrapfigure}where part $i$ connects to parts $i-1$ and $i+1$. Because $[e_i^-, e_i^+]$ depend only on the geometry of the part and its two connectors involved in the cycle, not on the connection parameters, we can precompute these bounds for all the parts defined in our Annotated Object Library. 
Consider the inset figure representing a design which has a cycle of 4 parts. In order for the cycle to close up, the following linear program in the distances $e_i$ between connectors must be feasible:
\begin{align}
\min_{e_i} 1 \quad \mathrm{s.t.} \quad &\begin{bmatrix}
    1 & -1 & -1 & -1\\
    -1 & 1 & -1 & -1\\
    -1 & -1 & 1 & -1\\
    -1 & -1 & -1 & 1
    \end{bmatrix}
    \begin{bmatrix}
        e_1 \\ e_2 \\ e_3 \\ e_4
    \end{bmatrix} \geq \vec{0}\\ &e_i^- \leq e_i \leq e_i^+,\quad  i \in [1..4].\notag
\end{align}
Checking the existence of a set of distances $e_i$ satisfying the bound constraints and triangle inequality then amounts to checking the feasibility of a set of linear inequality constraints, which can be solved in milliseconds by standard Python libraries, quickly rejecting impossible \texttt{connect} operations.

\paragraph{Stall Prevention} When we actually need to run the optimization, we put in measures for stall prevention. To halt optimization of $C$ when the solver stalls, we pass a custom callback function to \texttt{scipy.optimize} that performs linear regression on $C(\mathbf{x}_i)$ for a sliding window of the last ten DoF iterates $\mathbf{x}_i$. We abort the optimization in failure if the slope of the fit line is less than 0.1 (meaning not much progress is being made by the optimizer). This strategy has gained us an additional 1.4x speedup on average for the three examples with cycles.

\subsubsection{Solving the assembly}\label{subsubsec:twostepsolver}

After a valid assembly is constructed in FabHaL, the user can invoke the solver to find the values for the degrees of freedom in the system that bring the target part as close as possible to its specified configuration while being in static equilibrium under gravity and respecting all cycle-closure and critical-dimension constraints.

This solve is a constrained optimization problem: we wish to minimize the user objective subject to the balance of forces and torques on each non-environment part. Early experiments revealed that black-box nonlinear optimization was prohibitively slow at solving this problem, and moreover, often failed to converge to a feasible local minimum. Therefore, we propose instead a two-step solver that first minimizes the user objective subject to all constraints being satisfied, and then uses the optimized configuration as an initial guess for a simulation that relaxes the assembly to static equilibrium.

\paragraph{First Step: Minimizing the user objective} We use the Powell~\shortcite{2020SciPy-NMeth} method to find a feasible configuration of the assembly that minimizes the user objective:
\begin{equation*}
\mathbf{x}_{\mathrm{feas}} = \argmin_{\mathbf{x}} f_{\mathrm{obj}}(\mathbf{x}) + \sigma(C_m(\mathbf{x}) +  C(\mathbf{x})) \quad \mathrm{s.t.}\quad \mathbf{x}_{\mathrm{min}} \leq \mathbf{x} \leq \mathbf{x}_{\mathrm{max}},
\end{equation*}
where $C_m(\mathbf{x})$ are the multi-connection constraints described in Section~\ref{subsubsec:verifyattach}, $C(\mathbf{x})$ are the cycle-closure constraints as used in Section~\ref{subsubsec:cycle}, and $\sigma$ is a penalty parameter. We use a starting value of $\sigma=100$; if after optimization the constraint residual $C_m(\mathbf{x}_{\mathrm{feas}}) + C(\mathbf{x}_{\mathrm{feas}})$ is not below $10^{-6}$, we double $\sigma$ and repeat the optimization, using $\mathbf{x}_{\mathrm{feas}}$ as the initial guess. We repeat this process up to 5 times which is usually enough for finding $\mathbf{x}_{\mathrm{feas}}$; if the constraint residual is still not below $10^{-6}$ after 5 times, we pass the best configuration found to the second step. 

\paragraph{Second Step: Relaxing under gravity} We use a physics solver to relax the assembly to an equilibrium state under its self-load, starting from the guess $\mathbf{x}_{\mathrm{feas}}.$ Let $q_i\in SE(3)$ represent the configuration of the $i$th part, and $\mathbf{q} = \{q_i\}_{i=1}^n$ the configuration vector of the entire assembly. For an assembly with $c$ total pairs of primitives connected together, let $g_j(\mathbf{q}, \mathbf{x}) \in \mathbb{R}^6$ for $j=1,\ldots,c$ be constraint functions encoding that each pair of primitives are connected together with connection parameters $\mathbf{x}$.

To relax the assembly under gravity, we solve
\begin{align}
    &\argmin_{\mathbf{q}, \mathbf{x}}  E(\mathbf{q},\mathbf{x}) \quad \mathrm{s.t.}\quad \mathbf{x}_{\mathrm{min}} \leq \mathbf{x} \leq \mathbf{x}_{\mathrm{max}} \label{eq:physsim} \\
    &E(\mathbf{q},\mathbf{x}) = \sum_i P_i(\mathbf{q}) + \sigma \sum_{j=1}^c \|g_j(\mathbf{q},\mathbf{x})\|^2, \notag
\end{align}
where $P_i(\mathbf{q})$ measures part $i$'s gravitational potential energy and $\sigma$ is a penalty parameter enforcing that connectors stay attached: we use $\sigma=100$.
We optimize Equation~\eqref{eq:physsim} using an active-set Newton's method~\cite{NoceWrig06}.

\begin{figure} [ht]
\centering
\includegraphics[width=\linewidth]{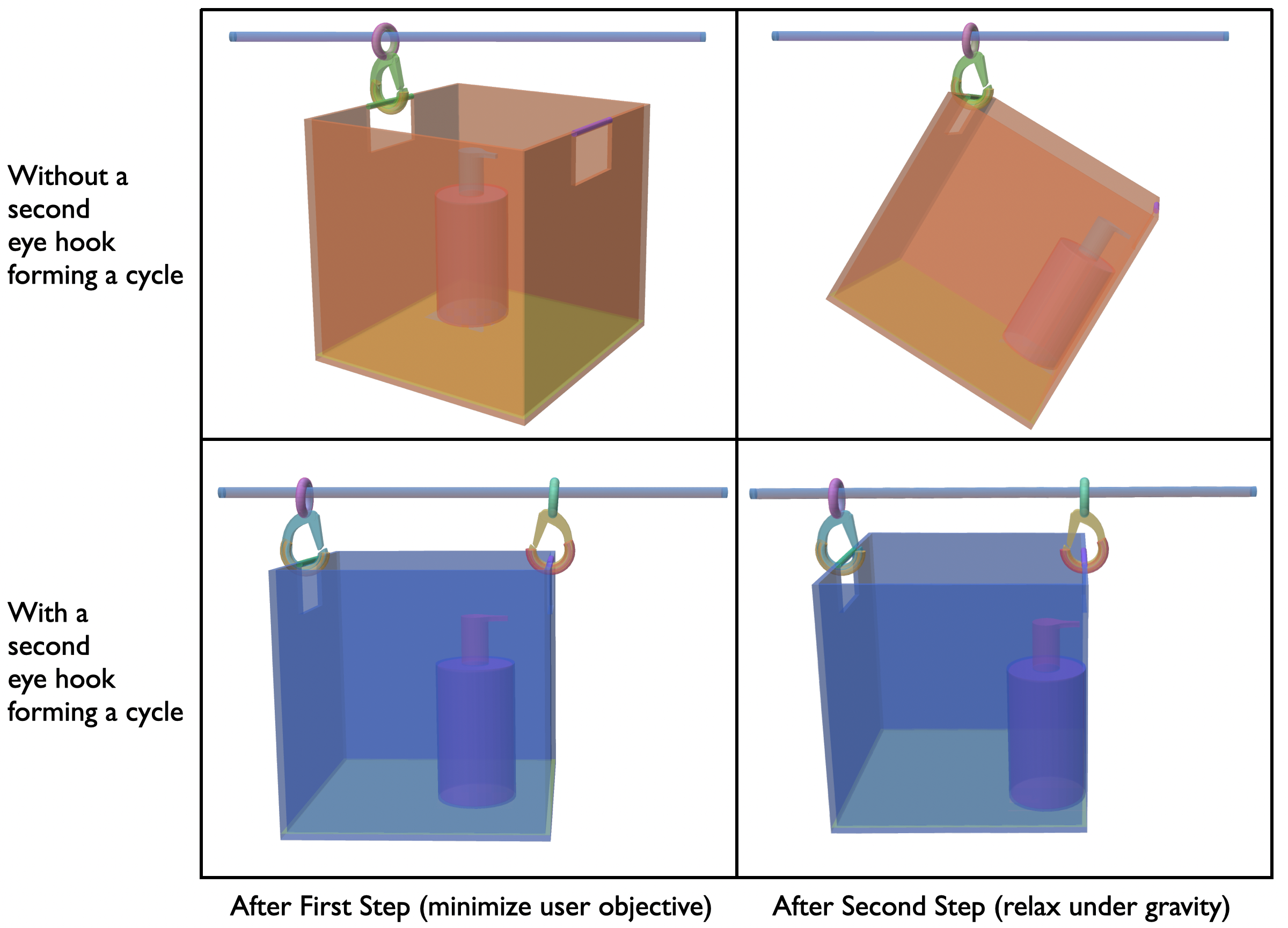}
\vspace{-15pt}
\caption{The top row shows the hack design without a second eyehook to balance the basket, and the bottom row shows the hack design with the second eyehook. The first column shows the intermediate results after first running the user objective minimization, and the second column shows the resulting configuration after the second step is run.}
\label{fig:twostepofbasket}
\end{figure}

To demonstrate the two-step process, we take the hack design from Figure~\ref{fig:loop} as an example, which hangs a soap bottle from a rod using eyehooks and a basket. Both programs visualized in Figure~\ref{fig:twostepofbasket} uses the same target configuration specification for the soap bottle and thus after the first step, the soap bottle is in a configuration that is closest to the target configuration. But after the second step of relaxing under gravity, without a second eyehook to balance, the top row's design falls under gravity into a less desirable configuration compared to the bottom row's design.

During the physics relaxation, we also predict whether the assembly will fall apart due to connectors slipping off of each other. To perform this analysis, we annotate each connection parameter for each primitive in our library with one of three tags:
\begin{itemize}
\itemsep0em
\item \texttt{UNBOUNDED} parameters are periodic and should be allowed to ``wrap around'' from $x_{\mathrm{max}}$ to $x_{\mathrm{min}}$ during optimization. For example, for a ring that can rotate $360$ degrees, the angle parameter specifying the rotation of the ring about its central axis is \texttt{UNBOUNDED}.
\item we tag parameters as \texttt{BOUNDED\_AND\_CLAMPED} if the geometry of the primitive prevents the parameter from ever leaving the interval $[x_{\mathrm{min}}, x_{\mathrm{max}}]$. The position parameter of a rod along the bottom of a clothes hanger is one example of such a parameter.
\item finally, a parameter is \texttt{BOUNDED\_AND\_OPEN} if exceeding the bounds of the parameter would cause the assembly to fall apart. The position parameter of a dowel rod, for example, is \texttt{BOUNDED\_AND\_OPEN}: hooks or rings that slide past the end of the dowel rod fall off the assembly.
\end{itemize}

At the end of optimization, for each \texttt{BOUNDED\_AND\_OPEN} parameter $i$ we check whether $x_i$ is in the inequality constraint active set, i.e. whether $x_i$ is equal to its maximum or minimum allowed value, and if so, whether $\nabla_{x_i} E$ points away from $x_i$'s feasible interval. If so, we report to the user that the assembly falls apart.

A complex assembly might have many different equilibrium states \begin{wrapfigure}{l}{80pt}
    \captionsetup{margin=0cm}
    \vspace{-12pt}
    \centering
    \includegraphics[width=100pt]{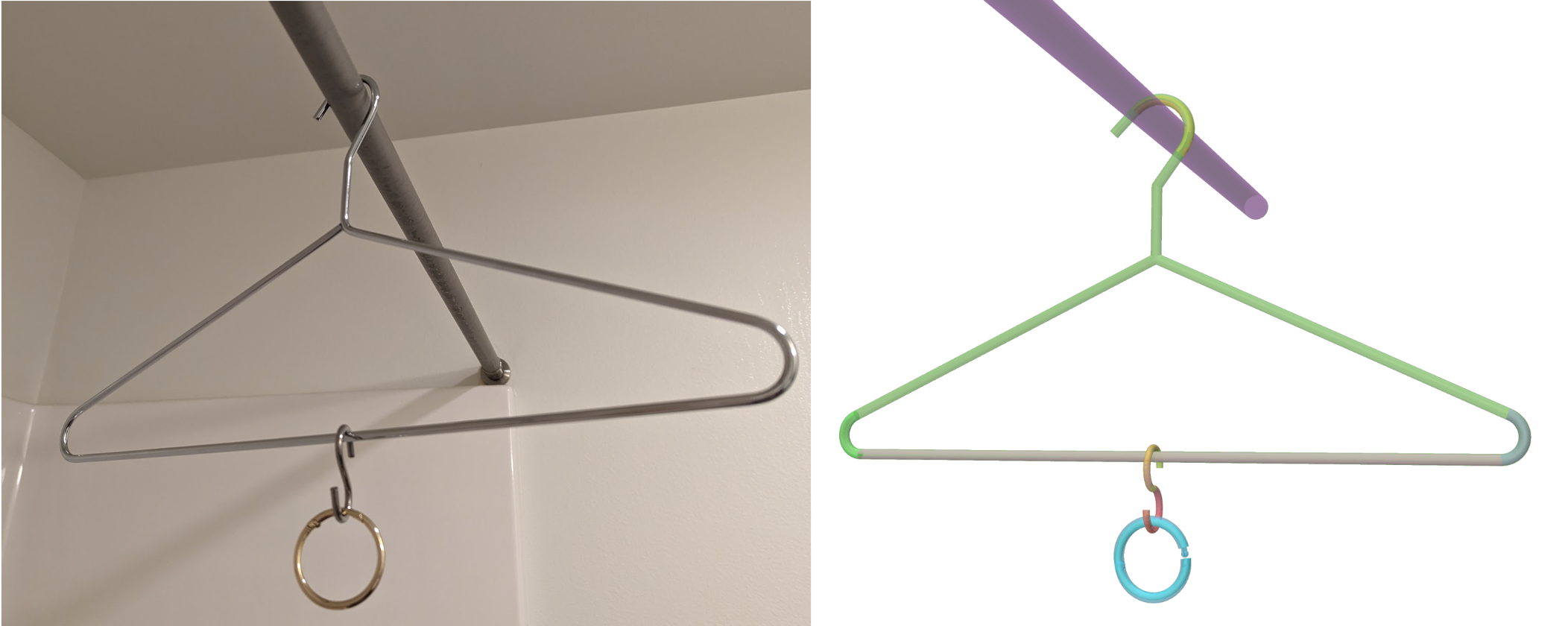}
    \vspace{-20pt}
    \caption{The ``Demo''.}
    \label{fig:ex0demo}
    \vspace{-13pt}
\end{wrapfigure}under gravity; our method above finds just one of them. For instance in the ``Demo'' assembly (see inset Figure~\ref{fig:ex0demo}) the S-hook and the ring could slide to either end of the hanger's rod depending on which side $\mathbf{x}_{\mathrm{feas}}$ encodes they are closer to.

\subsection{DSL Implementation and Parametric FabHaL Programs}\label{subsec:programmatic}

Being a DSL, FabHaL lends itself well to programmatic generation of families of programs if the design is constructed parametrically. We implemented FabHaL as a shallowly embedded DSL with Python. In other words, it is embedded in the host language Python without its own abstract syntax tree. This allows the DSL to be used as a Python library or package, and have access to common control structures like loops and conditionals from the host language. Thus, we can easily generate parametrized designs using the host language features, and use the solver to find the set of parameters from hundreds of variations that let the program best satisfy the given target part configuration.

\begin{figure*}[ht]
\centering
\includegraphics[width=0.9\linewidth]{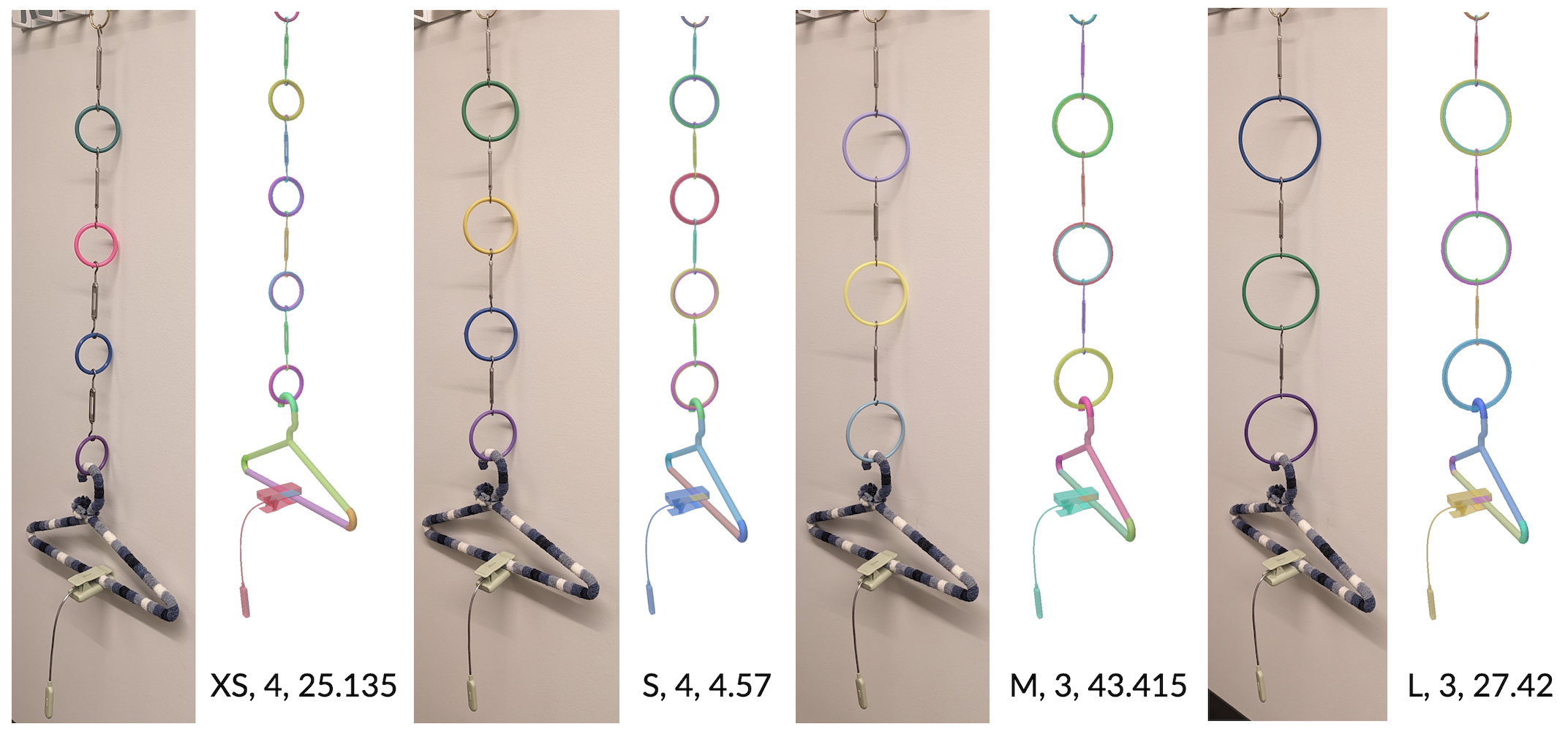}
\vspace{-10pt}
\caption{The four variations that most closely match the desired target part configuration given different ring sizes, with the photo on the left and simulated result on the right. The parameter combination is indicated below the simulated result.}
\vspace{-5pt}
\label{fig:fourcliplights}
\end{figure*}

As an example, suppose that we are preparing for a trip to a summer camp with some bunk beds. We would like to hang a clippable reading light to be at a certain distance from a hook on the top bunk bed so that it's far away enough to not affect others in the same space \begin{wrapfigure}{l}{45pt}
    \vspace{-12pt}
    \centering
    \includegraphics[width=65pt]{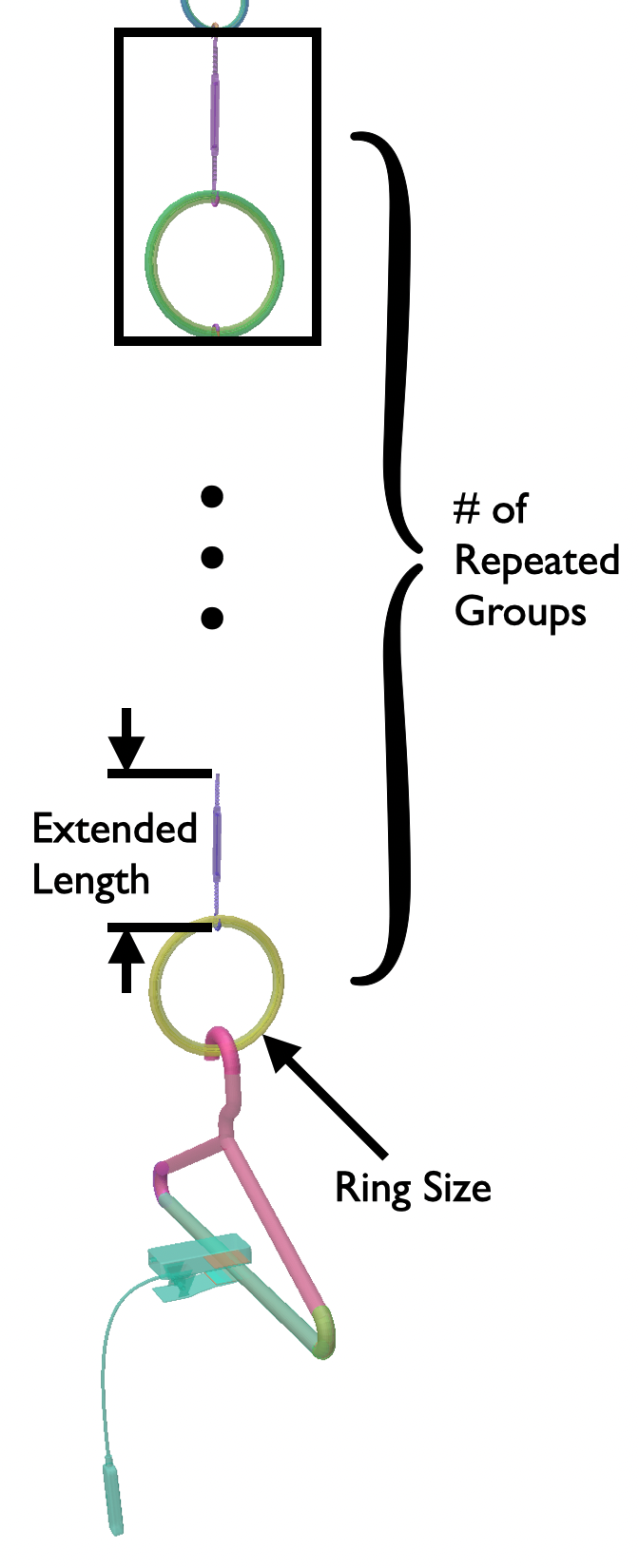}
    \vspace{-25pt}
\end{wrapfigure}and we can also reach the light's switch easily. The available parts here are a hanger, extendable M4 turnbuckles, and rings of different sizes. To design a fixture hack for this scenario, we might come up with a parametrized program consisting of a chain of $n$ turnbuckle-ring pairs, with each turnbuckle extended by $l$ millimeters and each ring of radius $X$ (see inset). We can programmatically generate a family of programs that represent potential hack designs for this scenario. If we already know the desired length and the size of rings that we have, we can use the solver to help find the best parameters of $n$ and $l$ for a given ring size. Figure~\ref{fig:fourcliplights} shows the four designs that match the target configuration, each corresponding to the four ring sizes ($X \in \{XS, S, M, L\}$) and selected out of the 80 program variations with $n \in [1..4]$, $l \in [0, 45.7]$ (discretized into 20 values).

\section{Interfacing with FabHaL}

FabHacks is built on top of the S-DSL FabHaL and in this section we introduce how users can interface with the DSL through a user interface, in addition to directly writing programs in FabHaL.

\subsection{The FabHacks Interface and User Workflow} \label{subsec:interactions}

As introduced in Section~\ref{subsec:langconstructs}, to construct a hack design, users start with specifying a starting environment with \texttt{add}, a target part's configuration relative to the environment with \texttt{end\_with}, and then create connections between connector primitives on two parts with \texttt{connect}. Then they can use \texttt{solve} to check design validity and solve for the configuration of their design under gravity. Based on feedback about whether a connection is valid and the visual feedback shown in the UI, users can choose to iterate on their design as needed.

\begin{figure}[ht]
\centering
\begin{subfigure}{\linewidth}
\includegraphics[width=\textwidth]{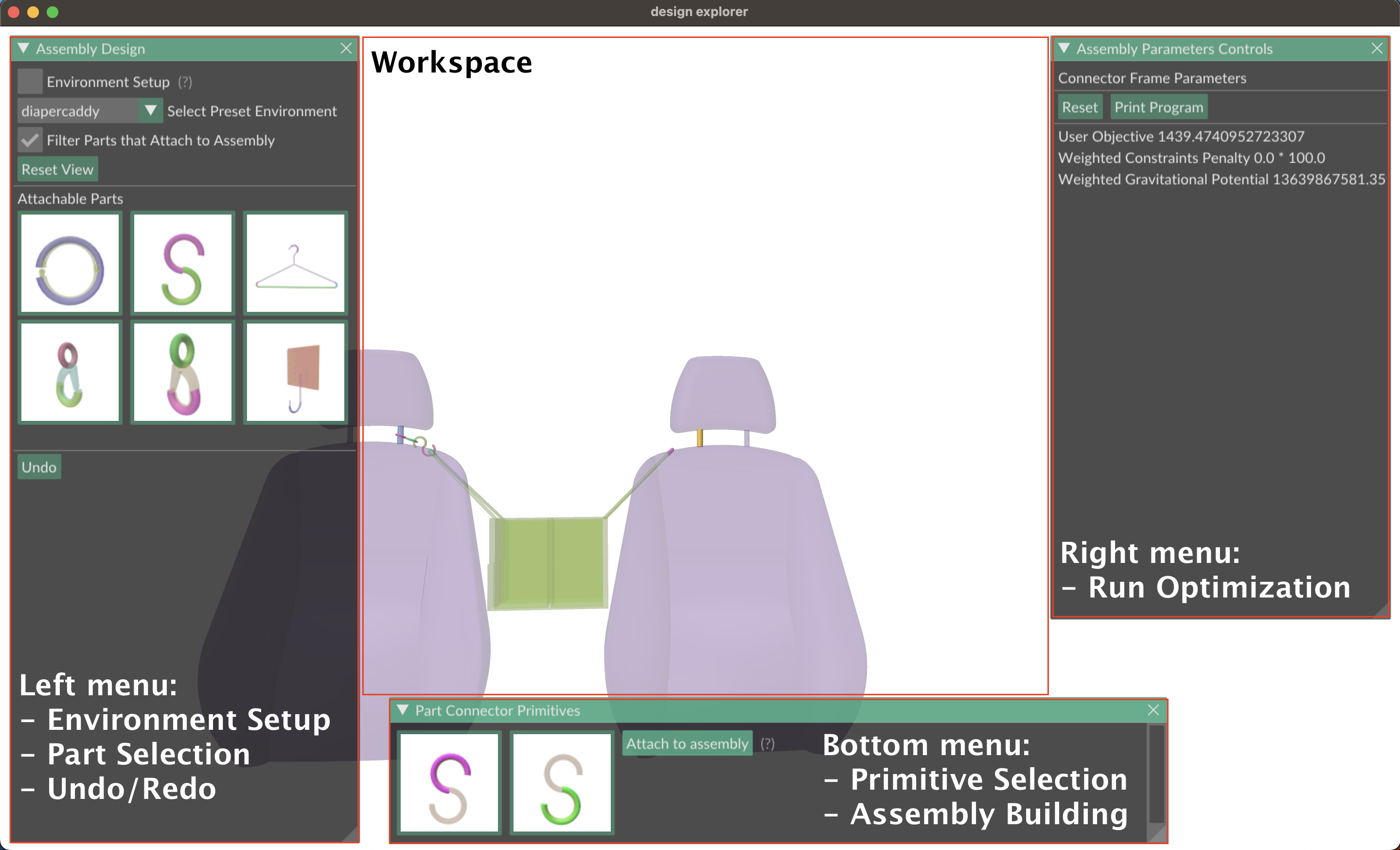}
\vspace{-15pt}
\caption{The user interface.}
\label{fig:uiblocks}
\end{subfigure}
\\
\begin{subfigure}{0.5\linewidth}
\includegraphics[width=\textwidth]
{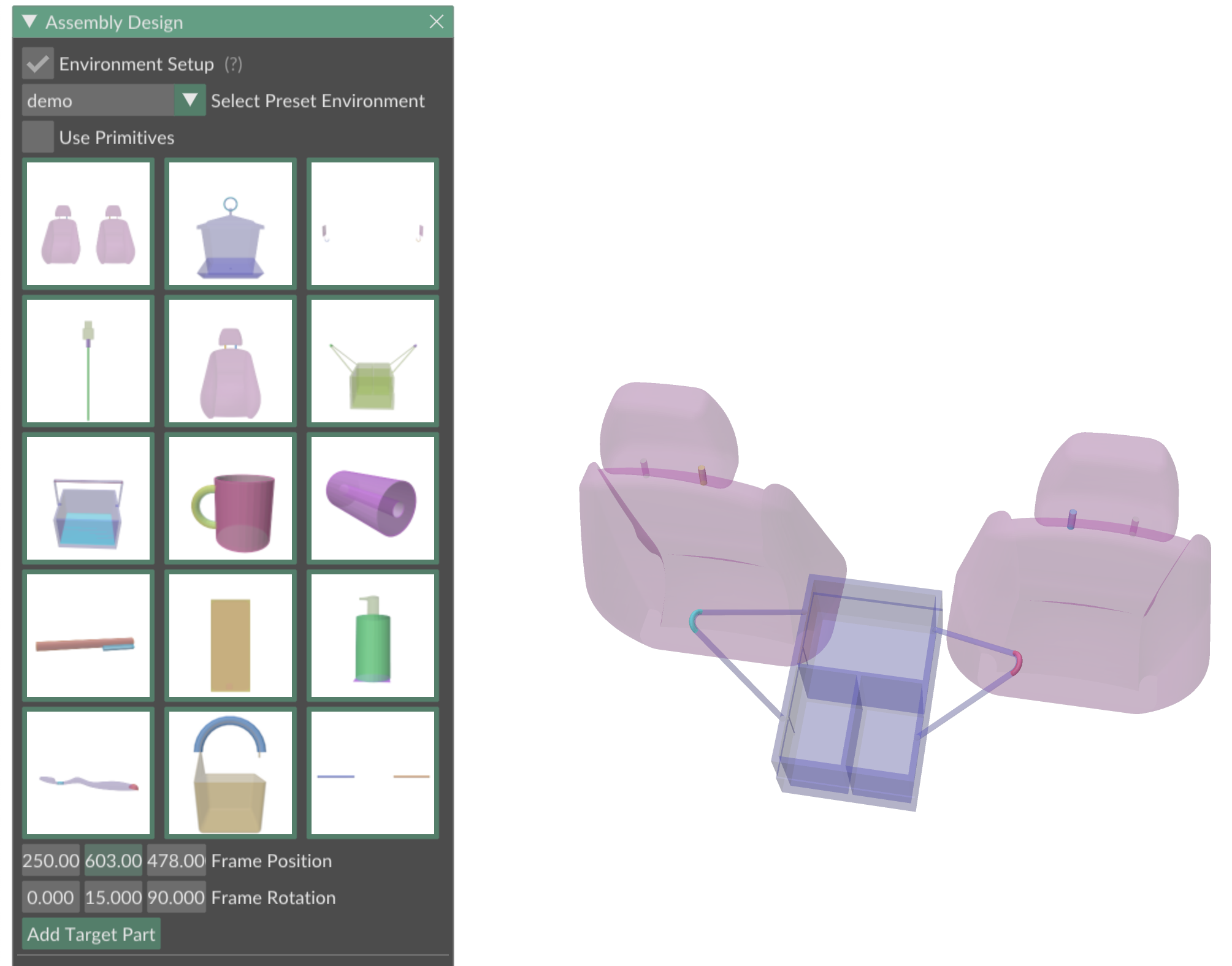}
\caption{Step 1: the UI in the process of setting up the target part (a diaper caddy) relative to the environment (car seats).}
\label{fig:mode1}
\end{subfigure}
\hfill
\begin{subfigure}{0.46\linewidth}
\includegraphics[width=\textwidth]{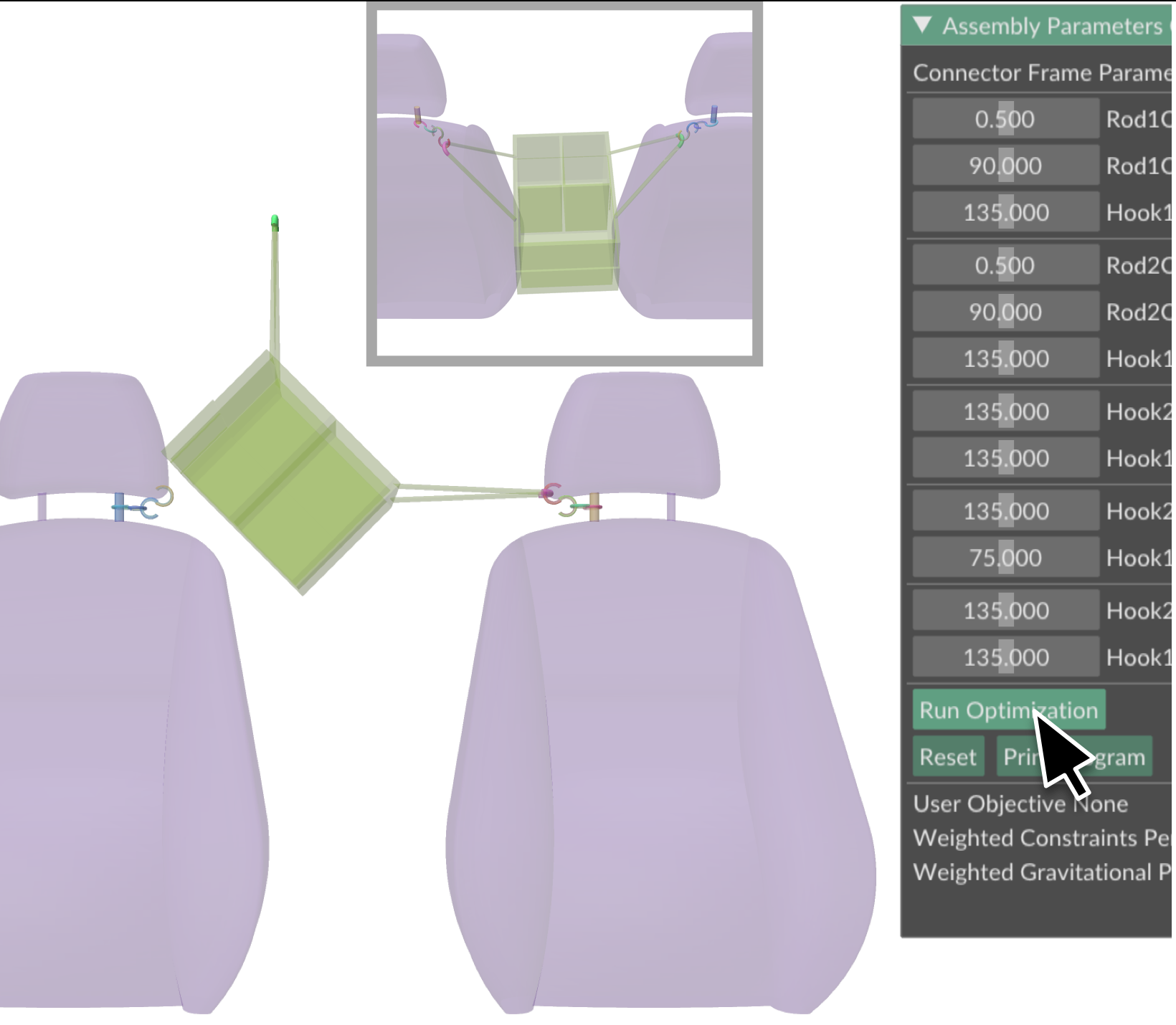}
\caption{Step 3: The user clicks on ``Run Optimization'' and  the inset shows the solved configuration.}
\label{fig:mode3}
\end{subfigure}
\vspace{-10pt}
\caption{Top: a screenshot of the user interface. Bottom: example interactions for Steps 1 (left) and 3 (right).}
\label{fig:interactions}
\end{figure}

The UI consists of the workspace region and three menus (see Figure~\ref{fig:uiblocks}). The left menu is mainly used during environment setup and for selecting parts to use in assembly design; the parts shown here are all from the ``Annotated Object Library''. The bottom menu is for selecting connector primitives of the part selected from the left menu and here is where the buttons for constructing the assembly will show up. The right menu is for solving for the assembly's final configuration.

We design the UI to have interactions that roughly correspond to the program construction process. 

\paragraph{Step 1: Environment Setup}
Users start by setting up the environments where this hack will be situated in. Taking the diaper caddy hanging hack (Figure~\ref{fig:ex13}) as an example, in Figure~\ref{fig:mode1}, we have already added the car seats as the starting environment, and are in the process of specifying the configuration of the target part (diaper caddy) relative to the environment. The desired configurations (position and orientation) of the environment and the target part can be changed with sliders. This finishes the environment setup, which corresponds to \texttt{add} and \texttt{end\_with} in the program.

\paragraph{Step 2: Assembly Design}\label{subsubsec:mode2}
Next, users construct the assembly by specifying which connections to make. The user can either select a part from the left menu to connect it to the assembly, or select two connector primitives already in the assembly and specify that they should be connected. As defined in our DSL, \texttt{connect} might introduce unsatisfiable constraints and thus needs to be verified before it can be done. To reduce some pre-checks that need to be done, we provide two-way filtering based on the connectivity table (Table~\ref{tab:connectivity}). For example, if a hook is selected in the menu, then only hook, hole, rod, and tube primitives will be enabled for selection in the workspace, and vice versa for a hook selected in the workspace. If a connection cannot be made because of failed pre-checks or that the solver cannot find a valid set of parameters that satisfy the constraints, specific feedback will be provided to the user.

\paragraph{Step 3: Solving}\label{subsubsec:optimize}
After the environment and the target part are fully connected, the assembly is considered ``valid'' and the physics solver can be run. The user can invoke \texttt{solve} with the button ``Run Optimization'' and the FabHaL representation of the assembly will be used to solve for a final configuration that minimizes changes in the configuration of the target part and is stable under gravity subject to any constraints in the design (Figure~\ref{fig:mode3}). We note that this is not an interactive-rate step because the full physics-based solving can take up to a few minutes for complicated assemblies.

After the user sees the visual or textual feedback on their design, they can choose to continue modifying it either with some backtracking via undo and redo buttons, or just adding more parts, and re-run the solve to view the updated design.

\section{Evaluation of FabHacks}\label{sec:results}

In this section, we discuss implementation details and describe how we evaluated our system with a user study, which shows that our UI provides an effective way to interface with FabHaL for users.

\subsection{Implementation}\label{subsec:impl}
As mentioned in Section~\ref{subsec:programmatic}, FabHaL is implemented using a shallow embedding strategy in Python. Using Python as the host language allows easy integration with existing optimization and geometry processing libraries in our implementation of the solver~\cite{2020SciPy-NMeth,polyscope,libigl}. The UI is also implemented in Python using polyscope~\shortcite{polyscope} with extended features from imgui. 

\subsection{User Study with FabHacks Interface}\label{subsec:userstudy}

We evaluate how useful our tool is for hack designs through a user study with ten participants. Participant ages ranged from 18--34 with existing CAD experience ranging from none to greater than 5 years. Participants reported their gender as Male (5); Female (3); Non-Binary (1) and N/A (1). The user study was conducted in our lab using a machine we provided and took about an hour in total. Audio and screen capture (with highlighted clicks) were recorded during the study; click events were also recorded in a data file.

\subsubsection{Method}

After consenting participants, we showed them a short tutorial in which we constructed the ``Demo'' assembly (see inset Figure~\ref{fig:ex0demo}) in FabHacks and in the CAD tool (OnShape) interface and the participants repeated the same steps in both tools. Participants were then asked to complete two tasks and think aloud in the process. After the study, participants were asked to answer three questions: ``Can you tell us up to three things you'd like to see us keep in the FabHacks tool?''; ``Can you tell us up to three things you'd like to see changed in the FabHacks tool?'' and ``Can you think of a change you'd like to make to your space in the office or at home that the FabHacks tool could help you with?''

\paragraph{Task 1} Participants were then instructed to replicate a given hack design in FabHacks and in the CAD tool. Specifically, participants were told to hang a basket between two towel rods using two hooks and were shown an example final assembly (see Figure~\ref{fig:dslExample}). In both the CAD tool and FabHacks, they were given an interface with two rods fixed in place (the environment) and a basket (target) that should be fixed in the space between them. In the CAD tool, the necessary parts to complete the assembly were already inserted into the same local coordinate space as the rods and basket, but not connected. In FabHacks, the user was expected to select those parts from the FabHacks menu. In both tools, the user was given about 10 minutes to replicate the assembly they had been shown, after which they could stop when they were frustrated even if they had not succeeded. 

\paragraph{Task 2} The second task was an open-ended design task where the user was asked to hang a bird feeder from two hooks but was not shown a solution. Again, the wall hooks (environment) and the bird feeder (target) were given. They were given up to 30 minutes to create a design and were asked to come up with additional designs if time remained.

\subsubsection{Measures}

For Task 1, we extracted \textit{Clicks}; \textit{Actions}; and \textit{Backtracks}. Clicks was simply the total number of recorded clicks from the start to the end of a task. However, since the number of raw clicks does not directly reflect the user experience, we also coded the data into Actions including \textbf{creating a connection}, \textbf{editing a connection}, \textbf{moving a part}, and \textbf{optimizing the solution}. The last two action categories (move a part and optimize the solution) only exist in the CAD and FabHacks data, respectively. Manually positioning parts is necessary in traditional CAD, whereas final positions are determined via simulation and optimization in FabHacks. Actions might include multiple clicks (such as when rotating a part). Finally, Backtracks include operations such as \textit{undo} or \textit{redo} or the equivalent action (such as deleting a connection). Repeated backtrack actions (without any other action between them) were counted as a single Backtrack data point.

For Task 2, we coded all designs as \textit{feasible} or \textit{infeasible}; and then grouped them into categories based on similarity. Finally, we grouped questionnaire responses into categories and discussed them until we reached a consensus. 

\subsubsection{Results and Discussions}\label{sec:studyresults}

Overall, our study demonstrated that FabHacks is an efficient and intuitive way to construct assemblies.

\begin{figure}[ht]
\centering
\includegraphics[width=\linewidth]{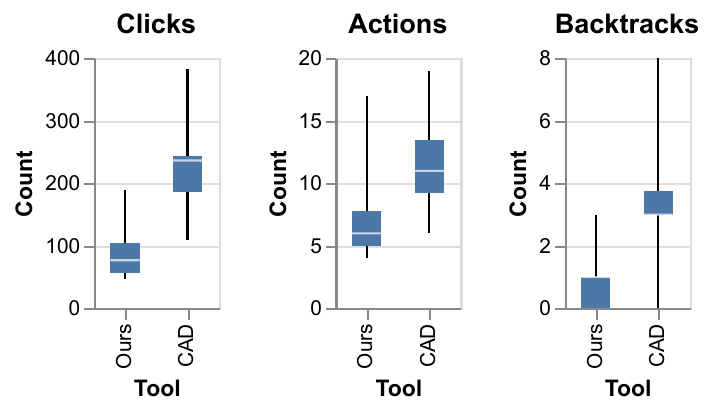}
\vspace{-10pt}
\caption{Quantitative comparison of assembly modeling in FabHacks versus traditional CAD (lower is better). Our tool is significantly more efficient, measured by total clicks and modeling actions, and also more intuitive, measured by the number of times each participant backtracked in the modeling process.}
\label{fig:task1_efficiency_and_intuitiveness}
\end{figure}

\begin{figure}[ht]
    \centering
    \includegraphics[width=\linewidth]{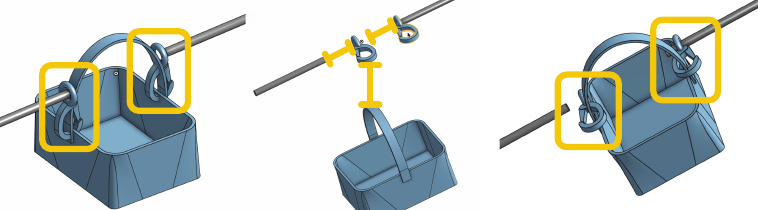}
    \caption{Common problems with CAD models produced in Task 1: part intersections, disconnected parts, non-physical positions and orientations.}
    \label{fig:cadfails}
\end{figure}

\paragraph{Task 1} We found that FabHacks assemblies could be constructed with fewer clicks, fewer actions, and fewer backtracks than OnShape assemblies, as summarized in  Figure~\ref{fig:task1_efficiency_and_intuitiveness}. Further, every participant created a valid and correct model of the target assembly in FabHacks, while none were able to do so in traditional CAD; all models were either missing degrees of freedom and/or were overconstrained and thus not valid for the CAD program's constraint solver. Some kinds of model errors seen in the CAD tool, illustrated in Figure~\ref{fig:cadfails}, were part intersections, disconnected parts, and non-physical positions and orientations.

\paragraph{Task 2} Between the 10 participants, 25 feasible designs were created, each of which was unique, though many used similar strategies. Twenty-three of these belong to one of four common strategies: (A) constructing symmetric chains of small objects to anchor the birdfeeder between the two hooks (7 instances), (B) constructing two short chains, hanging a coat hanger upside-down between them, and dangling the birdfeeder from the hook of the coat hanger (6 instances), (C) hanging a coat hanger from each wall hook and anchoring the bird feeder where they meet (8 instances), and (D) chaining two coat hangers from each wall hook and connecting the birdfeeder in the middle of them (2 instances). While several users discovered each pattern, no two were identical; users chose different types or numbers of parts to achieve similar constructions, or connected the same parts in different ways.

\begin{figure*}
    \centering
    \includegraphics[width=\linewidth]{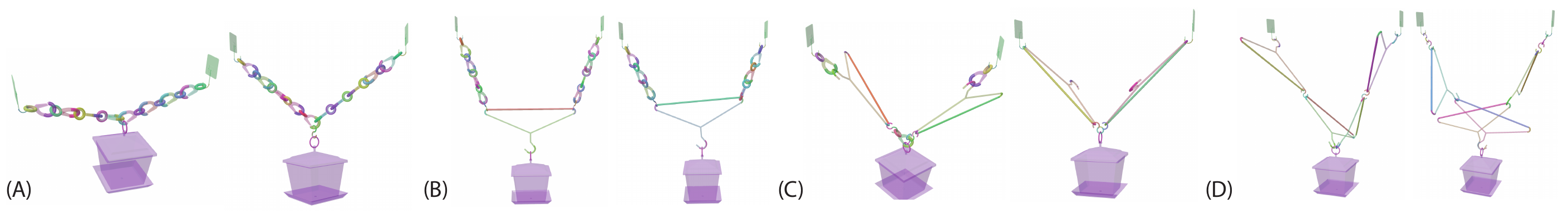}
    \caption{Examples of each of the 4 common design strategies (A-D) found by participants in Task 2.}
    \label{fig:task2categories}
\end{figure*}

Looking at participants' answers to our three questions, we saw several important themes arise. First, multiple participants liked how ``intuitive'' FabHacks was and praised the physics solver. One participant praised the ``real-time realistic feedback'' on connections and another praised the ``simplicity'' of making connections in FabHacks. At the same time, participants noted areas for improvement. For example, multiple users mentioned that ``not knowing the reason a [connection] is failing [when validation is run] can be frustrating'' and asked for a wider variety of undo and delete operations (a simple feature to add). Participants also made suggestions such as having a constraint on the number of available pieces; better support for orienting, panning, and zooming; a tree diagram showing the connections; and better feedback about what is selected. These critiques generally represent opportunities for improved user experience design rather than fundamental flaws with the mental model required to use our tool. For example, it would be possible to tell the user if a connection is failing because of a geometric flaw such as a part not being long enough to connect, or to visualize the best solution (connection state) found by the validation, highlighting failed connections.

We also found that 6/10 participants had concrete ideas for how they would use FabHacks in their everyday life, from a tree swing to outdoor lights to wall hangers to hang decorations or photos. Of four participants who did not see a use for FabHacks one felt that the library needed to be expanded and account for things like weight; one felt they could more easily make a plan in their head; and two just did not have an idea for how to use it. Although participant comments suggest that there is room for improvement, these mostly focus on things that can be solved with a larger library and iteration on the user experience. Future work could explore adding physical properties like weight to our physics solver or letting users choose from a library of materials for their parts.

\section{Limitations and Future Work}

This work introduces a design system that enables users to create fixture hacks built out of household items. Our solution is centered around a new Solver-aided DSL, FabHaL, which was inspired by the analysis of a collection of hacks. With the solver-aided paradigm, FabHaL allows users to create partial specifications of a hack, which simplifies the design of hacks to connecting primitives between parts. Our study showed that FabHacks is much better than existing CAD tools at supporting end-user construction of assemblies on quantitative measures and that participants found it intuitive to use. Participants identified potential opportunities for using FabHacks in everyday life, suggesting that the ideas presented in this work can inspire a new age of sustainable DIY design.

A limitation of our system is that it excludes any hacks using soft parts or examples where the shape of the parts can be altered during assembly, such as a piece of wood that can be cut to size. An important future direction of this work is to extend the proposed abstractions to handle these additional scenarios. For example, our parametric connections can be expanded to accommodate additional degrees of freedom, allowing for the representation of deformable objects or items that can change dimensions when cut. Additionally, more physical solvers can be incorporated to handle deformable shapes and more complex part-part interactions. In these scenarios, it may be valuable to analyze if exposing other degrees of freedom to the user in the UI, such as connection parameters, may support design exploration. 

Our system also presents opportunities for automating how parts that make up the library are created. For example, it would be interesting to explore automated recognition and fitting of connection primitives given a 3D model of a part. A step further would be to automatically add a part to the library from LIDAR data or multiple images of an object, which would further increase the modeling power of the system and help bridge the reality gap. 

Another promising opportunity is the complete automation of assembly design. By abstracting out eight common connector primitives and rules on their connection behaviors, our proposed DSL not only supports interactive design but has the potential to facilitate the generation of optimal designs under various objectives because it fundamentally reduces the search space. Automating the design of home hacks is a challenging task, as it involves searching through discrete combinations of parts and finding suitable continuous parameters that meet the specifications. Our abstractions enable us to decouple this problem into a program synthesis task nested with continuous optimization, which is performed by our solver. How to make program synthesis techniques usable in this context is an interesting research problem. 

FabHaL as a DSL can also benefit from the recent advances in large language models. Recent experiments that use LLMs for generating~\cite{jain_generating_2023, skreta_errors_2023} or completing programs~\cite{piereder_using_2024} in various DSLs show promising results. In preliminary experiments, we prompted GPT-4~\cite{chatgpt} to design a hack for hanging the birdfeeder with eyehooks, S-hooks, and hangers. While most attempts don't lead to a desirable design, \begin{wrapfigure}{l}{70pt}
    \vspace{-12pt}
    \centering
    \includegraphics[width=90pt]{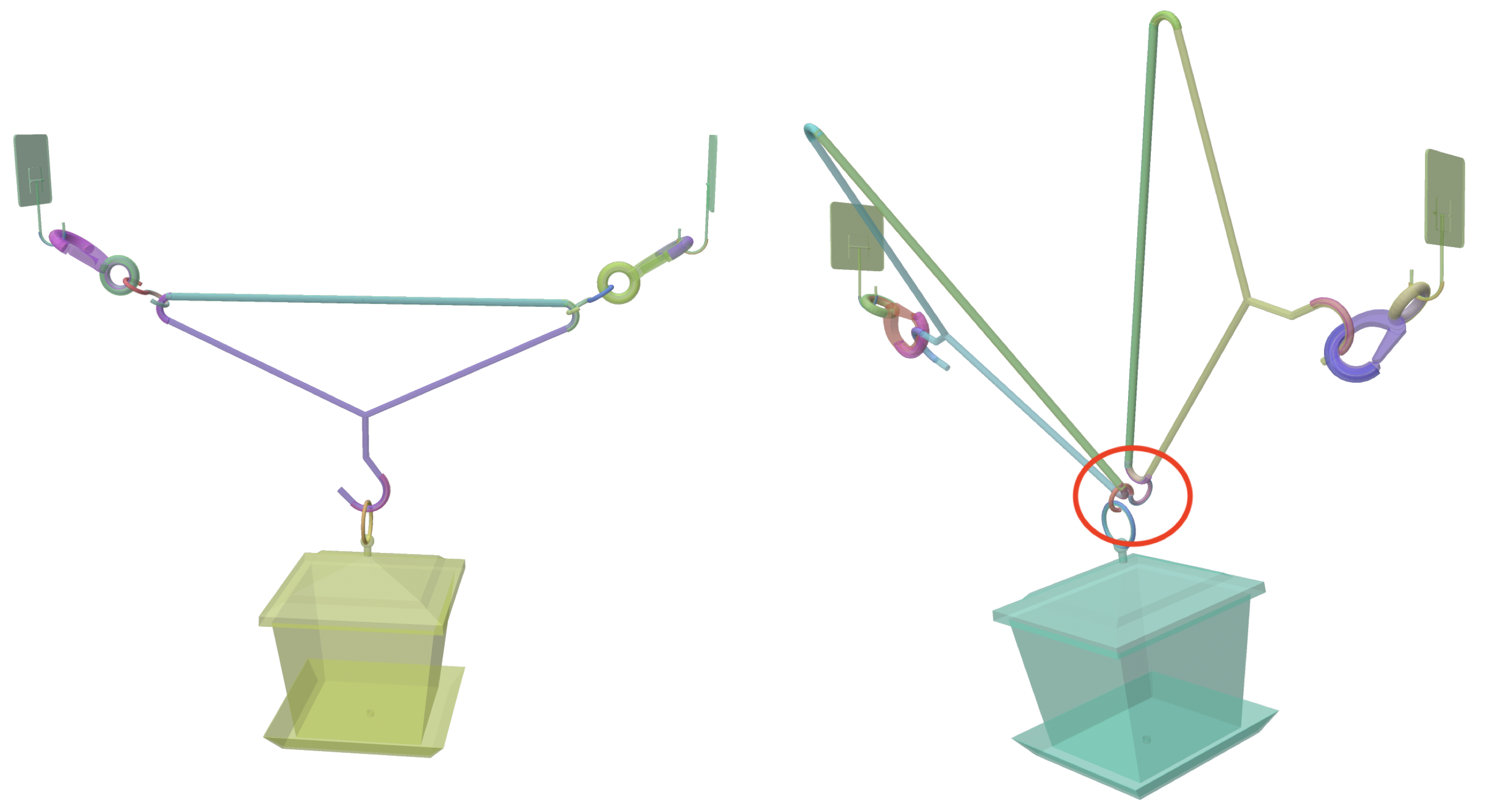}
    \vspace{-23pt}
\end{wrapfigure}GPT-4 was able to come up with valid and near-valid designs (see inset). Here the left of the inset shows a design created by GPT-4 that is very close to our design shown in Figure~\ref{fig:teaser}, and the right shows a design that is not physically valid when evaluated with our solver (the S-hook will disconnect and fall off) but is similar to the user-created designs using strategy C in Figure~\ref{fig:task2categories}. It is an exciting direction to explore how to enable LLMs to create physically valid hack designs with FabHaL, where the underlying solver can become useful in generating feedback based on the solve results to prompt LLMs to fix issues in invalid or undesired designs.

\section{Conclusion}
We create a solver-aided domain-specific language, FabHaL, to represent ``hacks'' repurposed from everyday objects. The language design is based on the analysis of a collection of hacks where we found that hacks are made up of common types of connections between common primitive shapes. The DSL can represent fixture designs compactly and at the same time verify and simulate the design with the help of a solver. We show with a user study that our system can help users create valid hack designs.

\bibliographystyle{ACM-Reference-Format}
\bibliography{citations}


\begin{thebibliography}{47}


\ifx \showCODEN    \undefined \def \showCODEN     #1{\unskip}     \fi
\ifx \showDOI      \undefined \def \showDOI       #1{#1}\fi
\ifx \showISBNx    \undefined \def \showISBNx     #1{\unskip}     \fi
\ifx \showISBNxiii \undefined \def \showISBNxiii  #1{\unskip}     \fi
\ifx \showISSN     \undefined \def \showISSN      #1{\unskip}     \fi
\ifx \showLCCN     \undefined \def \showLCCN      #1{\unskip}     \fi
\ifx \shownote     \undefined \def \shownote      #1{#1}          \fi
\ifx \showarticletitle \undefined \def \showarticletitle #1{#1}   \fi
\ifx \showURL      \undefined \def \showURL       {\relax}        \fi
\providecommand\bibfield[2]{#2}
\providecommand\bibinfo[2]{#2}
\providecommand\natexlab[1]{#1}
\providecommand\showeprint[2][]{arXiv:#2}

\bibitem[Arabi and Kim(2022)]%
        {arabi_augmenting_2022}
\bibfield{author}{\bibinfo{person}{Abul~Al Arabi} {and} \bibinfo{person}{Jeeeun Kim}.} \bibinfo{year}{2022}\natexlab{}.
\newblock \showarticletitle{Augmenting {Everyday} {Objects} into {Personal} {Robotic} {Devices}}. In \bibinfo{booktitle}{\emph{{SIGGRAPH} {Asia} 2022 {Emerging} {Technologies}}}. \bibinfo{publisher}{ACM}, \bibinfo{address}{Daegu Republic of Korea}, \bibinfo{pages}{1--2}.
\newblock
\showISBNx{9781450394727}
\urldef\tempurl%
\url{https://doi.org/10.1145/3550471.3564763}
\showDOI{\tempurl}


\bibitem[Arabi et~al\mbox{.}(2022)]%
        {arabi_mobiot:_2022}
\bibfield{author}{\bibinfo{person}{Abul~Al Arabi}, \bibinfo{person}{Jiahao Li}, \bibinfo{person}{Xiang~'Anthony Chen}, {and} \bibinfo{person}{Jeeeun Kim}.} \bibinfo{year}{2022}\natexlab{}.
\newblock \showarticletitle{Mobiot: {Augmenting} {Everyday} {Objects} into {Moving} {IoT} {Devices} {Using} {3D} {Printed} {Attachments} {Generated} by {Demonstration}}. In \bibinfo{booktitle}{\emph{{CHI} {Conference} on {Human} {Factors} in {Computing} {Systems}}}. \bibinfo{publisher}{ACM}, \bibinfo{address}{New Orleans LA USA}, \bibinfo{pages}{1--14}.
\newblock
\showISBNx{9781450391573}
\urldef\tempurl%
\url{https://doi.org/10.1145/3491102.3517645}
\showDOI{\tempurl}


\bibitem[Caylor(2019)]%
        {homehacksofficial_75_2019}
\bibfield{author}{\bibinfo{person}{Marilyn Caylor}.} \bibinfo{year}{2019}\natexlab{}.
\newblock \bibinfo{title}{75 super easy ways to organize your entire home}.
\newblock
\newblock
\urldef\tempurl%
\url{https://homehacks.co/easy-home-organizational-tips/}
\showURL{%
\tempurl}


\bibitem[Chen et~al\mbox{.}(2018)]%
        {chen_medley:_2018}
\bibfield{author}{\bibinfo{person}{Xiang~'Anthony' Chen}, \bibinfo{person}{Stelian Coros}, {and} \bibinfo{person}{Scott~E. Hudson}.} \bibinfo{year}{2018}\natexlab{}.
\newblock \showarticletitle{Medley: {A} {Library} of {Embeddables} to {Explore} {Rich} {Material} {Properties} for {3D} {Printed} {Objects}}. In \bibinfo{booktitle}{\emph{Proceedings of the 2018 {CHI} {Conference} on {Human} {Factors} in {Computing} {Systems}}}. \bibinfo{publisher}{ACM}, \bibinfo{address}{Montreal QC Canada}, \bibinfo{pages}{1--12}.
\newblock
\showISBNx{9781450356206}
\urldef\tempurl%
\url{https://doi.org/10.1145/3173574.3173736}
\showDOI{\tempurl}


\bibitem[Chen et~al\mbox{.}(2015)]%
        {chen_encore:_2015}
\bibfield{author}{\bibinfo{person}{Xiang~'Anthony' Chen}, \bibinfo{person}{Stelian Coros}, \bibinfo{person}{Jennifer Mankoff}, {and} \bibinfo{person}{Scott~E. Hudson}.} \bibinfo{year}{2015}\natexlab{}.
\newblock \showarticletitle{Encore: {3D} {Printed} {Augmentation} of {Everyday} {Objects} with {Printed}-{Over}, {Affixed} and {Interlocked} {Attachments}}. In \bibinfo{booktitle}{\emph{Proceedings of the 28th {Annual} {ACM} {Symposium} on {User} {Interface} {Software} \& {Technology}}} \emph{(\bibinfo{series}{{UIST} '15})}. \bibinfo{publisher}{ACM}, \bibinfo{address}{New York, NY, USA}, \bibinfo{pages}{73--82}.
\newblock
\showISBNx{978-1-4503-3779-3}
\urldef\tempurl%
\url{https://doi.org/10.1145/2807442.2807498}
\showDOI{\tempurl}
\newblock
\shownote{event-place: Charlotte, NC, USA}.


\bibitem[Chen et~al\mbox{.}(2016)]%
        {chen_reprise:_2016}
\bibfield{author}{\bibinfo{person}{Xiang~'Anthony' Chen}, \bibinfo{person}{Jeeeun Kim}, \bibinfo{person}{Jennifer Mankoff}, \bibinfo{person}{Tovi Grossman}, \bibinfo{person}{Stelian Coros}, {and} \bibinfo{person}{Scott~E. Hudson}.} \bibinfo{year}{2016}\natexlab{}.
\newblock \showarticletitle{Reprise: {A} {Design} {Tool} for {Specifying}, {Generating}, and {Customizing} {3D} {Printable} {Adaptations} on {Everyday} {Objects}}. In \bibinfo{booktitle}{\emph{Proceedings of the 29th {Annual} {Symposium} on {User} {Interface} {Software} and {Technology}}} \emph{(\bibinfo{series}{{UIST} '16})}. \bibinfo{publisher}{ACM}, \bibinfo{address}{New York, NY, USA}, \bibinfo{pages}{29--39}.
\newblock
\showISBNx{978-1-4503-4189-9}
\urldef\tempurl%
\url{https://doi.org/10.1145/2984511.2984512}
\showDOI{\tempurl}
\newblock
\shownote{event-place: Tokyo, Japan}.


\bibitem[Choi and Ishii(2021)]%
        {choi_therms-up!:_2021}
\bibfield{author}{\bibinfo{person}{Kyung~Yun Choi} {and} \bibinfo{person}{Hiroshi Ishii}.} \bibinfo{year}{2021}\natexlab{}.
\newblock \showarticletitle{Therms-{Up}!: {DIY} {Inflatables} and {Interactive} {Materials} by {Upcycling} {Wasted} {Thermoplastic} {Bags}}. In \bibinfo{booktitle}{\emph{Proceedings of the {Fifteenth} {International} {Conference} on {Tangible}, {Embedded}, and {Embodied} {Interaction}}}. \bibinfo{publisher}{ACM}, \bibinfo{address}{Salzburg Austria}, \bibinfo{pages}{1--8}.
\newblock
\showISBNx{9781450382137}
\urldef\tempurl%
\url{https://doi.org/10.1145/3430524.3442457}
\showDOI{\tempurl}


\bibitem[Crafts(2022)]%
        {noauthor_5-minute_nodate}
\bibfield{author}{\bibinfo{person}{5-Minute Crafts}.} \bibinfo{year}{2022}\natexlab{}.
\newblock \bibinfo{title}{5-{Minute} {Crafts} — {Learn}. {Create}. {Improve}.}
\newblock
\newblock
\urldef\tempurl%
\url{https://5minutecrafts.site/}
\showURL{%
\tempurl}


\bibitem[Davidoff et~al\mbox{.}(2011)]%
        {davidoff_mechanical_2011}
\bibfield{author}{\bibinfo{person}{Scott Davidoff}, \bibinfo{person}{Nicolas Villar}, \bibinfo{person}{Alex~S. Taylor}, {and} \bibinfo{person}{Shahram Izadi}.} \bibinfo{year}{2011}\natexlab{}.
\newblock \showarticletitle{Mechanical hijacking: how robots can accelerate {UbiComp} deployments}. In \bibinfo{booktitle}{\emph{Proceedings of the 13th international conference on {Ubiquitous} computing}}. \bibinfo{publisher}{ACM}, \bibinfo{address}{Beijing China}, \bibinfo{pages}{267--270}.
\newblock
\showISBNx{9781450306300}
\urldef\tempurl%
\url{https://doi.org/10.1145/2030112.2030148}
\showDOI{\tempurl}


\bibitem[Featherstone(1983)]%
        {featherstone_calculation_1983}
\bibfield{author}{\bibinfo{person}{R. Featherstone}.} \bibinfo{year}{1983}\natexlab{}.
\newblock \showarticletitle{The {Calculation} of {Robot} {Dynamics} {Using} {Articulated}-{Body} {Inertias}}.
\newblock \bibinfo{journal}{\emph{The International Journal of Robotics Research}} \bibinfo{volume}{2}, \bibinfo{number}{1} (\bibinfo{date}{March} \bibinfo{year}{1983}), \bibinfo{pages}{13--30}.
\newblock
\showISSN{0278-3649, 1741-3176}
\urldef\tempurl%
\url{https://doi.org/10.1177/027836498300200102}
\showDOI{\tempurl}


\bibitem[Fisher et~al\mbox{.}(2014)]%
        {fisher2014biology}
\bibfield{author}{\bibinfo{person}{Jasmin Fisher}, \bibinfo{person}{Nir Piterman}, {and} \bibinfo{person}{Rastislav Bodik}.} \bibinfo{year}{2014}\natexlab{}.
\newblock \showarticletitle{Toward Synthesizing Executable Models in Biology}.
\newblock \bibinfo{journal}{\emph{Frontiers in Bioengineering and Biotechnology}}  \bibinfo{volume}{2} (\bibinfo{year}{2014}), \bibinfo{pages}{1--8}.
\newblock
\showISSN{2296-4185}
\urldef\tempurl%
\url{https://doi.org/10.3389/fbioe.2014.00075}
\showDOI{\tempurl}


\bibitem[Fiyaa(2013)]%
        {noauthor_15_2013}
\bibfield{author}{\bibinfo{person}{Fiyaa}.} \bibinfo{year}{2013}\natexlab{}.
\newblock \bibinfo{title}{15 {Cord} {Management} {Life} {Hacks} for {No} {More} {Tangled} {Wires}}.
\newblock
\newblock
\urldef\tempurl%
\url{https://www.amazinginteriordesign.com/15-cord-management-life-hacks-for-no-more-tangled-wires/}
\showURL{%
\tempurl}


\bibitem[Guo et~al\mbox{.}(2017)]%
        {guo_facade:_2017}
\bibfield{author}{\bibinfo{person}{Anhong Guo}, \bibinfo{person}{Jeeeun Kim}, \bibinfo{person}{Xiang~'Anthony' Chen}, \bibinfo{person}{Tom Yeh}, \bibinfo{person}{Scott~E. Hudson}, \bibinfo{person}{Jennifer Mankoff}, {and} \bibinfo{person}{Jeffrey~P. Bigham}.} \bibinfo{year}{2017}\natexlab{}.
\newblock \showarticletitle{Facade: {Auto}-generating {Tactile} {Interfaces} to {Appliances}}. In \bibinfo{booktitle}{\emph{Proceedings of the 2017 {CHI} {Conference} on {Human} {Factors} in {Computing} {Systems}}}. \bibinfo{publisher}{ACM}, \bibinfo{address}{Denver Colorado USA}, \bibinfo{pages}{5826--5838}.
\newblock
\showISBNx{9781450346559}
\urldef\tempurl%
\url{https://doi.org/10.1145/3025453.3025845}
\showDOI{\tempurl}


\bibitem[Hofmann et~al\mbox{.}(2018)]%
        {hofmann_greater_2018}
\bibfield{author}{\bibinfo{person}{Megan Hofmann}, \bibinfo{person}{Gabriella Hann}, \bibinfo{person}{Scott~E. Hudson}, {and} \bibinfo{person}{Jennifer Mankoff}.} \bibinfo{year}{2018}\natexlab{}.
\newblock \showarticletitle{Greater than the {Sum} of its {PARTs}: {Expressing} and {Reusing} {Design} {Intent} in {3D} {Models}}. In \bibinfo{booktitle}{\emph{Proceedings of the 2018 {CHI} {Conference} on {Human} {Factors} in {Computing} {Systems}}}. \bibinfo{publisher}{ACM}, \bibinfo{address}{Montreal QC Canada}, \bibinfo{pages}{1--12}.
\newblock
\showISBNx{9781450356206}
\urldef\tempurl%
\url{https://doi.org/10.1145/3173574.3173875}
\showDOI{\tempurl}


\bibitem[Hottelier et~al\mbox{.}(2014)]%
        {hottelier_programming_2014}
\bibfield{author}{\bibinfo{person}{Thibaud Hottelier}, \bibinfo{person}{Ras Bodik}, {and} \bibinfo{person}{Kimiko Ryokai}.} \bibinfo{year}{2014}\natexlab{}.
\newblock \showarticletitle{Programming by manipulation for layout}. In \bibinfo{booktitle}{\emph{Proceedings of the 27th annual {ACM} symposium on {User} interface software and technology}}. \bibinfo{publisher}{ACM}, \bibinfo{address}{Honolulu Hawaii USA}, \bibinfo{pages}{231--241}.
\newblock
\showISBNx{9781450330695}
\urldef\tempurl%
\url{https://doi.org/10.1145/2642918.2647378}
\showDOI{\tempurl}


\bibitem[Jacobson et~al\mbox{.}(2018)]%
        {libigl}
\bibfield{author}{\bibinfo{person}{Alec Jacobson}, \bibinfo{person}{Daniele Panozzo}, {et~al\mbox{.}}} \bibinfo{year}{2018}\natexlab{}.
\newblock \bibinfo{title}{{libigl}: A simple {C++} geometry processing library}.
\newblock
\newblock
\newblock
\shownote{https://libigl.github.io/}.


\bibitem[Jain et~al\mbox{.}(2023)]%
        {jain_generating_2023}
\bibfield{author}{\bibinfo{person}{Rijul Jain}, \bibinfo{person}{Wode Ni}, {and} \bibinfo{person}{Joshua Sunshine}.} \bibinfo{year}{2023}\natexlab{}.
\newblock \showarticletitle{Generating {Domain}-{Specific} {Programs} for {Diagram} {Authoring} with {Large} {Language} {Models}}. In \bibinfo{booktitle}{\emph{Companion {Proceedings} of the 2023 {ACM} {SIGPLAN} {International} {Conference} on {Systems}, {Programming}, {Languages}, and {Applications}: {Software} for {Humanity}}} \emph{(\bibinfo{series}{{SPLASH} 2023})}. \bibinfo{publisher}{Association for Computing Machinery}, \bibinfo{address}{New York, NY, USA}, \bibinfo{pages}{70--71}.
\newblock
\showISBNx{9798400703843}
\urldef\tempurl%
\url{https://doi.org/10.1145/3618305.3623612}
\showDOI{\tempurl}


\bibitem[Jha et~al\mbox{.}(2010)]%
        {jha2010oracle}
\bibfield{author}{\bibinfo{person}{Susmit Jha}, \bibinfo{person}{Sumit Gulwani}, \bibinfo{person}{Sanjit~A. Seshia}, {and} \bibinfo{person}{Ashish Tiwari}.} \bibinfo{year}{2010}\natexlab{}.
\newblock \showarticletitle{Oracle-guided component-based program synthesis}. In \bibinfo{booktitle}{\emph{Proceedings of the 32nd {ACM}/{IEEE} {International} {Conference} on {Software} {Engineering} - {Volume} 1}}. \bibinfo{publisher}{ACM}, \bibinfo{address}{Cape Town South Africa}, \bibinfo{pages}{215--224}.
\newblock
\showISBNx{9781605587196}
\urldef\tempurl%
\url{https://doi.org/10.1145/1806799.1806833}
\showDOI{\tempurl}


\bibitem[Jones et~al\mbox{.}(2021)]%
        {jones_automate:_2021}
\bibfield{author}{\bibinfo{person}{Benjamin Jones}, \bibinfo{person}{Dalton Hildreth}, \bibinfo{person}{Duowen Chen}, \bibinfo{person}{Ilya Baran}, \bibinfo{person}{Vladimir~G. Kim}, {and} \bibinfo{person}{Adriana Schulz}.} \bibinfo{year}{2021}\natexlab{}.
\newblock \showarticletitle{{AutoMate}: a dataset and learning approach for automatic mating of {CAD} assemblies}.
\newblock \bibinfo{journal}{\emph{ACM Transactions on Graphics}} \bibinfo{volume}{40}, \bibinfo{number}{6} (\bibinfo{date}{Dec.} \bibinfo{year}{2021}), \bibinfo{pages}{1--18}.
\newblock
\showISSN{0730-0301, 1557-7368}
\urldef\tempurl%
\url{https://doi.org/10.1145/3478513.3480562}
\showDOI{\tempurl}


\bibitem[Jones et~al\mbox{.}(2020)]%
        {jones_shapeassembly:_2020}
\bibfield{author}{\bibinfo{person}{R.~Kenny Jones}, \bibinfo{person}{Theresa Barton}, \bibinfo{person}{Xianghao Xu}, \bibinfo{person}{Kai Wang}, \bibinfo{person}{Ellen Jiang}, \bibinfo{person}{Paul Guerrero}, \bibinfo{person}{Niloy~J. Mitra}, {and} \bibinfo{person}{Daniel Ritchie}.} \bibinfo{year}{2020}\natexlab{}.
\newblock \showarticletitle{{ShapeAssembly}: learning to generate programs for {3D} shape structure synthesis}.
\newblock \bibinfo{journal}{\emph{ACM Transactions on Graphics}} \bibinfo{volume}{39}, \bibinfo{number}{6} (\bibinfo{date}{Dec.} \bibinfo{year}{2020}), \bibinfo{pages}{1--20}.
\newblock
\showISSN{0730-0301, 1557-7368}
\urldef\tempurl%
\url{https://doi.org/10.1145/3414685.3417812}
\showDOI{\tempurl}


\bibitem[Karo(2019)]%
        {noauthor_25_2019}
\bibfield{author}{\bibinfo{person}{Karo}.} \bibinfo{year}{2019}\natexlab{}.
\newblock \bibinfo{title}{25 {IKEA} {Hacks} to {Keep} {Things} {Organized}}.
\newblock
\newblock
\urldef\tempurl%
\url{https://craftsyhacks.com/ikea-organizing/}
\showURL{%
\tempurl}


\bibitem[Koyama et~al\mbox{.}(2015)]%
        {koyama_autoconnect:_2015}
\bibfield{author}{\bibinfo{person}{Yuki Koyama}, \bibinfo{person}{Shinjiro Sueda}, \bibinfo{person}{Emma Steinhardt}, \bibinfo{person}{Takeo Igarashi}, \bibinfo{person}{Ariel Shamir}, {and} \bibinfo{person}{Wojciech Matusik}.} \bibinfo{year}{2015}\natexlab{}.
\newblock \showarticletitle{{AutoConnect}: computational design of {3D}-printable connectors}.
\newblock \bibinfo{journal}{\emph{ACM Transactions on Graphics}} \bibinfo{volume}{34}, \bibinfo{number}{6} (\bibinfo{date}{Nov.} \bibinfo{year}{2015}), \bibinfo{pages}{1--11}.
\newblock
\showISSN{0730-0301, 1557-7368}
\urldef\tempurl%
\url{https://doi.org/10.1145/2816795.2818060}
\showDOI{\tempurl}


\bibitem[Lamb et~al\mbox{.}(2019)]%
        {lamb_automated_2019}
\bibfield{author}{\bibinfo{person}{Nikolas Lamb}, \bibinfo{person}{Sean Banerjee}, {and} \bibinfo{person}{Natasha~Kholgade Banerjee}.} \bibinfo{year}{2019}\natexlab{}.
\newblock \showarticletitle{Automated {Reconstruction} of {Smoothly} {Joining} {3D} {Printed} {Restorations} to {Fix} {Broken} {Objects}}. In \bibinfo{booktitle}{\emph{Proceedings of the {ACM} {Symposium} on {Computational} {Fabrication}}} \emph{(\bibinfo{series}{{SCF} '19})}. \bibinfo{publisher}{ACM}, \bibinfo{address}{New York, NY, USA}, \bibinfo{pages}{3:1--3:12}.
\newblock
\showISBNx{978-1-4503-6795-0}
\urldef\tempurl%
\url{https://doi.org/10.1145/3328939.3329005}
\showDOI{\tempurl}
\newblock
\shownote{event-place: Pittsburgh, Pennsylvania}.


\bibitem[Li et~al\mbox{.}(2020)]%
        {li_romeo:_2020}
\bibfield{author}{\bibinfo{person}{Jiahao Li}, \bibinfo{person}{Meilin Cui}, \bibinfo{person}{Jeeeun Kim}, {and} \bibinfo{person}{Xiang~'Anthony' Chen}.} \bibinfo{year}{2020}\natexlab{}.
\newblock \showarticletitle{Romeo: {A} {Design} {Tool} for {Embedding} {Transformable} {Parts} in {3D} {Models} to {Robotically} {Augment} {Default} {Functionalities}}. In \bibinfo{booktitle}{\emph{Proceedings of the 33rd {Annual} {ACM} {Symposium} on {User} {Interface} {Software} and {Technology}}}. \bibinfo{publisher}{ACM}, \bibinfo{address}{Virtual Event USA}, \bibinfo{pages}{897--911}.
\newblock
\showISBNx{9781450375146}
\urldef\tempurl%
\url{https://doi.org/10.1145/3379337.3415826}
\showDOI{\tempurl}


\bibitem[Li et~al\mbox{.}(2019)]%
        {li_robiot:_2019}
\bibfield{author}{\bibinfo{person}{Jiahao Li}, \bibinfo{person}{Jeeeun Kim}, {and} \bibinfo{person}{Xiang~'Anthony' Chen}.} \bibinfo{year}{2019}\natexlab{}.
\newblock \showarticletitle{Robiot: {A} {Design} {Tool} for {Actuating} {Everyday} {Objects} with {Automatically} {Generated} {3D} {Printable} {Mechanisms}}. In \bibinfo{booktitle}{\emph{Proceedings of the 32nd {Annual} {ACM} {Symposium} on {User} {Interface} {Software} and {Technology}}} \emph{(\bibinfo{series}{{UIST} '19})}. \bibinfo{publisher}{Association for Computing Machinery}, \bibinfo{address}{New York, NY, USA}, \bibinfo{pages}{673--685}.
\newblock
\showISBNx{9781450368162}
\urldef\tempurl%
\url{https://doi.org/10.1145/3332165.3347894}
\showDOI{\tempurl}


\bibitem[Li et~al\mbox{.}(2022)]%
        {li_roman:_2022}
\bibfield{author}{\bibinfo{person}{Jiahao Li}, \bibinfo{person}{Alexis Samoylov}, \bibinfo{person}{Jeeeun Kim}, {and} \bibinfo{person}{Xiang~'Anthony' Chen}.} \bibinfo{year}{2022}\natexlab{}.
\newblock \showarticletitle{Roman: Making Everyday Objects Robotically Manipulable with 3D-Printable Add-on Mechanisms}. In \bibinfo{booktitle}{\emph{Proceedings of the 2022 CHI Conference on Human Factors in Computing Systems}} (New Orleans, LA, USA) \emph{(\bibinfo{series}{CHI '22})}. \bibinfo{publisher}{Association for Computing Machinery}, \bibinfo{address}{New York, NY, USA}, Article \bibinfo{articleno}{272}, \bibinfo{numpages}{17}~pages.
\newblock
\showISBNx{9781450391573}
\urldef\tempurl%
\url{https://doi.org/10.1145/3491102.3501818}
\showDOI{\tempurl}


\bibitem[Nocedal and Wright(2006)]%
        {NoceWrig06}
\bibfield{author}{\bibinfo{person}{Jorge Nocedal} {and} \bibinfo{person}{Stephen~J. Wright}.} \bibinfo{year}{2006}\natexlab{}.
\newblock \bibinfo{booktitle}{\emph{Numerical Optimization} (\bibinfo{edition}{2e} ed.)}.
\newblock \bibinfo{publisher}{Springer}, \bibinfo{address}{New York, NY, USA}.
\newblock


\bibitem[Onshape(2023)]%
        {business_onshape_nodate}
\bibfield{author}{\bibinfo{person}{Onshape}.} \bibinfo{year}{2023}\natexlab{}.
\newblock \bibinfo{title}{Onshape {\textbar} {Product} {Development} {Platform}}.
\newblock
\newblock
\urldef\tempurl%
\url{https://www.onshape.com/en/}
\showURL{%
\tempurl}


\bibitem[OpenAI(2024)]%
        {chatgpt}
\bibfield{author}{\bibinfo{person}{OpenAI}.} \bibinfo{year}{2024}\natexlab{}.
\newblock \bibinfo{title}{{ChatGPT}}.
\newblock
\newblock
\urldef\tempurl%
\url{https://chat.openai.com}
\showURL{%
\tempurl}


\bibitem[Phothilimthana et~al\mbox{.}(2019)]%
        {mangpo2019swizzle}
\bibfield{author}{\bibinfo{person}{Phitchaya~Mangpo Phothilimthana}, \bibinfo{person}{Archibald~Samuel Elliott}, \bibinfo{person}{An Wang}, \bibinfo{person}{Abhinav Jangda}, \bibinfo{person}{Bastian Hagedorn}, \bibinfo{person}{Henrik Barthels}, \bibinfo{person}{Samuel~J. Kaufman}, \bibinfo{person}{Vinod Grover}, \bibinfo{person}{Emina Torlak}, {and} \bibinfo{person}{Rastislav Bodik}.} \bibinfo{year}{2019}\natexlab{}.
\newblock \showarticletitle{Swizzle {Inventor}: {Data} {Movement} {Synthesis} for {GPU} {Kernels}}. In \bibinfo{booktitle}{\emph{Proceedings of the {Twenty}-{Fourth} {International} {Conference} on {Architectural} {Support} for {Programming} {Languages} and {Operating} {Systems}}}. \bibinfo{publisher}{ACM}, \bibinfo{address}{Providence RI USA}, \bibinfo{pages}{65--78}.
\newblock
\showISBNx{9781450362405}
\urldef\tempurl%
\url{https://doi.org/10.1145/3297858.3304059}
\showDOI{\tempurl}


\bibitem[Piereder et~al\mbox{.}(2024)]%
        {piereder_using_2024}
\bibfield{author}{\bibinfo{person}{Christina Piereder}, \bibinfo{person}{G{\"{u}}nter Fleck}, \bibinfo{person}{Verena Geist}, \bibinfo{person}{Michael Moser}, {and} \bibinfo{person}{Josef Pichler}.} \bibinfo{year}{2024}\natexlab{}.
\newblock \showarticletitle{Using {AI}-{Based} {Code} {Completion} for {Domain}-{Specific} {Languages}}. In \bibinfo{booktitle}{\emph{Product-{Focused} {Software} {Process} {Improvement}}} \emph{(\bibinfo{series}{Lecture {Notes} in {Computer} {Science}})}, \bibfield{editor}{\bibinfo{person}{Regine Kadgien}, \bibinfo{person}{Andreas Jedlitschka}, \bibinfo{person}{Andrea Janes}, \bibinfo{person}{Valentina Lenarduzzi}, {and} \bibinfo{person}{Xiaozhou Li}} (Eds.). \bibinfo{publisher}{Springer Nature Switzerland}, \bibinfo{address}{Cham}, \bibinfo{pages}{227--242}.
\newblock
\showISBNx{9783031492662}
\urldef\tempurl%
\url{https://doi.org/10.1007/978-3-031-49266-2_16}
\showDOI{\tempurl}


\bibitem[Piro(2015)]%
        {piro_8_2015}
\bibfield{author}{\bibinfo{person}{Lauren Piro}.} \bibinfo{year}{2015}\natexlab{}.
\newblock \bibinfo{title}{8 {Clutter} {Problems} {Solved} by {Shower} {Rings}}.
\newblock
\newblock
\urldef\tempurl%
\url{https://www.goodhousekeeping.com/home/decorating-ideas/shower-curtain-rings-organizing}
\showURL{%
\tempurl}


\bibitem[Ramakers et~al\mbox{.}(2016)]%
        {ramakers_retrofab:_2016}
\bibfield{author}{\bibinfo{person}{Raf Ramakers}, \bibinfo{person}{Fraser Anderson}, \bibinfo{person}{Tovi Grossman}, {and} \bibinfo{person}{George Fitzmaurice}.} \bibinfo{year}{2016}\natexlab{}.
\newblock \showarticletitle{{RetroFab}: {A} {Design} {Tool} for {Retrofitting} {Physical} {Interfaces} using {Actuators}, {Sensors} and {3D} {Printing}}. In \bibinfo{booktitle}{\emph{Proceedings of the 2016 {CHI} {Conference} on {Human} {Factors} in {Computing} {Systems}}}. \bibinfo{publisher}{ACM}, \bibinfo{address}{San Jose California USA}, \bibinfo{pages}{409--419}.
\newblock
\showISBNx{9781450333627}
\urldef\tempurl%
\url{https://doi.org/10.1145/2858036.2858485}
\showDOI{\tempurl}


\bibitem[Sharp et~al\mbox{.}(2019)]%
        {polyscope}
\bibfield{author}{\bibinfo{person}{Nicholas Sharp} {et~al\mbox{.}}} \bibinfo{year}{2019}\natexlab{}.
\newblock \bibinfo{title}{Polyscope}.
\newblock
\newblock
\newblock
\shownote{www.polyscope.run}.


\bibitem[Skreta et~al\mbox{.}(2023)]%
        {skreta_errors_2023}
\bibfield{author}{\bibinfo{person}{Marta Skreta}, \bibinfo{person}{Naruki Yoshikawa}, \bibinfo{person}{Sebastian Arellano-Rubach}, \bibinfo{person}{Zhi Ji}, \bibinfo{person}{Lasse~Bj{\o}rn Kristensen}, \bibinfo{person}{Kourosh Darvish}, \bibinfo{person}{Al{\'{a}}n Aspuru-Guzik}, \bibinfo{person}{Florian Shkurti}, {and} \bibinfo{person}{Animesh Garg}.} \bibinfo{year}{2023}\natexlab{}.
\newblock \bibinfo{title}{Errors are {Useful} {Prompts}: {Instruction} {Guided} {Task} {Programming} with {Verifier}-{Assisted} {Iterative} {Prompting}}.
\newblock
\newblock
\urldef\tempurl%
\url{https://doi.org/10.48550/arXiv.2303.14100}
\showDOI{\tempurl}
\newblock
\shownote{arXiv:2303.14100 [cs]}.


\bibitem[SOLIDWORKS(2023)]%
        {noauthor_3d_nodate}
\bibfield{author}{\bibinfo{person}{SOLIDWORKS}.} \bibinfo{year}{2023}\natexlab{}.
\newblock \bibinfo{title}{{3D} {CAD} {Design} {Software} {\textbar} {SOLIDWORKS}}.
\newblock
\newblock
\urldef\tempurl%
\url{https://www.solidworks.com/}
\showURL{%
\tempurl}


\bibitem[Stanley(2021)]%
        {stanley_33_2021}
\bibfield{author}{\bibinfo{person}{Jenny Stanley}.} \bibinfo{year}{2021}\natexlab{}.
\newblock \bibinfo{title}{33 {Brilliant} {Home} {Hacks} {Using} {Our} 3 {Favorite} {Items}}.
\newblock
\newblock
\urldef\tempurl%
\url{https://www.familyhandyman.com/list/20-home-hacks-hangers-rubber-bands-and-cardboard-tubes/}
\showURL{%
\tempurl}


\bibitem[Sullivan and Heumann(2019)]%
        {noauthor_karen_nodate}
\bibfield{author}{\bibinfo{person}{Karen Sullivan} {and} \bibinfo{person}{Jim Heumann}.} \bibinfo{year}{2019}\natexlab{}.
\newblock \bibinfo{title}{Karen and {Jim}'s {Excellent} {Adventure}: {Fiddly} {Bits}: {Making} life on a small boat safer and more comfortable}.
\newblock
\newblock
\urldef\tempurl%
\url{http://karenandjimsexcellentadventure.blogspot.com/p/fiddly-bits.html}
\showURL{%
\tempurl}


\bibitem[Teibrich et~al\mbox{.}(2015)]%
        {teibrich_patching_2015}
\bibfield{author}{\bibinfo{person}{Alexander Teibrich}, \bibinfo{person}{Stefanie Mueller}, \bibinfo{person}{FranÃ§ois GuimbretiÃšre}, \bibinfo{person}{Robert Kovacs}, \bibinfo{person}{Stefan Neubert}, {and} \bibinfo{person}{Patrick Baudisch}.} \bibinfo{year}{2015}\natexlab{}.
\newblock \showarticletitle{Patching {Physical} {Objects}}. In \bibinfo{booktitle}{\emph{Proceedings of the 28th {Annual} {ACM} {Symposium} on {User} {Interface} {Software} \& {Technology}}} \emph{(\bibinfo{series}{{UIST} '15})}. \bibinfo{publisher}{ACM}, \bibinfo{address}{New York, NY, USA}, \bibinfo{pages}{83--91}.
\newblock
\showISBNx{978-1-4503-3779-3}
\urldef\tempurl%
\url{https://doi.org/10.1145/2807442.2807467}
\showDOI{\tempurl}
\newblock
\shownote{event-place: Charlotte, NC, USA}.


\bibitem[Torlak and Bodik(2013)]%
        {rosette2013}
\bibfield{author}{\bibinfo{person}{Emina Torlak} {and} \bibinfo{person}{Rastislav Bodik}.} \bibinfo{year}{2013}\natexlab{}.
\newblock \showarticletitle{Growing Solver-Aided Languages with Rosette}. In \bibinfo{booktitle}{\emph{Proceedings of the 2013 ACM International Symposium on New Ideas, New Paradigms, and Reflections on Programming \& Software}} (Indianapolis, Indiana, USA) \emph{(\bibinfo{series}{Onward! 2013})}. \bibinfo{publisher}{Association for Computing Machinery}, \bibinfo{address}{New York, NY, USA}, \bibinfo{pages}{135–152}.
\newblock
\showISBNx{9781450324724}
\urldef\tempurl%
\url{https://doi.org/10.1145/2509578.2509586}
\showDOI{\tempurl}


\bibitem[Virtanen et~al\mbox{.}(2020)]%
        {2020SciPy-NMeth}
\bibfield{author}{\bibinfo{person}{Pauli Virtanen}, \bibinfo{person}{Ralf Gommers}, \bibinfo{person}{Travis~E. Oliphant}, \bibinfo{person}{Matt Haberland}, \bibinfo{person}{Tyler Reddy}, \bibinfo{person}{David Cournapeau}, \bibinfo{person}{Evgeni Burovski}, \bibinfo{person}{Pearu Peterson}, \bibinfo{person}{Warren Weckesser}, \bibinfo{person}{Jonathan Bright}, \bibinfo{person}{St{\'e}fan~J. {van der Walt}}, \bibinfo{person}{Matthew Brett}, \bibinfo{person}{Joshua Wilson}, \bibinfo{person}{K.~Jarrod Millman}, \bibinfo{person}{Nikolay Mayorov}, \bibinfo{person}{Andrew R.~J. Nelson}, \bibinfo{person}{Eric Jones}, \bibinfo{person}{Robert Kern}, \bibinfo{person}{Eric Larson}, \bibinfo{person}{C~J Carey}, \bibinfo{person}{{\.I}lhan Polat}, \bibinfo{person}{Yu Feng}, \bibinfo{person}{Eric~W. Moore}, \bibinfo{person}{Jake {VanderPlas}}, \bibinfo{person}{Denis Laxalde}, \bibinfo{person}{Josef Perktold}, \bibinfo{person}{Robert Cimrman}, \bibinfo{person}{Ian Henriksen}, \bibinfo{person}{E.~A. Quintero},
  \bibinfo{person}{Charles~R. Harris}, \bibinfo{person}{Anne~M. Archibald}, \bibinfo{person}{Ant{\^o}nio~H. Ribeiro}, \bibinfo{person}{Fabian Pedregosa}, \bibinfo{person}{Paul {van Mulbregt}}, {and} \bibinfo{person}{{SciPy 1.0 Contributors}}.} \bibinfo{year}{2020}\natexlab{}.
\newblock \showarticletitle{{{SciPy} 1.0: Fundamental Algorithms for Scientific Computing in Python}}.
\newblock \bibinfo{journal}{\emph{Nature Methods}}  \bibinfo{volume}{17} (\bibinfo{year}{2020}), \bibinfo{pages}{261--272}.
\newblock
\urldef\tempurl%
\url{https://doi.org/10.1038/s41592-019-0686-2}
\showDOI{\tempurl}


\bibitem[Wu et~al\mbox{.}(2019)]%
        {wu_carpentry_2019}
\bibfield{author}{\bibinfo{person}{Chenming Wu}, \bibinfo{person}{Haisen Zhao}, \bibinfo{person}{Chandrakana Nandi}, \bibinfo{person}{Jeffrey~I. Lipton}, \bibinfo{person}{Zachary Tatlock}, {and} \bibinfo{person}{Adriana Schulz}.} \bibinfo{year}{2019}\natexlab{}.
\newblock \showarticletitle{Carpentry compiler}.
\newblock \bibinfo{journal}{\emph{ACM Transactions on Graphics}} \bibinfo{volume}{38}, \bibinfo{number}{6} (\bibinfo{date}{Dec.} \bibinfo{year}{2019}), \bibinfo{pages}{1--14}.
\newblock
\showISSN{0730-0301, 1557-7368}
\urldef\tempurl%
\url{https://doi.org/10.1145/3355089.3356518}
\showDOI{\tempurl}


\bibitem[Wu and Devendorf(2020)]%
        {wu_unfabricate_2020}
\bibfield{author}{\bibinfo{person}{Shanel Wu} {and} \bibinfo{person}{Laura Devendorf}.} \bibinfo{year}{2020}\natexlab{}.
\newblock \showarticletitle{Unfabricate: {Designing} {Smart} {Textiles} for {Disassembly}}.
\newblock In \bibinfo{booktitle}{\emph{Proceedings of the 2020 {CHI} {Conference} on {Human} {Factors} in {Computing} {Systems}}}. \bibinfo{publisher}{Association for Computing Machinery}, \bibinfo{address}{New York, NY, USA}, \bibinfo{pages}{1--14}.
\newblock
\showISBNx{978-1-4503-6708-0}
\urldef\tempurl%
\url{https://doi.org/10.1145/3313831.3376227}
\showURL{%
\tempurl}


\bibitem[Yan et~al\mbox{.}(2023)]%
        {yan_future_2023}
\bibfield{author}{\bibinfo{person}{Zeyu Yan}, \bibinfo{person}{Tingyu Cheng}, \bibinfo{person}{Jasmine Lu}, \bibinfo{person}{Pedro Lopes}, {and} \bibinfo{person}{Huaishu Peng}.} \bibinfo{year}{2023}\natexlab{}.
\newblock \showarticletitle{Future {Paradigms} for {Sustainable} {Making}}. In \bibinfo{booktitle}{\emph{Adjunct {Proceedings} of the 36th {Annual} {ACM} {Symposium} on {User} {Interface} {Software} and {Technology}}} \emph{(\bibinfo{series}{{UIST} '23 {Adjunct}})}. \bibinfo{publisher}{Association for Computing Machinery}, \bibinfo{address}{New York, NY, USA}, \bibinfo{pages}{1--3}.
\newblock
\showISBNx{9798400700965}
\urldef\tempurl%
\url{https://doi.org/10.1145/3586182.3617433}
\showDOI{\tempurl}


\bibitem[Ye et~al\mbox{.}(2020)]%
        {ye_penrose:_2020}
\bibfield{author}{\bibinfo{person}{Katherine Ye}, \bibinfo{person}{Wode Ni}, \bibinfo{person}{Max Krieger}, \bibinfo{person}{Dor Ma'ayan}, \bibinfo{person}{Jenna Wise}, \bibinfo{person}{Jonathan Aldrich}, \bibinfo{person}{Joshua Sunshine}, {and} \bibinfo{person}{Keenan Crane}.} \bibinfo{year}{2020}\natexlab{}.
\newblock \showarticletitle{Penrose: from mathematical notation to beautiful diagrams}.
\newblock \bibinfo{journal}{\emph{ACM Transactions on Graphics}} \bibinfo{volume}{39}, \bibinfo{number}{4} (\bibinfo{date}{Aug.} \bibinfo{year}{2020}), \bibinfo{pages}{144:144:1--144:144:16}.
\newblock
\showISSN{0730-0301}
\urldef\tempurl%
\url{https://doi.org/10.1145/3386569.3392375}
\showDOI{\tempurl}


\bibitem[Zhao et~al\mbox{.}(2020)]%
        {zhao_robogrammar:_2020}
\bibfield{author}{\bibinfo{person}{Allan Zhao}, \bibinfo{person}{Jie Xu}, \bibinfo{person}{Mina Konaković-Luković}, \bibinfo{person}{Josephine Hughes}, \bibinfo{person}{Andrew Spielberg}, \bibinfo{person}{Daniela Rus}, {and} \bibinfo{person}{Wojciech Matusik}.} \bibinfo{year}{2020}\natexlab{}.
\newblock \showarticletitle{{RoboGrammar}: graph grammar for terrain-optimized robot design}.
\newblock \bibinfo{journal}{\emph{ACM Transactions on Graphics}} \bibinfo{volume}{39}, \bibinfo{number}{6} (\bibinfo{date}{Dec.} \bibinfo{year}{2020}), \bibinfo{pages}{1--16}.
\newblock
\showISSN{0730-0301, 1557-7368}
\urldef\tempurl%
\url{https://doi.org/10.1145/3414685.3417831}
\showDOI{\tempurl}


\bibitem[Zhao et~al\mbox{.}(2022)]%
        {zhao_co-optimization_2022}
\bibfield{author}{\bibinfo{person}{Haisen Zhao}, \bibinfo{person}{Max Willsey}, \bibinfo{person}{Amy Zhu}, \bibinfo{person}{Chandrakana Nandi}, \bibinfo{person}{Zachary Tatlock}, \bibinfo{person}{Justin Solomon}, {and} \bibinfo{person}{Adriana Schulz}.} \bibinfo{year}{2022}\natexlab{}.
\newblock \showarticletitle{Co-{Optimization} of {Design} and {Fabrication} {Plans} for {Carpentry}}.
\newblock \bibinfo{journal}{\emph{ACM Transactions on Graphics}} \bibinfo{volume}{41}, \bibinfo{number}{3} (\bibinfo{date}{March} \bibinfo{year}{2022}), \bibinfo{pages}{32:1--32:13}.
\newblock
\showISSN{0730-0301}
\urldef\tempurl%
\url{https://doi.org/10.1145/3508499}
\showDOI{\tempurl}


\end{thebibliography}

\appendix
\section{Analysis of the Design Space of Home Hacks}\label{appendix:analysis}

The concept of ``home hacking'' covers a variety of topics. To better understand this design space, we first analyzed a collection of hacks and defined our problem domain, which informed our DSL design. Next, we go over research work relevant to home hacks design and existing tools that can be used to model home hacks.

We first gathered over 400 examples of hacks across 17 sources (including DIY blogs, videos, and individual designs from colleagues). After eliminating hacks that are repeated, or essentially the same but used under different scenarios, we selected 48 distinct hacks for further analysis.

\begin{figure*}
\centering
\includegraphics[width=\linewidth]{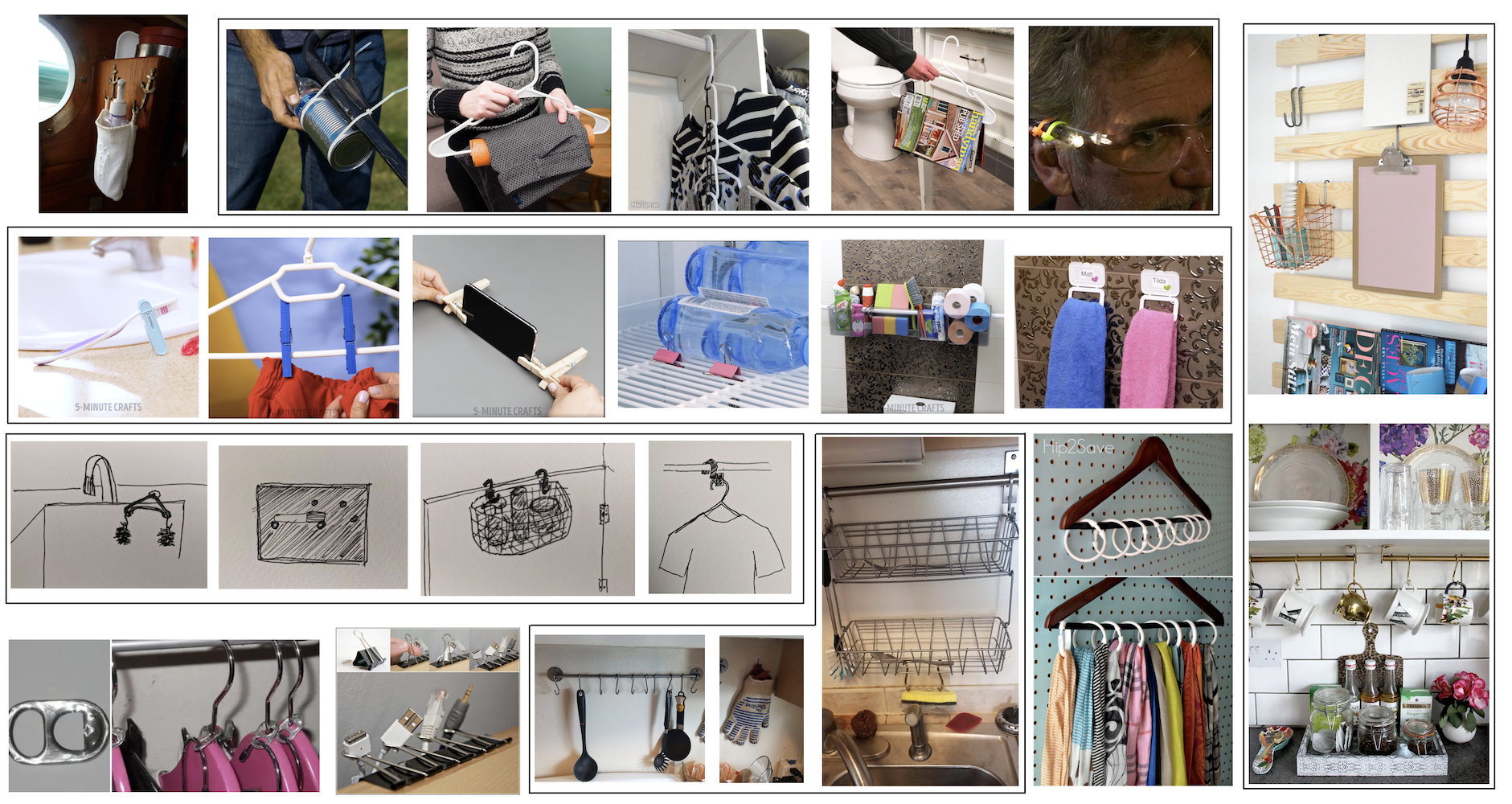}
\caption{As ordered in the image, the hacks or hack groups in black boxes are from~\shortcite{noauthor_karen_nodate, stanley_33_2021, noauthor_5-minute_nodate}, one of the colleagues, ~\shortcite{homehacksofficial_75_2019,noauthor_15_2013}, another one of the colleagues, ~\shortcite{piro_8_2015,noauthor_25_2019}. We provide a short description for each hack and its source in Table~\ref{tab:24hacks}.}
\label{fig:24hacks}
\end{figure*}

Our analysis started with identifying each hack's functionality. This gives us two main categories: hacks that hold or fix some objects in a specific location and orientation -- or \emph{fixtures} (27 out of 48), and hacks which typically make creative reuse of a single item to change the shape or feel (material property) of an existing object to allow for better grasping or easier interaction, e.g., a pool noodle used to organize wires and hide them, cover sharp saw edges, or create padding on furniture corners for a baby-proof environment. In contrast to the latter category, fixture hacks usually involve multiple parts, e.g., several wire baskets chained together with doublehooks to organize items above a kitchen sink. And since their goal is to hold a part at a specific location and orientation relative to its environment, gravity will affect the design's stability. Thus, fixture design can be hard to reason with intuition and is well-suited as a computational design problem.
Since deformable objects are difficult to model and simulate efficiently compared to rigid bodies, and it is unclear how end users can accurately specify manual modifications, we further limit the domain to rigid fixtures with only undeformed constituting objects because they are the majority of fixtures and present a well-scoped subset (24 out of 27).
We show the complete set of 24 rigid (or can be seen as rigid) fixture hacks in Figure~\ref{fig:24hacks}.

\begin{table}[ht]
\centering
\begin{tabular}{c|p{7em}|p{14em}}
No. & Hack Source & Short Description\\
\hline
\hline
1 & \cite{noauthor_karen_nodate} &soap bottle bag \\
\hline
2 & \cite{stanley_33_2021} &bottle holder on mower \\
3&&nonslip hanger\\
4&&hangers chained with rings\\
5&&magazine on a hanger\\
6&&glass light holder\\
\hline 
7&\cite{noauthor_5-minute_nodate} & toothbrush holder \\
8&& pants hanger with clothes clips \\
9&& phone holder from clothes clips \\
10&& binder clips stoppers in fridge\\
11&& bathroom organizer\\
12&& tissue box towel hanger\\
\hline
13&Colleague A & dish sponge hanger\\
14&&pen over pins on a board\\
15&&shower essentials holder\\
16&&hang clothes parallel to wall\\
\hline
17&\cite{homehacksofficial_75_2019} & hangers chained with soda can tabs\\
\hline
18&\cite{noauthor_15_2013} & charger holder \\
\hline
19&Colleague B & kitchen tools rack\\
20&& oven mitten holder\\
21&& wire baskets chained with S-hooks\\
\hline
22&\cite{piro_8_2015} & scarf organizer\\
\hline
23&\cite{noauthor_25_2019}& bed slat as rack\\
24& & cup hanger\\
\end{tabular}
\caption{Hacks and their sources.}
\label{tab:24hacks}
\end{table}

We then analyzed how the individual objects, which we call ``parts'', were connected in the subset of rigid undeformed fixtures. Although many different parts are involved in the hack examples, connections typically form between common types of \emph{connector primitives}. These connector primitives connect in ways that are independent of the objects that they are part of. From our analysis, we extract the following categories of connector primitives (Figure~\ref{fig:hacks_primitives}): \emph{rod}, \emph{hook}, \emph{hole}, \emph{tube}, \emph{clip}, \emph{edge}, \emph{surface}, and \emph{hemisphere}.

\begin{table}[ht]
\centering
\begin{tabular}{ c | c c c c c c c c }
 & hook & hole & hemi. & edge & rod & tube & clip & surf. \\
 \hline
 hook & 1 & 5 & \cellcolor[HTML]{C0C0C0}&\cellcolor[HTML]{C0C0C0} & 12 & 1 & \cellcolor[HTML]{C0C0C0}& \cellcolor[HTML]{C0C0C0}\\
 hole & \cellcolor[HTML]{5A5A5A} & \cellcolor[HTML]{C0C0C0} &\cellcolor[HTML]{C0C0C0} & \cellcolor[HTML]{C0C0C0}& 4 & * & \cellcolor[HTML]{C0C0C0}& \cellcolor[HTML]{C0C0C0}\\
 hemi. & \cellcolor[HTML]{5A5A5A}& \cellcolor[HTML]{5A5A5A}& \cellcolor[HTML]{C0C0C0} & \cellcolor[HTML]{C0C0C0}&\cellcolor[HTML]{C0C0C0} & \cellcolor[HTML]{C0C0C0}&\cellcolor[HTML]{C0C0C0} & 2\\
 edge & \cellcolor[HTML]{5A5A5A}& \cellcolor[HTML]{5A5A5A}& \cellcolor[HTML]{5A5A5A}& \cellcolor[HTML]{C0C0C0}& \cellcolor[HTML]{C0C0C0}& \cellcolor[HTML]{C0C0C0}& 3 & \cellcolor[HTML]{C0C0C0}\\
 rod & \cellcolor[HTML]{5A5A5A}& \cellcolor[HTML]{5A5A5A}& \cellcolor[HTML]{5A5A5A}& \cellcolor[HTML]{5A5A5A}& \cellcolor[HTML]{C0C0C0} & 2 & 5 & \cellcolor[HTML]{C0C0C0}\\
 tube & \cellcolor[HTML]{5A5A5A}& \cellcolor[HTML]{5A5A5A}& \cellcolor[HTML]{5A5A5A}& \cellcolor[HTML]{5A5A5A}& \cellcolor[HTML]{5A5A5A}& \# & * & \cellcolor[HTML]{C0C0C0}\\
 clip & \cellcolor[HTML]{5A5A5A}& \cellcolor[HTML]{5A5A5A}& \cellcolor[HTML]{5A5A5A}& \cellcolor[HTML]{5A5A5A}& \cellcolor[HTML]{5A5A5A}& \cellcolor[HTML]{5A5A5A}& \cellcolor[HTML]{C0C0C0}& \cellcolor[HTML]{C0C0C0}\\
 surf. & \cellcolor[HTML]{5A5A5A}& \cellcolor[HTML]{5A5A5A}& \cellcolor[HTML]{5A5A5A}& \cellcolor[HTML]{5A5A5A}& \cellcolor[HTML]{5A5A5A}& \cellcolor[HTML]{5A5A5A}& \cellcolor[HTML]{5A5A5A}& 7
\end{tabular}
\caption{This table shows the number of hacks each connection type appeared in from the 24 rigid undeformed fixture hacks (Figure~\ref{fig:24hacks}). * means this connection type didn't appear but we deduced that it is compatible based on similar connection types. \# means this connection type didn't appear in rigid fixture hacks but appeared in a non-rigid fixture hack. Light grey means this connection type did not appear or cannot be deduced. We ignore the lower-triangular region (dark grey) as it is redundant with the upper-triangular region.}
\label{tab:connectors}
\end{table}

Further analyzing the hack designs by looking at the parts and the connecting shapes of the parts, we determined the relationship between these connector primitives on whether they connect and how they align with each other when they connect. We summarized the common design patterns of how objects can be connected in Table~\ref{tab:connectors}. The eight connector primitives and the relationships between them become the basis for the design of the DSL and the underlying logic for assembly building in our system, which is introduced in Section~\ref{sec:DSL}.

\section{Programs for Gallery Examples}\label{appendix:programs}

\subsection{Toothbrush Holder}
\begin{tiny}
\begin{verbatim}
ASSEMBLY_toothbrush_holder = Assembly()

surface = Surface({"width": 400, "length": 400})
ENV_start = Environment({"surface": surface})
ENV_end = Environment({"toothbrush": Toothbrush()})

start_frame = Frame()
ASSEMBLY_toothbrush_holder.add(ENV_start, start_frame)

PART_clip = PlasticClip()
ASSEMBLY_toothbrush_holder.connect(PART_clip.hemisphere1, ENV_start.surface)
ASSEMBLY_toothbrush_holder.connect(PART_clip.hemisphere2, ENV_start.surface)
ASSEMBLY_toothbrush_holder.connect(ENV_end.rod, PART_clip.clip)
ASSEMBLY_toothbrush_holder.connect(ENV_end.hemisphere, ENV_start.surface)

end_frame = Frame([-10, -62.5, 50], [-65,0,0])
ASSEMBLY_toothbrush_holder.end_with(ENV_end, end_frame)
\end{verbatim}
\end{tiny}

\subsection{Charger Holder}
\begin{tiny}
\begin{verbatim}
ASSEMBLY_cable_holder = Assembly()

edge = Edge({"width": 100, "length": 200, "height": 1.5})
ENV_start = Environment({"edge": edge})
ENV_end = Environment({"cable": Cable()})

start_frame = Frame([0,0,150],[0,0,0])
ASSEMBLY_cable_holder.add(ENV_start.edge, start_frame)

PART_binderclip = BinderClip()
ASSEMBLY_cable_holder.connect(PART_binderclip.clip, ENV_start.edge, is_fixed=True)
ASSEMBLY_cable_holder.connect(ENV_end.rod1, PART_binderclip.hole1)
ASSEMBLY_cable_holder.connect(ENV_end.rod1, PART_binderclip.hole2)

end_frame = Frame([0,57.5,163], [0,0,0])
ASSEMBLY_cable_holder.end_with(ENV_end.rod2, end_frame)
\end{verbatim}
\end{tiny}

\subsection{Soap Bottle Holder}
\begin{tiny}
\begin{verbatim}
ASSEMBLY_soapbottle_holder = Assembly()

rod = Rod({"length": 500, "radius": 5})
ENV_start = Environment({"door": rod})
ENV_end = Environment({"soapbottle": SoapBottle()})

start_frame = Frame([0,0,500], [90,0,90])
ASSEMBLY_soapbottle_holder.add(ENV_start.door, start_frame)

PART_hookeye1 = HookEyeLeftS()
ASSEMBLY_soapbottle_holder.connect(PART_hookeye1.hole, ENV_start.door)
PART_basket = Basket()
ASSEMBLY_soapbottle_holder.connect(PART_basket.rod1, PART_hookeye1.hook)
PART_hookeye2 = HookEyeLeftS()
ASSEMBLY_soapbottle_holder.connect(PART_hookeye2.hole, ENV_start.door, alignment="flip")
ASSEMBLY_soapbottle_holder.connect(PART_hookeye2.hook, PART_basket.rod2)
ASSEMBLY_soapbottle_holder.connect(ENV_end.surface, PART_basket.surface)

end_frame = Frame([0,0,253], [0,0,180])
ASSEMBLY_soapbottle_holder.end_with(ENV_end, end_frame)
\end{verbatim}
\end{tiny}

\subsection{Mug Hanger}
\begin{tiny}
\begin{verbatim}
ASSEMBLY_mug_hanger = Assembly()

rod = Rod({"length": 500, "radius": 2})
ENV_start = Environment({"rod": rod})
surface = Surface({"length": 800, "width": 600})
ENV_wall = Environment({"wall": surface})
ENV_end = Environment({"mug": Mug()})

start_frame = Frame([0,0,200], [90,0,90])
ASSEMBLY_mug_hanger.add(ENV_start.rod, start_frame)
wall_frame = Frame([0,50,0], [90,0,0])
ASSEMBLY_mug_hanger.add(ENV_wall.wall, wall_frame)

PART_doublehook1 = DoubleHook()
PART_doublehook2 = DoubleHook()
PART_doublehook3 = DoubleHook()
ASSEMBLY_mug_hanger.connect(PART_doublehook1.hook2, ENV_start.rod)
ASSEMBLY_mug_hanger.connect(PART_doublehook2.hook2, PART_doublehook1.hook1)
ASSEMBLY_mug_hanger.connect(PART_doublehook3.hook1, PART_doublehook2.hook1)
ASSEMBLY_mug_hanger.connect(ENV_end.hook, PART_doublehook3.hook2)

end_frame = Frame([0,0,50], [-35,0,-90])
ASSEMBLY_mug_hanger.end_with(ENV_end.hook, end_frame)
\end{verbatim}
\end{tiny}

\subsection{Paper Towel Holder}
\begin{tiny}
\begin{verbatim}
ASSEMBLY_paper_towel_holder = Assembly()

ENV_start = Environment({"env": TowelHangingEnv()})
ENV_end = Environment({"paper_towel_roll": PaperTowelRoll()})

wall_frame = Frame([0,0,300], [0,0,0])
ASSEMBLY_paper_towel_holder.add(ENV_start, wall_frame)

PART_hookeye1 = HookEyeLeft()
PART_hookeye2 = HookEyeLeft()
PART_broomrod = BroomRod()
ASSEMBLY_paper_towel_holder.connect(PART_hookeye1.hole, ENV_start.hook1)
ASSEMBLY_paper_towel_holder.connect(PART_hookeye2.hole, ENV_start.hook2)
ASSEMBLY_paper_towel_holder.connect(PART_broomrod.tube, PART_hookeye1.hook, is_fixed=True)
ASSEMBLY_paper_towel_holder.connect(ENV_end.tube, PART_broomrod.tube)
ASSEMBLY_paper_towel_holder.connect(PART_hookeye2.hook, PART_broomrod.tube, is_fixed=True)

end_frame = Frame([53,0,160], [-90,-60,0])
ASSEMBLY_paper_towel_holder.end_with(ENV_end.tube, end_frame)
\end{verbatim}
\end{tiny}

\subsection{Diaper Caddy}
\begin{tiny}
\begin{verbatim}
ASSEMBLY_diaper_caddy = Assembly()

ENV_start = Environment({"backseat": BackSeats()})
ENV_end = Environment({"diaper_caddy": DiaperCaddy()})

start_frame = Frame([0,0,0], [0,0,0])
ASSEMBLY_diaper_caddy.add(ENV_start, start_frame)

PART_doublehook1 = DoubleHook()
PART_doublehook2 = DoubleHook()
PART_doublehook3 = DoubleHook()
PART_doublehook4 = DoubleHook()
ASSEMBLY_diaper_caddy.connect(PART_doublehook1.hook1, ENV_start.rod1)
ASSEMBLY_diaper_caddy.connect(PART_doublehook2.hook2, ENV_start.rod2)
ASSEMBLY_diaper_caddy.connect(PART_doublehook3.hook1, PART_doublehook1.hook2)
ASSEMBLY_diaper_caddy.connect(PART_doublehook4.hook1, PART_doublehook2.hook1)
ASSEMBLY_diaper_caddy.connect(ENV_end.hook2, PART_doublehook3.hook2)
ASSEMBLY_diaper_caddy.connect(ENV_end.hook1, PART_doublehook4.hook2)

end_frame = Frame([124.3,580,717.1], [-135.5,-40,20.5])
ASSEMBLY_diaper_caddy.end_with(ENV_end.hook2, end_frame)
\end{verbatim}
\end{tiny}

\section{Additional Image Credits}
We used the following online images for three figures in this work:
\begin{enumerate}
\item Figure~\ref{fig:teaser}: hula hoop image from \href{https://www.walmart.com/ip/kids-hula-hoop-body-building-plastic-children-gymnastics-toys-diameter-25-inch-blue/286284542}{Walmart}.
\item First inset figure in Section~\ref{sec:intro}: Image by \href{https://pixabay.com/users/peggy_marco-1553824/?utm_source=link-attribution&utm_medium=referral&utm_campaign=image&utm_content=1019808}{Peggy und Marco Lachmann-Anke} from \href{https://pixabay.com//?utm_source=link-attribution&utm_medium=referral&utm_campaign=image&utm_content=1019808}{Pixabay}, \href{https://www.freepik.com/free-vector/realistic-vector-icon-set-dark-dress-cupboard-with-two-doors-open-closed-isolated-white_43092619.htm#page=2&query=open%20closet&position=30&from_view=search&track=ais&uuid=a68e94f8-b1c1-4ac8-8906-f346dfc47038}{Image} by user15245033 on Freepik, \href{https://www.freepik.com/free-vector/illustration-shopping-online_2606536.htm#query=basket&position=36&from_view=search&track=ais&uuid=38425ea8-e4b3-4945-b42e-d9e6ad19b01a}{Image} by rawpixel.com on Freepik, \href{https://www.freepik.com/free-psd/basket-isolated-transparent-background_81702435.htm#page=3&query=round%20handle%20basket%20empty%20cartoon&position=31&from_view=search&track=ais&uuid=859be015-3a18-4f58-bfe1-5496e7ed42b5}{Image} by tohamina on Freepik, binder clip image from \href{https://media.officedepot.com/images/f_auto,q_auto,e_sharpen,h_450/products/561339/561339}{Office Depot}.
\item Inset Figure~\ref{fig:rodhook}: middle from \href{https://www.okfarmhousedecor.com/product/idesign-curved-metal-shower-curtain-rod-adjustable-customizable-curtain-rod-for-bathtub-stall-closet-doorway-41-72-inches-matte-black79077/}{okfarmhousedecor}, right from \href{https://www.architectureartdesigns.com/wp-content/uploads/2014/11/62.jpg}{architectureartdesigns}.
\end{enumerate}

\end{document}